\documentclass{article}

\usepackage{emulateapj}
\usepackage{onecolfloat}
\usepackage{graphicx}

\def\slantfrac#1#2{\hbox{$\,^#1\!/_#2$}}

\def\sqarcsec{arcsec$^2$}
\def\ion#1#2{\rm{#1}{\sc{#2}}\relax}

\def\tto#1{$\times10^{#1}$}
\def\rms{{\it rms}}

\def\escsa{ergs s$^{-1}$ cm$^{-2}$ sr$^{-1}$ \AA$^{-1}$}

\def\esca{ergs s$^{-1}$ cm$^{-2}$ \AA$^{-1}$}

\def\tto#1{$\times10^{#1}$}

\def\micron{\hbox{$\mu$m}}

\def\farcm{\hbox{.\kern -0.7ex\raisebox{.9ex}{\scriptsize$\prime$}}}
\def\farcs{\hbox{\kern 0.13ex.\kern -0.95ex%
  \raisebox{.9ex}{\scriptsize$\prime\prime$}\kern -0.1ex}}

\def\spose#1{\hbox to 0pt{#1\hss}}
\def\lesssim{\mathrel{\hbox{\rlap{\hbox{\lower4pt\hbox{$\sim$}}}\hbox{$<$}}}}
\def\gtrsim{\mathrel{\hbox{\rlap{\hbox{\lower4pt\hbox{$\sim$}}}\hbox{$>$}}}}
\def\lta{\mathrel{\spose{\lower 3pt\hbox{$\mathchar"218$}}
     \raise 2.0pt\hbox{$\mathchar"13C$}}}
\def\gta{\mathrel{\spose{\lower 3pt\hbox{$\mathchar"218$}}
     \raise 2.0pt\hbox{$\mathchar"13E$}}}
\def\etal{{\it et\thinspace al.\ }}

\lefthead{Bernstein, Freedman, \& Madore}
\righthead{Detections of the Optical EBL: Measurements of Zodiacal Light}

\submitted{Submitted: 13 June 2000 ; 
Revised:  2 November 2001}

\begin{document}
\twocolumn[

\title{	The First Detections of the Extragalactic Background
	Light at 3000, 5500, and 8000\AA\ (II): 
        Measurement of Foreground Zodiacal Light }

\author{Rebecca A. Bernstein\altaffilmark{1,2,3}}
\author{Wendy L. Freedman\altaffilmark{2}}
\author{Barry F. Madore\altaffilmark{2,4}}
 
\affil{\footnotesize 1) Division of Math, Physics, and Astronomy,
		California Institute of Technology,
                Pasadena, CA 91125}
\affil{\footnotesize 2) Carnegie Observatories,
                813 Santa Barbara St, 
                Pasadena, CA 91101}
\affil{\footnotesize 3) rab@ociw.edu, Hubble Fellow}
\affil{\footnotesize 4) NASA/IPAC Extragalactic Database, 
                California Institute of Technology, 
                Pasadena, CA 91125}

\setcounter{footnote}{0}

\begin{abstract}

We present a measurement of the absolute surface brightness of the
zodiacal light (3900--5100\AA) toward a fixed extragalactic target at
high ecliptic latitude based on moderate resolution ($\sim$1.3\AA\ per
pixel) spectrophotometry obtained with the du Pont 2.5m telescope at
Las Campanas Observatory in Chile.  This measurement and
contemporaneous Hubble Space Telescope data from WFPC2 and FOS
comprise a coordinated program to measure the mean flux of the diffuse
extragalactic background light (EBL). The zodiacal light at optical
wavelengths results from scattering by interplanetary dust, so that
the zodiacal light flux toward any extragalactic target varies
seasonally with the position of the Earth. This measurement of
zodiacal light is therefore relevant to the specific observations
(date and target field) under discussion. To obtain this result, we
have developed a technique that uses the strength of the zodiacal
Fraunhofer lines to identify the absolute flux of the zodiacal light
in the multiple--component night sky spectrum.  Statistical
uncertainties in the result are 0.6\% ($1\sigma$).  However, the
dominant source of uncertainty is systematic errors, which we estimate
to be 1.1\% ($1\sigma$).  We discuss the contributions included in
this estimate explicitly.  The systematic errors in this result
contribute 25\% in quadrature to the final error in our coordinated
EBL measurement, which is presented in the first paper of this series.



\end{abstract}

]

\setcounter{footnote}{0}


\section{Introduction} \label{intro}

This is the second in a series of three papers in  which we present a
measurement of the mean flux of the optical extragalactic background
light and the cosmological implications of that result (see Bernstein,
Freedman, \& Madore 2002a \& 2002c).  The extragalactic background
light (EBL) is the spatially averaged surface brightness of all
extragalactic sources, resolved and unresolved.  As such, the absolute
flux of the EBL is a powerful and fundamental cosmological constant
which can significantly constrain galaxy formation and evolution
scenarios. Like all diffuse backgrounds, however, the optical EBL is
very difficult to isolate from foreground sources, which are two
orders of magnitude brighter.  At high Galactic and ecliptic latitudes
($>30^\circ$), the sky flux observed from the ground is dominated by
terrestrial airglow and zodiacal light (ZL), each with a surface
brightness of $\sim 23$ AB mag arcsec$^{-2}$. The Hubble Space
Telescope (HST), which orbits at an altitude of 600 km, avoids
atmospheric emission, but the total sky flux is still dominated by
ZL. An accurate measurement of the ZL is therefore crucial to a
successful detection of the diffuse EBL.

Our measurement of the EBL involves simultaneous HST and ground--based
observations. From HST we measure the total flux of the night sky,
including ZL.  Using spectrophotometry over the range 3860--5150\AA\
(1.25\AA\ per pixel) taken with the Boller \& Chivens Spectrograph on
the duPont 2.5m Telescope at Las Campanas Observatory in Chile, we
measure the absolute flux of the ZL contributing to the HST
observations, which we can then subtract.  In Bernstein, Freedman, \&
Madore (2002a, henceforth Paper I), we present the full details of the
coordinated program to measure the EBL.  In this paper, we present the
ground--based measurement of the absolute flux of the ZL.  As
calibration of these data is crucial to the scientific goals, the data
acquisition, reduction, and flux calibration are discussed here in detail.

Background regarding the nature of the ZL is given in \S\ref{backg}.
The observations, data reduction, and flux calibration are discussed
in \S\ref{lco.obser}. In \S\ref{nightsky}, we briefly described the
complications which arise due to atmospheric scattering, which
redirects off--axis flux into and on--axis flux out of the line of
sight.  Detailed calculations of the atmospheric scattering relevant
to precisely our observing situation (defined by the observatory
location and positions of the Sun, Galaxy, and target relative to
eachother and the horizon) are relegated to the Appendix, and
summarized in \S\ref{nightsky}.  In \S\ref{lco.analy}, we describe the
technique used to measure the zodiacal light flux in reduced
spectra. The results are summarized in \S\ref{lco.resul}.

\section{Background}\label{backg}

\begin{figure}[t] 
\begin{center}
\includegraphics[width=1.75in,angle=-90]{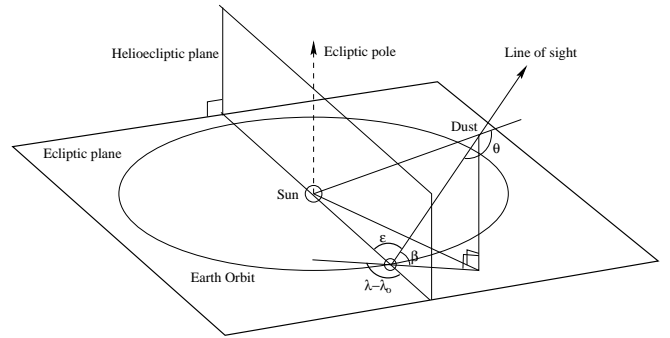}
\caption{\footnotesize The geometry of zodiacal light scattering
(adapted from Matsuura, Matsumoto, \& Matsuhara 1995).  The viewing
line of sight is defined by ecliptic latitude ($\beta$), and
geocentric ecliptic longitude, ($\lambda-\lambda_\odot$), or
alternatively, by $\beta$ and the elongation angle, $\epsilon$, which
is defined as $\cos\epsilon=\cos(\lambda-\lambda_\odot) \cos\beta$.
The scattering angle is $\theta$. }
\label{fig:zodgeom}
\end{center}
\end{figure}

Zodiacal light (ZL) is sunlight scattered off of dust grains in the
solar system. Dust column densities are largest toward the ecliptic
plane, causing the highest ZL intensities there.  The scattering
geometry is illustrated in Figure \ref{fig:zodgeom} for a high
latitude field viewed near the anti-solar direction. The figure shows
the scattering angle and line of sight as defined by geocentric
ecliptic longitude ($\lambda-\lambda_\odot$, in which $\lambda$ is
ecliptic longitude) and ecliptic latitude ($\beta$).  At ecliptic
latitudes greater than 30 degrees and large scattering angles, the
zodiacal light can be as faint as $\sim23$ mag arcsec$^{-2}$ at
5000\AA\ ($\sim$1\tto{-7} \escsa) with a solar--type spectrum. In the
ecliptic plane and near the Sun, the ZL can be as much as 20 times
brighter than at high ecliptic latitudes and significantly reddened
relative to the incident solar spectrum.  The positional dependence of
the ZL flux and color is a function of the density and composition of
the scattering particles and of scattering geometry such that the
zodiacal light is redder and brighter at smaller elongation angles
(see Leinert \etal 1998). In general, the solar spectrum is preserved
in the spectrum of the ZL with less than 30\% deviation in the
broad--band spectral shape from the UV to the near--IR ($0.2-2\mu$m).

Such weakly wavelength--dependent scattering can be characterized by
simple Mie scattering theory, which generally describes the
interaction of photons with solid particles larger than the wavelength
of the incident light (R\"oser \& Staude 1978).  The interplanetary
dust (IPD) cloud at the orientations of interest to us is known to be
composed predominantly of particles larger than 10\micron\ with
moderate surface roughness and layered composition. These conclusions
are based on IR observations of the thermal properties of the IPD,
dynamical arguments, and laboratory work on dust captured in the
upper atmosphere and on the moon (see Reach \etal 1996; Berriman \etal
 1994, Dermott \etal 1996; Brownlee 1978; Fechtig \etal 1974;
Leinert \etal 1998 and references therein).  In the absence of free
electrons, the cross--sections for non-linear scattering, such as
Raman scattering or two--photon processes, are too low to be a
significant effect for dust particles with large dielectric constants.

Indeed, Mie scattering models for rough particles with the size and
composition of the IPD successfully describe the weakly
wavelength--dependent scattering characteristic of the ZL: the
incident spectrum is slightly reddened over very broad band--passes,
while narrow--band spectral features are preserved to high precision
(Weiss--Wrana 1983, Schiffer 1985).  Both these narrow-- and
broad--band characteristics have been empirically demonstrated, as the
ZL spectrum is seen to be roughly 5-10\% redder per 1000\AA\ over the
range from 3000-8000\AA\ than the Sun (see Leinert \etal 1998 and
references therein), and Beggs \etal (1964) have shown that the
Fraunhofer lines in the ZL show no measurable deviation from their
solar equivalent widths to the accuracy of their calibration (2\%).
The broad--band reddening is characterized as the color of the ZL
relative to the solar spectrum as a function of wavelength, 
\begin{equation}
C(\lambda,\lambda_0) = \frac{I_{ZL}(\lambda)/ I_\odot(\lambda)}
{I_{ZL}(\lambda_0)/I_\odot(\lambda_0)}.
\label{eq:color}
\end{equation}

While Mie scattering explains the spectrum and intensity of the ZL in
general, the exact flux of the ZL cannot be modeled to the accuracy
we require here, simply due to uncertainties in the exact composition
and column density of the IPD.  Empirically, the 
surface brightness of the ZL is known to 
roughly 10\% accuracy as a function of scattering geometry
and ecliptic latitude alone.  However, uncertainty in flux
calibrations of those surface brightness measurements and spatial
variability in the IPD cloud (such as cometary trails) preclude more
accurate predictions (Levasseur--Regourd \& Dumont 1980; Richter \etal
 1982; Leinert \etal 1998 and references therein).  In addition,
most published measurements of the ZL flux include the background EBL
in the reported surface brightness of the ZL.  In order to detect the
EBL, we must explicitly measure the zodiacal contribution to the sky
surface brightness along a specific line of sight  at a particular
wavelength. 

In order to uniquely define the ZL spectrum, both the mean flux in a
narrow--bandpass and the broad--band color over the wavelength range
of interest must be measured for the line of sight in question.  We
can do so by making use of the fact that the equivalent widths of
solar Fraunhofer lines, reproduced in the ZL spectrum, are known to
very high accuracy: we can then determine the continuum level
(mean flux) of the ZL at a given wavelength by measuring the apparent
equivalent width of the Fraunhofer in spectrum of the night sky.  For
example, if the equivalent width of the Fraunhofer lines in the night
sky around 4500\AA\ are only one half the strength of the same
absorption features in the solar spectrum, then the ZL contributes
only 50\% of the night sky flux at that wavelength.  A practical
complication of this approach is, of course, identifying the continuum
level for accurate measurement of the equivalent widths.  Because the
ZL and atmospheric emission (airglow) contribute almost
equally to the night sky flux in the visible wavelength range at the
ecliptic orientation of our observations, this becomes prohibitively
difficult where airglow lines become strong (above 5500\AA).  The
problem is further complicated by the fact that the airglow spectrum
is continually changing, both in the strength of particular lines and
in its mean, wide--band flux.  In order to accomplish this
measurement, we have therefore developed a new technique for
deconvolving one known signal from a variable, multicomponent
spectrum.  The details of this technique are discussed in
\S\ref{lco.analy}.

\section{Observations and Data Reduction}\label{lco.obser}

\begin{figure}[t]
\begin{center}
\includegraphics[width=3in,angle=0]{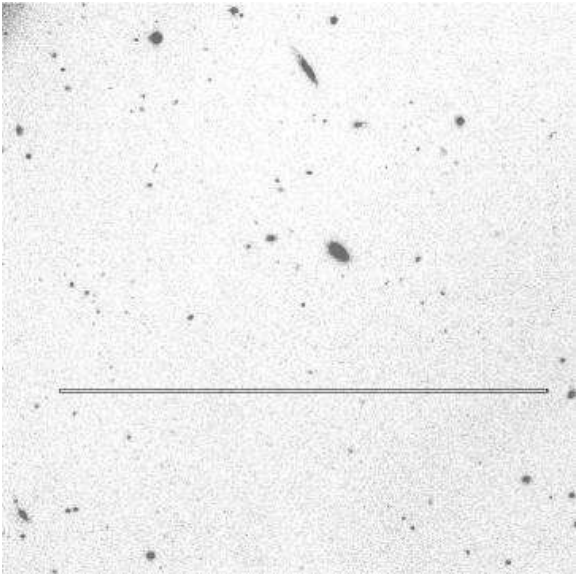}
\caption{\footnotesize 
 The slit position during program observations.  The image is
a $4\times4$ arcmin$^2$, $r$-band exposure taken the 2.5m duPont
telescope at Las Campanas Observatory.  WFPC2/HST observations were
taken in parallel with the spectroscopy presented here for the purpose
of measuring the EBL.  For comparison, the WFPC2/HST field of view is
also contained within this image, located towards the lower left
quadrant (see Figure 7, Paper I).
}
\label{fig:lco.slitpic}
\end{center}
\end{figure}

We obtained long--slit spectra of the night sky using the Boller \&
Chivens Spectrograph on the 2.5m duPont Telescope at Las Campanas
Observatory (LCO) on the nights of 1995 November 26--29.  The first
night of the run was lost to clouds.  On the third night, a
thermal short in the dewar caused poor temperature regulation of the
CCD which resulted in low and variable charge transfer efficiency (see
\S\ref{lco.cte}).  The data from these two nights were therefore
dropped from further analysis. The dewar was repaired for the
remaining nights of the run.   As discussed further in
\S\ref{lco.dark} and \S\ref{lco.fluxc}, both detector performance and
photometric conditions were stable to better than 1\% on the two
useful nights of the run.

As described in Paper I, these spectra were taken within the field of
view of our HST/WFPC2 observations, which executed in consecutive
orbits on 1995 November 27--28.  Because the HST and LCO observations
executed simultaneously and along exactly the same pointing (see Paper
I),  the ZL contribution to both data sets is identical.
The slit position is shown overlayed on a $r$-band image of the field
in Figure \ref{fig:lco.slitpic}.  The exact coordinates of the
spectroscopic observations were selected from ground--based imaging to
avoid objects brighter than $r=26$ mag arcsec$^{-2}$.  This was done
for convenience in the data reduction; extragalactic sources need not
be rigorously avoided, as spectral lines from extragalactic sources
fainter than $r\sim24$ AB mag will be significantly redshifted, on
average, and should not align with solar features. In addition, such
objects will have low enough surface number density ($<0.2$
arcsec$^{-2}$) that they will not significantly impact the average
extracted spectrum.

We used a 600 l/mm grating to obtain spectra over the wavelength range
3860--5150\AA\ with a dispersion of $\sim1.3$\AA\ per pixel.  The
wavelength range was chosen to include a maximum number of strong
Fraunhofer lines while avoiding strong airglow features.  The strong
\ion{Mg}{i} Fraunhofer lines near $\sim5170$\AA\ were identified in an
earlier run to be affected by rapidly variable airglow features and
were therefore excluded from our 1995 spectral coverage.  Ca H \& K
were included in our observations at the blue end.  A slit--width of
$\sim1.5$\,arcsec (see \S\ref{lco.solid}) produced roughly 2.6\AA\
resolution in the program observations.  Even though the ZL has a
surface brightness of $\mu_V\sim$23.2 mag arcsec$^{-2}$, we obtained a
signal--to--noise ratio of $\sim 40$ per spectral resolution element
from a single, 30 minute exposure by integrating over the total slit
surface area ($\sim$300\,arcsec$^2$).  To minimize read-noise, we
binned the data on--chip by four pixels in the spatial direction and
averaged over the full 3.4 arcmin spatial extent of the slit in the
data reduction.

\begin{figure*}[t]
\begin{center}
\includegraphics[width=5in,angle=0]{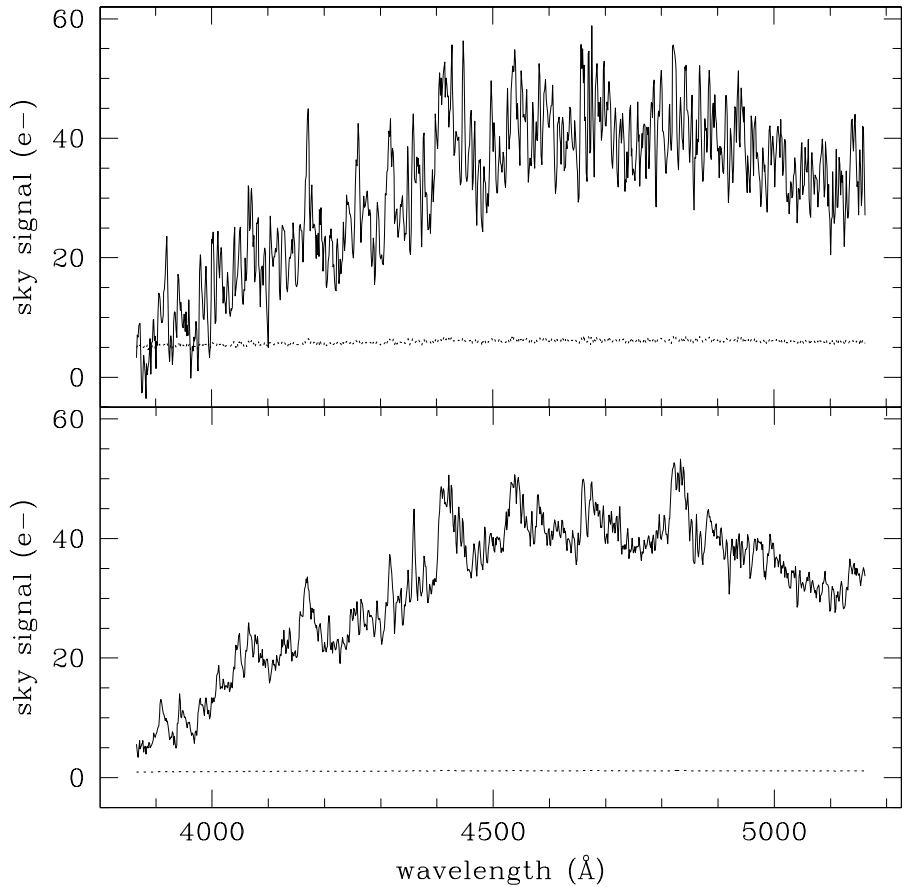}
\caption{\footnotesize 
Signal-to-noise of program data.
The upper plot shows the wavelength--calibrated spectrum in
electrons per pixel obtained from one spatial resolution element
(column) of an 1800 second exposure of the night sky.  The lower plot
shows the spectrum produced by averaging 87 columns.  The
corresponding error spectra, both read--noise dominated, are shown
with a dotted line at the bottom of each plot.
}
\label{fig:lco.rawspec}
\end{center}
\end{figure*}

Spectrophotometric standards were observed through a slit 10.8\,arcsec
wide.  The detector we used is a thinned Tektronics/SITe CCD with
$1024\times1024\times24$\micron\ pixels, 1.15e$^-$/DN gain, and
8.6e$^-$ read--noise in the $4\times1$ binned configuration.  The
quantum efficiency of the CCD is near 50\% over the spectral range of
these observations; however, the spectrograph throughput drops by a
factor of two between 5000\AA\ and 3800\AA, as can be seen from the
sensitivity curve plotted in Figure \ref{fig:lco.sensfunc}.  The upper
panel of Figure \ref{fig:lco.rawspec} shows the wavelength calibrated
spectrum obtained from one spatial resolution element (1 column) of
one of the $4\times1$ binned exposures; the lower panel shows the
averaged spectrum from the full image (87 columns).  Above 4100\AA, the
count rate from sky is more than twice the dark rate.  The maximum
error is 10\% per resolution element at the blue end of the spectra,
simply due to the low count--rate at bluer wavelengths.  Above
4100\AA, the error per resolution element is roughly 1\%.

Final calibration errors are summarized in Table
\ref{tab:lco.errorbudget}. The data reduction steps 
are described below in the order which they were performed.

\begin{deluxetable}{l c c}
\tablewidth{30pc}
\tablecaption{Error Budget for Zodiacal Light Flux
       \label{tab:lco.errorbudget}}
\tablehead{
\colhead{Step}  & \colhead{Statistical Uncertainty}     & \colhead{Systematic Uncertainty}}
\startdata
Bias level removal (\S\ref{lco.bias})                   &  	$<0.01$\%	& $\cdots$ \nl
Dark current removal (\S\ref{lco.dark})                 &        $<0.01$\%	& $\cdots$ \nl
Pixel--to--pixel flat--fielding (\S\ref{lco.flatf})     &       $<0.01$\%	& $\cdots$ \nl
Slit illumination (\S\ref{lco.flatf})                   &        $\cdots$	& $<0.1$\% \nl
Point source flux cal. (\S\ref{lco.ptsrc}) 		&       $\cdots$	& 0.6\% \nl
Aperture correction (\S\ref{lco.apcor})                 &        $\cdots$	& 0.2\% \nl
Solid angle (\S\ref{lco.solid})                         &	$\cdots$  	& 0.6\%  \nl
$I_{\rm scat}({\rm ZL})$ estimate (\S\ref{append.zl}) 	&      $\cdots$ 	& 1.2\% \nl
$I_{\rm scat}({\rm ISL})$ estimate (\S\ref{append.isl}) &   	$\cdots$	& 0.5\%   \nl
{\it rms} scatter in ZL correlation 
		(\S\ref{lco.analy},\S\ref{lco.resul}) 	&         0.6\%		& $\cdots$ \nl
\cline{2-3} \hfil & \hfil & \hfil \nl
Combined (1$\sigma$)\tablenotemark{a}                   &        0.6\%          & 1.1\% \nl
\enddata
\tablenotetext{a}{\footnotesize Statistical errors have been combined in
quadrature to obtain a cumulative, one-sigma error. 
Systematic errors have been combined assuming a flat 
probability distribution for each contributing source of error.
The resulting systematic error is roughly Gaussian distributed,
and the quoted value is the 68\% confidence interval. For a detailed
discussion see Paper I.}
\end{deluxetable}

\subsection{Detector Linearity}\label{lco.cte}

Because our observations of the ZL have total count levels in the
range 20--50 DN pixel$^{-1}$, and the standard star observations have close
to 5000 DN pixel$^{-1}$, it is crucial to verify that the CCD is linear over
this broad range.  In these data, 16 rows are read off beyond the
physical extent of the chip, averaged together, and recorded as a bias
row, which can be used as an accurate diagnostic of the charge
transfer efficiency (CTE) of the CCD.  Because the slit only
illuminates the central third of the chip, any residual charge which
is not passed through the parallel gates (due to low CTE) will appear
as a jump in the charge level of the bias row at the boundary between
the exposed and unexposed regions of the chip.  It is evident from
this diagnostic that the temperature regulation of the chip became
erratic during the third night of the run, causing an increase in the
spurious charge and causing the CTE to drop to an unacceptable level
($\sim 99.995$\% per transfer, or 95.0\% over 1024 rows).  The data
from this night were excluded from the analysis because charge shared
between rows due to incomplete charge transfer will affect the
apparent strength of spectral features.  On the nights during which
the temperature of the CCD remained stable, the mean level in the bias
row was not detectably higher in columns illuminated by the
slit than in those which were not illuminated. This was true
even for dome flats images, in which the mean level of the illuminated
columns was $\sim$7000\,DN pixel$^{-1}$.  In
addition, our final results are based on data imaged in rows 1--700 of
the CCD.  The data which contribute to our final results were obtained
with a minimum CTE of 99.9995\% per transfer, or 99.7\% over 700 rows.

Another possible cause of non--linearity at low count levels is
deferred charge.  We performed the standard tests for deferred charge
as described in Gilliland (1992) and find that only 0.3\% of the
pixels showed deviations from linearity greater than four times the
read--noise. Pixels exhibiting non-linearity were flagged in all
images and excluded from analysis.  We also performed a standard
linearity test by taking dome flat--field exposures with integration
times between 0.5 and 200 seconds and looking for variation in the
detected count--rate between 20--20,000 DN pixel$^{-1}$. The influence
of lamp instability was minimized by taking several series of
exposures and averaging the results.  The detector response was linear
to the limits of the sampled range which more than brackets the signal
level of the standard star observations (peaking at roughly 5,000\,DN)
and the program observations (20--50\,DN on average).  Non--repeating
deviations of less than 1\% were attributed to the instability of the
lamp.

\subsection{Bias Subtraction}\label{lco.bias}

As discussed in \S\ref{backg} above, we use the strength of the solar
Fraunhofer lines in the ZL spectrum to measure the ZL flux. This is a
differential measurement in the dispersion direction; uniform,
additive offsets due to bias or dark current will not affect our
results beyond the small effect on flux calibration.  Structure in
either bias or dark current, however, will increase random errors in
the results.  For example, sharp features in the dispersion direction
will increase the {\it rms} errors in the average strength of the Fraunhofer
absorption lines. Also, because we extract a single, one dimensional
spectrum from every two dimensional image by averaging over the full
spatial extent of the spectrum, any fluctuations in bias level in the
spatial direction add random errors to the averaged flux found at any
wavelength.

To remove spatial variations in both directions, the bias correction
was done in three steps. Variations in the dispersion direction were
subtracted using the 150 column overscan region.  The overscan was
fitted with a Savitsky--Golay routine (Press \etal 1992), which
follows rapid jumps in the mean level, and the fit was then subtracted
from each column.  The mean bias level was then removed by subtracting
a single mean bias value, which is the average of the bias row (row
1024) in each frame.  Finally, bias variations over the chip were
found to be very stable after the overscan column and mean bias level
were subtracted from every image.  These variations were thus removed
by subtracting a ``superbias'' image, which is the average of 250,
overscan-- and bias--subtracted bias frames.  To verify the accuracy
of this procedure, we test-reduced 50 bias frames which had not been
included in the superbias. After bias subtraction, these test frames
had an average value of 0.005 DN with an \rms\ scatter of $0.03$ DN
and no residual systematic structure.

\subsection{Dark Current Subtraction}\label{lco.dark}

Dark frames were taken immediately before and after observations on
each night, and 20 darks were taken at the start and end of the run,
which were combined to make a ``superdark,'' the mean level of which
drifted by $\sim 1.0$ DN over the extent of the frame. Unfortunately,
the 20--frame superdark is read--noise dominated and cannot provide a
pixel--to--pixel correction. The superdark was therefore smoothed
using a sliding $3\times3$ boxcar median filter, in order to avoid
adding noise and to allow the removal of the mean dark level. The
frames taken during the run were used to test the accuracy of the
superdark: after bias subtraction (as described above) and dark
subtraction, the test-reduced darks taken at the beginning and end of
the observations on 1995 November 27 and 29 had a mean level of $\pm
0.25$ DN with no coherent pattern.

\subsection{Flat Fielding and Illumination Correction}\label{lco.flatf}

We used a 1.5 arcsec slit for the program observations in order to
preserve resolution. However, we used a 10.8 arcsec slit for the
standard star observations in order to collect as much light as
possible. Different sets of flat--field and illumination corrections
were therefore required for two reasons.  First, microscopic roughness
on the edges of the slit jaws caused shadowing which changed as a
function of slit--width. Second, the slit--jaw mechanism on this
spectrograph is such that the jaws are parallel for separations up to
$\sim 5.4$\,arcsec ($500$\micron)  but are not parallel when the jaws
are set to 10.8\,arcsec in the center.  Variation in slit--width is
almost 10\% from end--to--end when the width at the center is
10.8\,arcsec.  To compensate for the variable slit--width for
standards, the illumination corrections for both slit--widths were
normalized to the spatial center of the slit. Standard stars were all
observed within two pixels of the central column used for
normalization, which places them within 99.98\% of the nominal value
for the wide--slit illumination correction.  Flat--field and
slit--illumination corrections were created from dome and twilight sky
flats, respectively, using the tasks RESPONSE and ILLUMINATION in the
IRAF SPECRED package.

\subsection{Wavelength Calibration}\label{lco.wavec}

Wavelength solutions were based on He-Ar comparison spectra taken
before and after each 1800 second program exposure and immediately
after each standard star observation.  At the beginning of the run,
great care was taken to align the dispersion axis with the pixel rows:
the \rms\ variation in the centroid position of arc lines over the full
3.4 arcmin spatial extent of the images is typically 0.05 -- 0.12
pixels.  This is true over the full wavelength range.  Consequently,
it was not necessary to rectify the two--dimensional program spectra
of the zodiacal light (night sky).  Each program spectrum was simply
averaged along rows (with $5\sigma$ cosmic ray rejection) to obtain a
one--dimensional spectrum, and a wavelength solution based on the
spatial center of the image was applied afterward.  Standard stars
were first extracted to produce one-dimensional spectra and then
wavelength calibrated in the usual way.

We identified 21--25 features in each comparison spectrum visually and
fitted a third order (four term) Legendre polynomial to the
pixel-wavelength solution to obtain a dispersion curve.  This was done
using the IDENTIFY task in IRAF.  The \rms\ residuals in the wavelength
solution were 0.016--0.04\AA. Shifts in the wavelength solution
between program observations were less than $\pm1.5$ pixels from the
start to the end of the night.  Linear dispersion solutions were
applied using the task DISPCOR.

\subsection{Flux Calibration}\label{lco.fluxc}

\begin{figure}[t]
\begin{center}
\includegraphics[width=3in,angle=0]{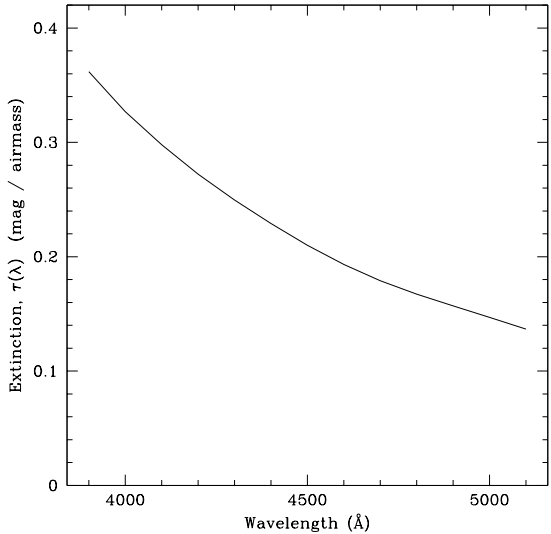}
\caption{\footnotesize 
The extinction function derived from observations of standard stars
taken on 1995 November 27 and 29. Extinction solutions for the
two nights individually were found to be identical to within the 
one-sigma statistical errors ($\lta0.01$\,mag). The function shown is a
sixth--order Legendre polynomial fit to extinction coefficients
obtained in 50\AA\ wide bins.}
\label{fig:lco.extin}
\end{center}
\end{figure}

Three independent components of the surface brightness calibration
will affect its final accuracy: point source calibration (which
includes the sensitivity and extinction corrections); aperture
correction (which compensates for the loss of light from point source
observations which does not occur in observation of a uniform
aperture--filling source); and the fiducial standard star system.  The
flux calibrated spectrum for a uniform source can be expressed as
\begin{equation}
I(\lambda) = \frac{ C(\lambda)\; S(\lambda)\; T({\rm A})\; 
10^{0.4\chi\tau(\lambda)} }{ \Omega } ,
\end{equation}
in which $C(\lambda)$, is the wavelength calibrated spectrum in DN
sec$^{-1}$\AA$^{-1}$ per pixel, $S(\lambda)$ is the sensitivity
function in ergs DN$^{-1}$, $T$(A) is the aperture correction for the
slit size in question, $\chi$ is the airmass of the observation,
$\tau(\lambda)$ is the extinction correction expressed in mag
airmass$^{-1}$, and $\Omega$ is the solid angle of each pixel in
steradians per pixel.  We discuss each component of the calibration
separately below.

\subsubsection{Point Source Calibration}\label{lco.ptsrc}

\begin{figure}[t]
\begin{center}
\includegraphics[width=3in,angle=0]{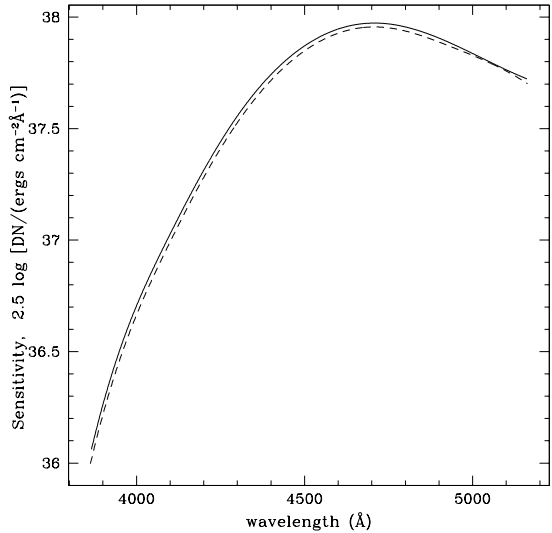}
\caption{\footnotesize  
Two sensitivity functions obtained from observations of
tertiary standards on different nights.  The solutions differ by
1-2\%, well within the fluctuations expected to arise from small
temperature variations of the CCD.
}
\label{fig:lco.sensfunc}
\end{center}
\end{figure}

We observed Hamuy \etal\ (1992, hereafter H92) tertiary
spectrophotometric standards roughly 15 times each night.  For each
standard observation, we first centered the star in a narrow slit
($\sim 2$\,arcsec), and opened the slit to 10.8 arcsec after guiding
was established.  Nine different standards were observed throughout
the run, with colors $0.0<(B-V)<0.6$\,mag.  This range represents as
broad a distribution as is available from the H92 standards and does
bracket the color of the night sky. All standards were observed with
the slit aligned along the parallactic angle (Filippenko 1982).

One--dimensional spectra were extracted from the two--dimensional
images and wavelength calibrated in the usual way using the APSUM and
DISPCOR tasks in IRAF.  Extinction corrections were calculated from
standard star observations themselves in 50\AA\ bins for each night
individually.  Typical residuals in the extinction solution in each
50\AA\ bin were 0.009\,mag (\rms) in the data taken on 1995 November 27
and 29.  As no difference was found between the two nights, the final
extinction solution was constructed from the data taken on both nights
together using a sixth--order Legendre polynomial fit to the
extinction as function of wavelength (see Figure \ref{fig:lco.extin}),
with an uncertainty of 0.2\% as a function of wavelength.  The
resulting extinction curve is in excellent agreement with the $r$--
and $g$--band extinction terms which we obtained from images taken
simultaneously with the 1m Swope telescope at the same site.

After the extinction corrections were applied to the standard star
spectra, the sensitivity curve for each night was determined using the
task SENSFUNC in the IRAF SPECRED package. The agreement between the
sensitivity curves for the two nights is excellent: the variations
between the sensitivity curves found for 1995 November 27 and 29 are
3\% at 4100\AA, and 2\% red-ward of 4500\AA\ (see Figure
\ref{fig:lco.sensfunc}).  Variations in the quantum efficiency of the
CCD on this level are expected to result from temperature changes of a
few degrees (M. Blouk, personal communication), so we do allow the
sensitivity function solutions to be slightly different from night to
night.  The standard deviation in the SENSFUNC solution for both
nights is 0.011\,mag, which translates into a 0.3\% error in the mean
sensitivity as a function of wavelength from 15 standard star
observations.  To be conservative, we adopt a systematic uncertainty
in the point source calibration of 0.6\%.  Systematic errors in the
tertiary standard star system of Hamuy \etal are discussed in
\S\ref{lco.stand}. They are not included in the accuracy of the
zodiacal light measurement discussed here.  Those uncertainties are,
however, relevant to the EBL detection for which we have used this
zodiacal light measurement.  That error is explicitly included in
final accuracy of the EBL detections discussed in Paper I.

\subsubsection{Aperture Correction}\label{lco.apcor}
\begin{figure}[t] 
\begin{center}
\includegraphics[width=3in,angle=0]{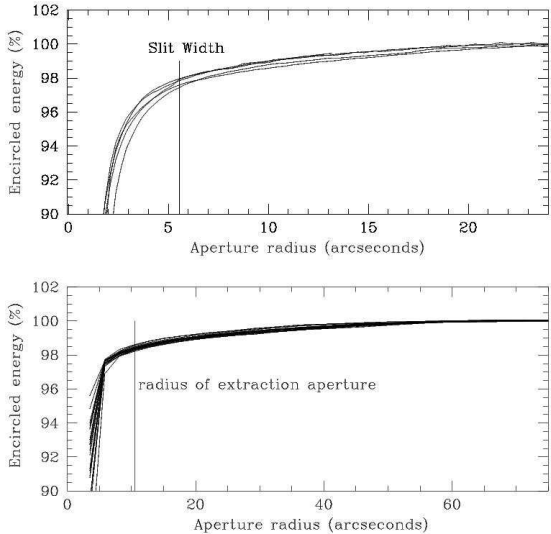}
\caption{\footnotesize 
The upper plot shows encircled energy curves for stars
imaged directly in the focal plane of the du Pont 2.5m telescope.  The
half--width of the slit (5.8\,arcsec) used for standard star
observations is indicated by the solid vertical line.  The percentage
flux from a point source which passes through this aperture is 97.9
($\pm$0.2)\%.  The lower plot shows the encircled energy curve along
the spatial extent of a spectroscopic image, showing that 98.4
($\pm0.1$)\% of the light entering the spectrograph from a point
sources is recovered when a $\pm 10$ arcsec extraction aperture is
used.
}
\label{fig:lco.apcor}
\end{center}
\end{figure}

The aperture correction compensates for flux which is lost from the
point source observations in two distinct ways.  First, light is lost
in the focal plane of the telescope if the radius of 100\% encircled
energy is larger than the half--width of the slit.  Second, when a one
dimensional spectrum is extracted from the two dimensional, dispersed
image, some light will lie outside the extraction aperture in the
spatial direction.

To measure flux lost in the aperture plane, we measured the PSF from
images taken on the last night of the run (1995 November 30).  We
plot the encircled flux with radius (growth curve) for five stars in
Figure \ref{fig:lco.apcor}.  To be certain that focus does not affect
the PSF at a 5\,arcsec radius, the stars used for this plot were taken
with widely varying focus.  As 5\,arcsec in the focal plane of the
duPont 2.5m Telescope corresponds to 0.45 mm (almost 19 pixels) it is
difficult to have the telescope out of focus enough to affect the
enclosed flux at a radius of 5\,arcsec.  This is clear from the lack
of variation in the shape of the growth curves shown in the figure.
The fractional flux enclosed by the half--width of the slit used for
standard star observations is 0.979 ($\pm$0.002).

To determine the percentage flux lost once through the slit, we have
mapped the growth curves along the spatial extent of the two
dimensional spectra for all 45 standards taken during the run.  The
sky level for this test was taken from pixels further than 80\,arcsec
from the peak of the star. We averaged 400 rows near the peak
sensitivity of the spectrograph to increase the signal--to--noise.  To
confirm that the PSF has negligible wavelength dependence, we
calculated the aperture correction at both the blue and red ends of
the spectral range and found no variation in the growth curve.  The
growth curve along the slit for the 18 highest signal--to--noise
spectra are shown in Figure \ref{fig:lco.apcor}.  The differences in
the encircled energy at the inner--most radii plotted are a result of
differences in sub-pixel centering for spectra observed with
$4\times1$ binning. This does not affect our results for spectra
extracted to an aperture of $\pm4$ (binned) pixels ($\pm 10.6$
arcsec). As this plot shows, the extraction aperture is well within
the region of good signal--to--noise in the star, and includes 98.4
($\pm0.1$)\% of the flux from a point source which passed through the
spectrograph slit.  Sky background was measured outside an annulus of
25 pixels (60\,arcsec) from the peak of the star. This aperture is
small enough to ensure adequate signal--to--noise in the extracted
stellar spectrum, and also ensures minimal error due to sky
subtraction, as the stellar signal is much brighter than the sky
background in the inner 10.6\,arcsec.

The total aperture correction for a uniform surface brightness,
aperture--filling source is then the fractional flux recovered for a
point source, or $T=0.963 (\pm 0.002$).

\subsubsection{Solid Angle of the Program Observations}\label{lco.solid}

The solid angle is a function of both the angular pixel scale (spatial
direction) and the angular slit--width (dispersion direction).  The
pixel scale was measured empirically by taking spectra of two stars
with known angular separation while they were simultaneously aligned
in the slit.  The measured separation in pixels was then compared to
the known angular separation of the stars.  Four pairs of stars were
observed in this way with angular separations in the range
44--82\,arcsec.  Each pair of stars was observed with the slit at 3
positions differing by less than 1 degree, in an attempt to obtain
truly parallel slit alignment.  The stars used for this purpose are in
the field of M67, for which the relative astrometry of members is
known to better than 0.3\,mas (Girard \etal 1989).  The pixel scale
was found to be $0.5843\pm0.0035$\,arcsec pixel$^{-1}$ ($1\sigma$ error).

The slit--width of the spectrograph is adjusted by a manual micrometer
while the instrument is on the telescope.  The calibration and
repeatability of the micrometer was verified using a microscope.  The
repeatability of the width setting was tested by opening the jaws to
their maximum extent between each of several sets of measurements.  At
the micrometer setting used for our program observations, the slit width was
measured to be $1.536 \pm 0.002$\,arcsec ($1\sigma$ error) at the
center of the jaws.  The total solid angle of each pixel is therefore
$0.8975 \pm 0.0054$\,\sqarcsec ($1\sigma$ error).

\subsection{Accuracy of Tertiary Standards}\label{lco.stand}

\begin{figure}[t] 
\begin{center}
\includegraphics[width=3in]{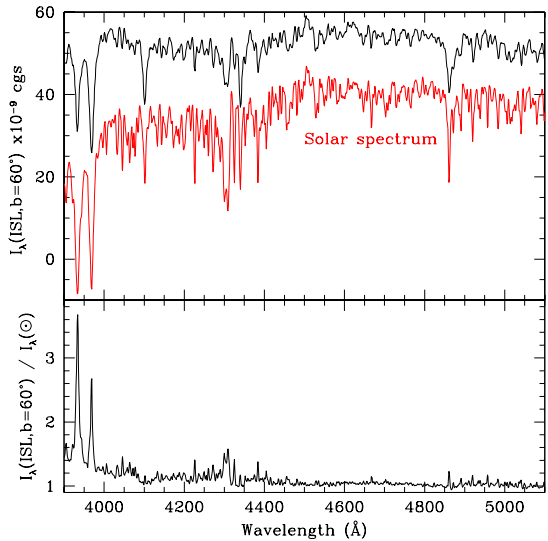}
\caption{\footnotesize In the upper plot, the spectrum of the
integrated starlight (ISL) at $|b|=60^\circ$ is compared to the solar
spectrum, which has been scaled to the same flux and offset to allow
visual comparison of the spectral features.  The lower plot shows the
ratio of the two, normalized at 4600\AA. Both plots show visually and
quantitatively that the absorption lines in the ISL spectrum are
weaker than the same features in the solar (and therefore zodiacal
light) spectrum.}
\label{fig:dglat60}
\end{center}
\end{figure}

Hamuy \etal (1992) quote the internal precision of their tertiary
spectrophotometric system to be better than 0.01\,mag at all
wavelengths, a claim which is well corroborated by the small
statistical errors we find in our own spectrophotometric flux
calibration (see below).  An additional concern for our purposes,
however, is the agreement between the H92 system and the primary
calibration for Vega in the spectral range of our observations
(approximately the Cousins $B$--band, 4200--5100\AA).

The tertiary system of H92 is calibrated based on the equatorial
secondary standards of Taylor (1984), which is in turn calibrated to
the primary calibration of Vega by Hayes \& Latham (1975).  As
described in H92, they recalibrate the Taylor (1984) secondary
standards to the now widely--accepted Hayes (1985, hereafter H85)
calibration of Vega. Details of that process are documented in H92.
H92 estimate that their internal consistency in converting Taylor
(1984) to the Hayes (1985) Vega system is 0.009\,mag in the wavelength
range of the Cousins $B$.  They also compare synthetic photometry of
the adjusted Vega spectrum with Johnson \& Harris (1954) photometry of
Vega and find an offset of $-0.016$ ($\pm 0.009$)\,mag at $B$ (in the
sense of Taylor minus Johnson).  This offset seems to be intrinsic to
the Vega calibration, as Hayes also finds that offset between his
observations of Vega and the original Johnson observations.  We
therefore conclude that the statistical accuracy of the tertiary
system is roughly 1\% and the systematic uncertainty is roughly 1.5\%.

\section{Components of the Night Sky Spectrum}\label{nightsky}

The spectrum of the night sky from Earth includes integrated starlight
(ISL), zodiacal light (ZL), EBL, and diffuse Galactic light (DGL) from
above the atmosphere, as well as atmospheric emission, or airglow.
Light from all sources is scattered by molecules and particulates in
the atmosphere, causing considerable redirection from one line of
sight to another. Absorption by particulates is a minor effect
compared to the scattering by both components.  The net effect of
atmospheric scattering on a source with small angular extent
(e.g. stars and galaxies) is atmospheric ``extinction.'' The net
scattering out of the line of sight can be measured in the typical way
using standard stars, as in \S\ref{lco.fluxc}.  Light from a very
extended source (e.g. ZL, ISL) will not suffer the same
``extinction''; rather, it will appear to be smoothed out over the
sky as light from different regions scatters into and out of the line
of sight. The efficiency of scattering in the atmosphere is
conveniently described in the familiar way as the extinction along the
line of sight $\tau_{\rm obs}(\lambda)$. 

If the scattering angles were small, and the diffuse, extended source
were uniform over the sky, then the scattering into and out of the
line of sight would roughly cancel.  However, Rayleigh scattering
occurs over very broad angles and the relative surface brightness of
ZL and ISL changes strongly over the sky.  We must therefore
explicitly calculate the net effects of scattering in our observing
situation (determined by the observatory location and positions of the
Sun, Galaxy, and target), as they are not intuitive.  This is done in
detail in the Appendix. We summarize the results here.

We can describe the observed spectrum of the night sky in the target
field , $I_{\rm NS}$, as follows:
\begin{eqnarray}
I_{\rm NS}(\lambda,t,\chi) &= &
	I_{\rm (3h,-20d)}(\lambda) e^{-\tau_{\rm obs}(\lambda)\chi} \nonumber \\
	& & + I_{\rm scat}(\lambda,t,\chi)+ I_{\rm air}(\lambda,t,\chi),
\label{eq:lco.obssky1}
\end{eqnarray}
in which $I_{\rm (3h,-20d)}(\lambda)$ is the flux from the target
field (coordinates $\alpha=3.00$h, $\delta=-20.18$d), $\tau_{\rm
obs}(\lambda)$ is the extinction for a point source, $\chi$ is the
airmass of the target field at the time of observations, $I_{\rm
scat}(\lambda,t,\chi)$ is the light scattered into the line of sight
as a function of time and wavelength, and $I_{\rm
air}(\lambda,t,\chi)$ is the effective airglow along the line of sight
(including any scattering effects which redistribute airglow over the
sky).  The flux from the target field, $I_{\rm (3h,-20d)}(\lambda)$,
includes the EBL, DGL and ZL flux within the solid angle of the slit.
There is no ISL component coming directly from the target field
because the slit simply contains no stars to $V=24$\,mag. The slit
also provides an extremely effective pupil stop which prevents
contributions from discrete sources off--axis.\footnote{No stars with
$V<12$ mag are within $12$ arcmin of the slit, and no stars with
$V<7$ are within 1.5 degrees. We have carefully characterized the
scattered light properties of the duPont telescope by positioning a
$V=4$ mag star around the field from on-axis to $20$ arcmin off-axis
in 4 directions at $1$ arcmin intervals.  The stray light entering the
slit from discrete off-axis sources is more than $10^{-6}$
fainter than surface brightness of the ZL in the field. See Paper I
for further discussion.}  $I_{\rm
scat}(\lambda,t,\chi)$ can be expressed as
\begin{eqnarray}
I_{\rm scat}(\lambda, t, \chi)&  = &
	I^{\rm R}_{\rm scat}(\lambda, t, \chi, {\rm ZL}) +  
	I^{\rm M}_{\rm scat}(\lambda, t, \chi, {\rm ZL}) + \nonumber \\
    & & I^{\rm R}_{\rm scat}(\lambda, t, \chi, {\rm ISL}) + 
	I^{\rm M}_{\rm scat}(\lambda, t, \chi, {\rm ISL}) ,
\label{eq:iscat}
\end{eqnarray}
where the superscript R or M denotes Rayleigh scattering (due to
molecules) or Mie scattering (due to particulates), the parenthetical
ZL or ISL denotes the source being scattered, and the parenthetical
$\lambda$, t, and $\chi$ denote dependence on those variables.  We do not
include the DGL in Equation \ref{eq:iscat} explicitly because the
total DGL is at least a factor of 50 times fainter than the direct ISL
and is therefore a trivial component ($<0.2$\% at 2$<$UT$<$6.5) of the
scattered light. The EBL is not included as it does not have strong
spectral features (see \S\ref{lco.analy}).

At any altitude, Rayliegh scattering is the dominant effect.  Because
the particulate density is concentrated at low altitudes, this is
especially true at high altitude observatories.  The total extinction,
as measured for a point source, is equal to the sum of the molecular and
particulate extinction, $\tau_{\rm obs}(\lambda) = \tau_{\rm
M}(\lambda) +\tau_{\rm R}(\lambda)$.  Rayleigh extinction, $\tau_{\rm
R}$, can be calculated from the well--known density distribution of
the atmosphere for any observatory. Mie extinction, which varies with
time and geography, can be inferred from the difference between the
observed and Rayleigh extinction.  At LCO, the extinction due to Mie
scattering is 20--40\%\ of the Rayleigh extinction.

We have calculated the scattered light from all terms in Equation
\ref{eq:iscat} in the Appendix at 30 minute intervals throughout the
nights of our observations.  To briefly summarize the results of our
calculations, the ZL total flux scattered into the line of sight ({\it
gained}) at any time during our observations is less than the total
flux scattered out of the line of sight ({\it lost}).  Thus, the net
result of atmospheric scattering for the case of ZL in our situation
is still a net {\it extinction} of order 2--8\%, which we can
conveniently describe by an effective extinction, $\tau_{\rm eff}$,
which we use in place of $\tau_{\rm obs}$ for ZL (see Figures
\ref{fig:net_wavel} and \ref{fig:eff_tau_ut2} in the Appendix).  We
can check the scattering predictions of our calculations in our ZL
analysis itself by looking for changes in the ZL solution with
time. We estimate that our calculation of the scattered ZL has an
average uncertainty of 8\%, which translates into a systematic
uncertainty in our ZL measurement of 1.2\%.

In the case of the ISL, the total flux gained due to scattering into
the line of sight is 12--24\% of the total ZL flux from low to high
airmass. However, the crucial issue is not the total mean
flux, but rather the strength of the spectral features which are in
common with the Sun (see Figure \ref{fig:isl_ut2_sun} in the
Appendix). The net influence of the scattered ISL on observations is
to increase the strength of the Fraunhofer lines over the night by
$0.6-4$\% redward of 4100\AA, and 5--35\% blueward.  Because the effect
is a strong function of wavelength, it is straightforward to identify
inconsistencies between the predicted ISL flux and our observations by
looking for changes in the ZL solution with wavelength.  We estimate
that our calculation of the scattered ISL has an uncertainty of 13\%,
which translates into a systematic uncertainty in our ZL measurement
of 0.5\% over the majority of our wavelength range.

\section{Analysis}\label{lco.analy}

\begin{figure*}[t] 
\begin{center}
\includegraphics[width=5in,angle=-90]{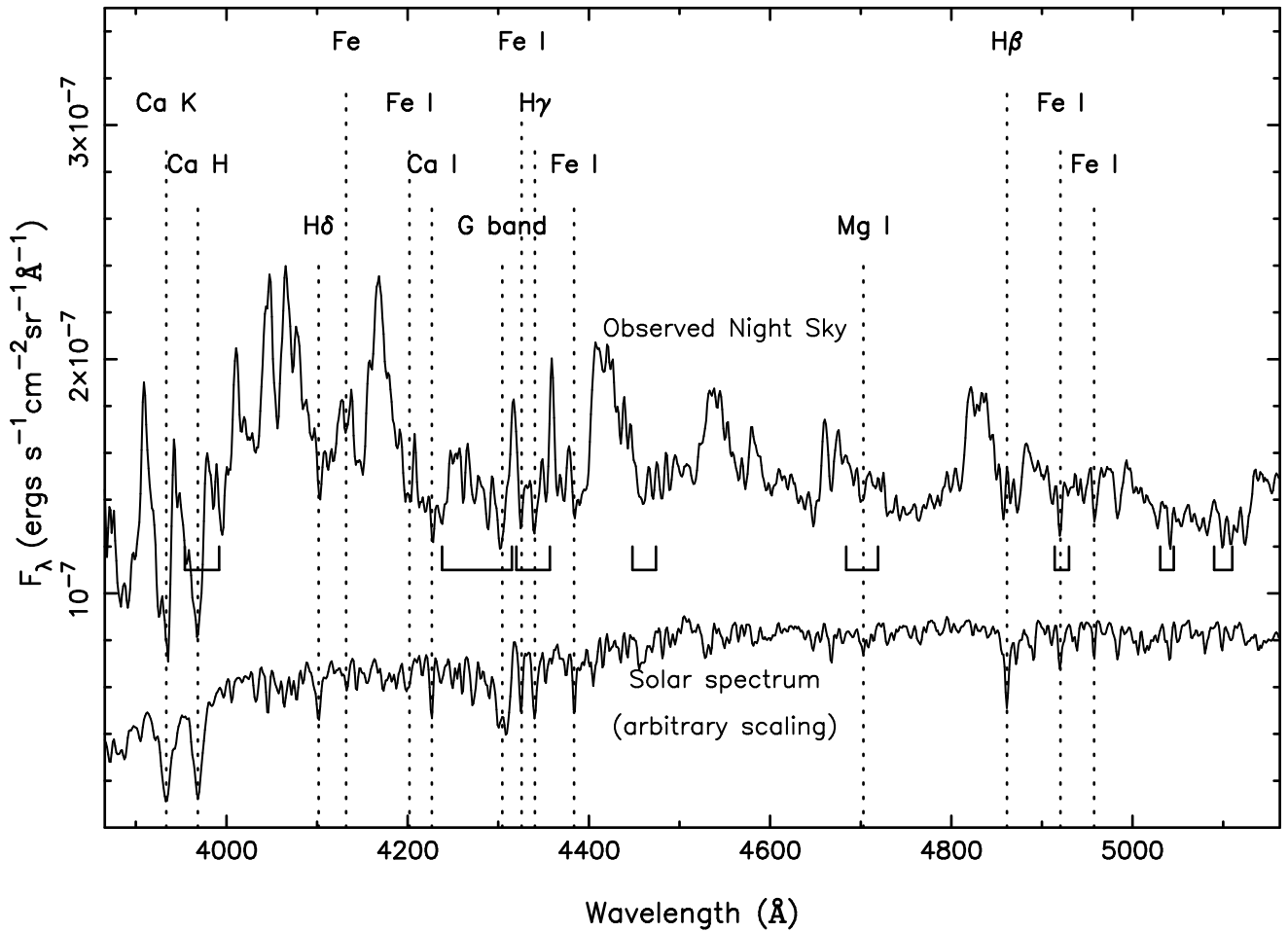}
\caption{\footnotesize 
The observed spectrum of the night sky compared to a solar
spectrum at arbitrary absolute flux.  The solar Fraunhofer absorption
lines which are preserved in the ZL are clearly visible in the
spectrum of the night sky.  The strongest of these lines are marked;
several blended solar features are also seen in the night sky
spectrum.  Square brackets indicate the wavelength regions used in the
final analysis.  
}
\label{fig:lco.nsky.and.sun}
\end{center}
\end{figure*}

\begin{figure*}[t] 
\begin{center}
\includegraphics[width=5in,angle=-90]{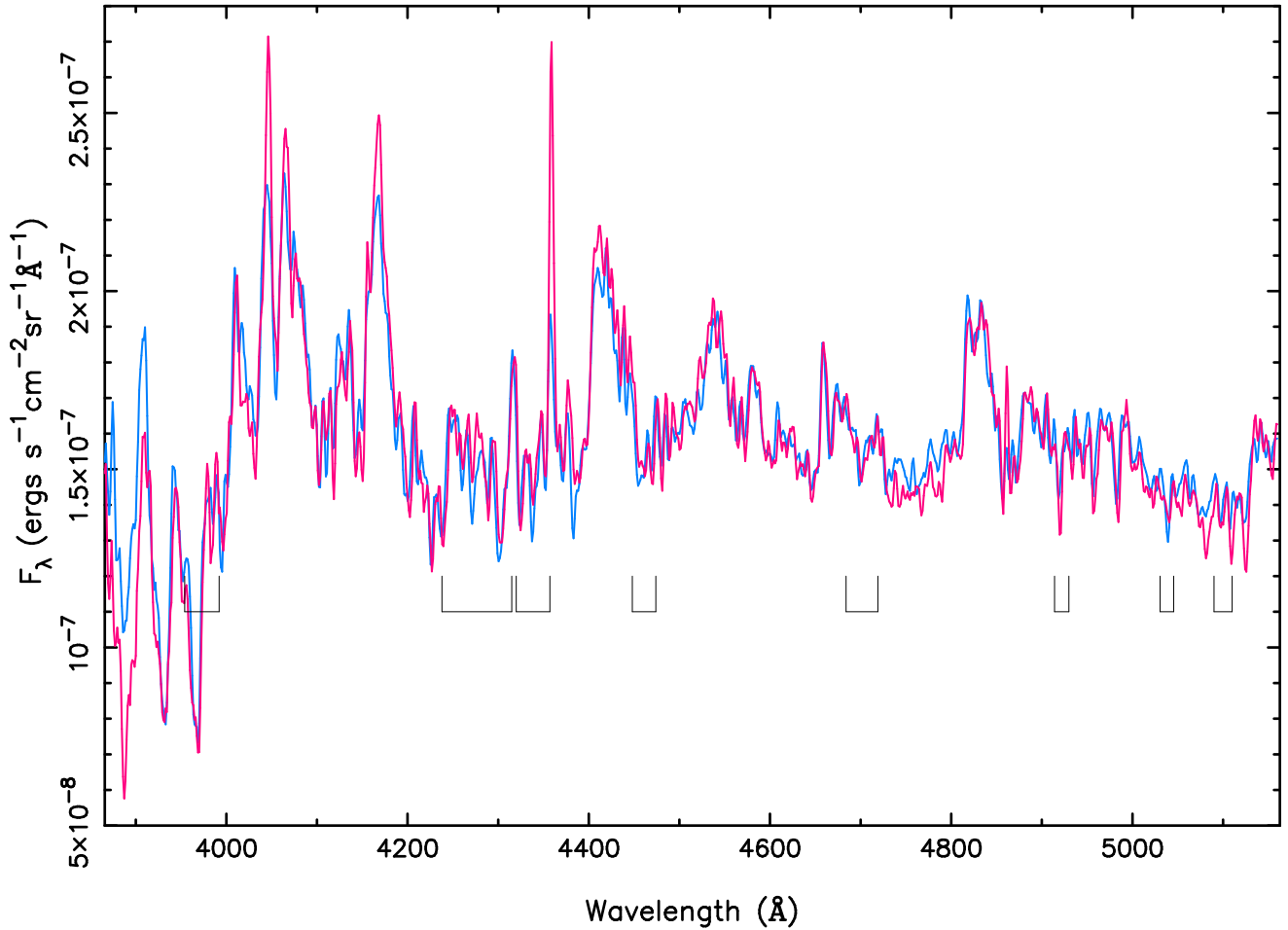}
\caption{\footnotesize 
Two spectra of the night sky taken on the same night,
several hours from twilight. Rapid fluctuations are evident in
the strength of many of the airglow features.
}
\label{fig:lco.2nsky}
\end{center}
\end{figure*}

Based on the discussion in the previous section, we can characterize
the observed spectrum of the night sky, $I_{\rm NS}$, at any time as a
combination of airglow, zodiacal light (ZL), diffuse Galactic light
(DGL), and scattered interstellar light (ISL) as follows:
\begin{eqnarray}
I_{\rm NS}(\lambda, t, \chi) &=&
  	I_{\rm air}(\lambda,t,\chi) + I_{\rm scat}(\lambda,t,\chi,{\rm ISL}) \nonumber \\
&& + I_{\rm ZL}(\lambda) e^{-\tau_{\rm eff}(\lambda,t)\chi}   \nonumber\\
&& + [I_{\rm EBL}(\lambda) + I_{\rm DGL}(\lambda)] 
	e^{-\tau_{\rm obs}(\lambda)\chi}  ,
\label{eq:lco.obssky}
\end{eqnarray}
in which $\tau_{\rm eff}(\lambda,t)$ is now the effective extinction
for ZL discussed in \S\ref{nightsky} and $I_{\rm scat}(\lambda,t,\chi,{\rm ISL})$
is the total scattered ISL flux due to Rayleigh and Mie scattering
in the atmosphere.

Sharp spectral features are not expected in the EBL because
redshifting will blur any distinct spectral features.  The EBL will
therefore not affect our measurements of the ZL from the strength of
the observed Fraunhofer lines.  The diffuse Galactic light, which
results from scattering of the ambient interstellar radiation field by
interstellar dust, is very weak in our field (0.8\% of the ZL flux,
see Paper I). In addition, as we have already discussed regarding the
scattered ISL, the strength of the spectral features we use in our
analysis is roughly 1.5 to 3.8 times weaker in the DGL than in the ZL
spectrum. The DGL from the target field thus contributes at most
0.2-0.5\% percent to the final result (see Figure \ref{fig:dglat60}).
We subtract this contribution from our ZL measurement after the fact
at a level of 0.3\%.  Emission lines due to ionized gas in the DGL do
not contribute in the spectral range of these observations (see Paper
I, Martin \etal 1991, Dube \etal 1979, and Reynolds 1990).

The observed spectrum can therefore be expressed as the sum of four
components: (a) an unstable emission line spectrum, due to airglow;
(b) a stable and featureless component, due to EBL; and (c) a stable,
absorption line component, due to ZL; and (d) a time variable
absorption component due to scattered ISL.  The component (c) can be
ignored, and (d) has been calculated.  The portion of the night sky
spectrum which has variable spectral features can therefore be
expressed as
\begin{equation}
I_{\rm obs}(\lambda,t,\chi) = I_{\rm air}(\lambda,t,\chi) 
 + I_{\rm scat}(\lambda, t, \chi, {\rm ISL})
 + c(\lambda) I_{\odot}(\lambda) e^{-\tau_{\rm eff}(\lambda,t)\chi},
\end{equation}
in which $I_{\odot}(\lambda)$ is the solar spectrum and $c(\lambda)$ is a
scaling factor which relates the mean surface brightness of the ZL to
the mean flux of the Sun.

Identifying the appropriate scaling spectrum, $c(\lambda)$, is
complicated by the fact that the ZL --- a pure absorption line
spectrum --- is greatly obscured by the emission line spectrum of the
airglow in the night sky.  Airglow features do overlap with some of
the solar Fraunhofer lines, as can be seen in the comparison of a
night sky spectrum and a scaled solar spectrum shown in Figure
\ref{fig:lco.nsky.and.sun}. The strength of particular airglow lines
can vary by several percent during a single night, as can be seen in
the comparison of two night sky spectra shown in Figure
\ref{fig:lco.2nsky}.  The airglow spectrum is composed of an effective
continuum due to O+NO (NO$_2$) recombination, broad
rotation--vibration transition bands, scattered light, and blended
lines, making a continuum level impossible to identify.  Not only do
rapid temporal variations occur with (de)ionization, but airglow is
also a complex function of airmass, observatory location, and local
atmospheric conditions such as volcanic activity (van Rhijn 1924,
1925; Roach \& Meinel 1955).  In short, a stable, fiducial airglow
spectrum with meaningful absolute or relative flux cannot be defined.
As a result, it is not possible to measure the equivalent width of
individual ZL Fraunhofer lines because the ZL continuum level is well
hidden.  Instead, we have developed a conceptually simple approach to
the problem of determining the scaling factors $c(\lambda)$ which does
not involve measuring the equivalent widths explicitly.

We begin with the assumption that the intrinsic airglow spectrum,
time-- and airmass--dependent though it may be, does not have spectral
features in common with the ZL spectrum. This is borne out by the fact
that we get consistent solutions using eight distinct spectral regions
spread over 1100\AA\ (see \S\ref{lco.analy}). When we subtract $I_{\rm
scat}(\lambda, t, \chi,{\rm ISL})$ and scaled solar spectrum with the
correct value for $c(\lambda)$, what remains is a pure airglow
spectrum, free of solar features:
\begin{eqnarray}
I_{\rm air}(\lambda,t,\chi) &=& I_{\rm obs}(\lambda,t,\chi)-
	I_{\rm  scat}(\lambda, t, \chi,{\rm ISL}) \nonumber \\
	&& - c(\lambda) I_{\odot}(\lambda) e^{-\tau_{\rm eff}(\lambda,t)\chi}.
\end{eqnarray}
We use a linear correlation function to determine when the difference
(residual airglow) spectrum is uncorrelated with the solar spectrum
and is consequently free of solar features.  When the correlation
between the difference spectrum and solar spectrum is minimized, the
correct ZL surface brightness has been subtracted from the observed
night sky.

The only available, high--resolution spectrum of the Sun is the
National Solar Observatory Solar Flux Atlas of the integrated solar
disk at 0.01\AA\ resolution.  The statistical error in the flux
calibration of this spectrum is 0.25\% as estimated by agreement in
overlapping sections of the normalized spectrum.  The effects of
atmospheric absorption by H$_2$O or O$_2$ are negligible below
6500\AA, as described in the published Atlas (Kurucz \etal 1984).  In
the optical, the normalized spectrum can be converted to absolute
solar irradiance using the Neckel \& Labs (1984) (NL84) absolute
calibration. As the NL84 is the standard with respect to which the ZL
color is defined, the absolute accuracy of the fiducial solar spectrum
is not a source of error in this work.  The Solar Flux Atlas,
calibrated to NL84, was obtained in digitized form from
R.L. Kurucz. It was convolved with a variable width Gaussian to match
the resolution of the observed spectra as a function of wavelength.
The wavelength--dependent resolution of each program spectrum was
determined from the arc lamp spectra which were used for wavelength
calibration.

The execution of this method is complicated by the fact that the
relative color of the airglow and solar spectra will dominate the
strength of the diagnostic correlation if the continuum shapes of both
spectra are not properly removed.  In order for the strength of the
linear correlation of $I_{\rm air}(\lambda)$ and $I_{\odot}(\lambda)$
to reflect the strength of coincident spectral features, both spectra
must have stationary mean values as a function of wavelength, as can
be seen clearly in the generic expression for a linear correlation:
\begin{equation}
R(x,y) = \frac{\sum_n (x_n - \overline{x})(y_n - \overline{y})}
{\sqrt{\sum_n (x_n - \overline{x})^2} \sqrt{\sum_n (y_n - \overline{y})^2}}.
\label{eq:lco.corr}
\end{equation}
In this case, $x$ and $y$ are $I_{\rm air}(\lambda)$ and $I_{\odot}
(\lambda)$ respectively, the subscript $n$ runs over wavelength.  It
is clear from this expression that the mean flux drops out of the
correlation, while differences from the mean are crucial. 

Of the 47 strongest solar features, we find that 39 give ZL solutions
which vary with time by more than 18\% over the night.  In all cases,
this variation is  correlated with the strength of adjacent
airglow lines.  The remaining eight solar features vary by less than
10\% with time.  The results discussed below are based on these eight
spectral regions, indicated in Figures \ref{fig:lco.nsky.and.sun} and
\ref{fig:lco.2nsky}. In these regions, the continuum can be well
approximated by a simple second order polynomial fit and easily
subtracted, however our results are quite insensitive to the method of
continuum fitting. Boxcar smoothing with scales between 75\AA\ and
199\AA\ (at least twice the width of the widest spectral region used
in the analysis), second or third order polynomial fitting, and
Savitsky--Golay smoothing (Press \etal 1992) all produce identical
results.

\section{Results}\label{lco.resul}
\begin{figure}[t] 
\begin{center}
\includegraphics[width=3in,angle=0]{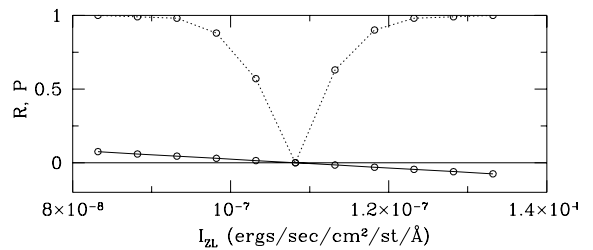}
\caption{\footnotesize 
The value of the correlation parameters used to define the
strength of the correlation between the residual ``airglow'' spectrum
(the zodiacal light-subtracted night sky spectrum) and the ZL
spectrum for different assumed contributions of ZL. The points
connected by the dotted line indicate the probability that the two
spectra are from the same parent set as a function of the ZL flux
assumed. The points connected by the solid line indicate the
correlation strength, zero being no correlation.  These parameters
describe a simple linear correlation as defined in equation
\ref{eq:lco.corr} (see Press \etal\ 1992). The ZL surface brightness
identified by this method for the observation shown here is
1.08\tto{-7} \escsa.
}
\label{fig:lco.corr+prob}
\end{center}
\end{figure}

In Figure \ref{fig:lco.corr+prob}, we show an example of the
correlation strength, $R$, and correlation probability, $P$, between a
ZL spectrum and the spectra which result when we assume a range of
values (8\tto{-8} to 1.4\tto{-7} \escsa) for the mean ZL flux
contributing to one region of one observed night sky spectrum.  Where
$R$ and $P$ go to zero, the correct ZL surface brightness
has been removed from the observed night sky spectrum.

\begin{figure}[t] 
\begin{center}
\includegraphics[width=3in,angle=00]{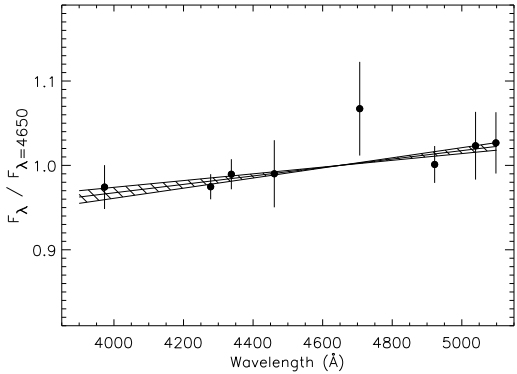}
\caption{\footnotesize The points show $c(\lambda)$ for the eight
spectral regions indicated in Figure \ref{fig:lco.nsky.and.sun} as
measured from 16 spectra taken on 1995 November 27 and 29, normalized
to $1.0$ at 4600--4700\AA.  The error bar on each point indicates the
standard deviation in the 16 measurements.  The hatched region shows
the color of the ZL relative to the solar spectrum determined by a
linear least squares fit to the points, excluding the point at
$\sim4700$\AA.  The best fit color is $C(3900,5100)=1.05\pm0.01$,
where the error corresponds to the one--sigma error in the fitted
slope.  The mean flux of the ZL at 4600--4700\AA\ from this fit is
$109.1(\pm0.5)\times10^{-9}$ \escsa.  }
\label{fig:lco.results}
\end{center}
\end{figure}

\begin{figure}[t] 
\begin{center}
\includegraphics[width=3in,angle=00]{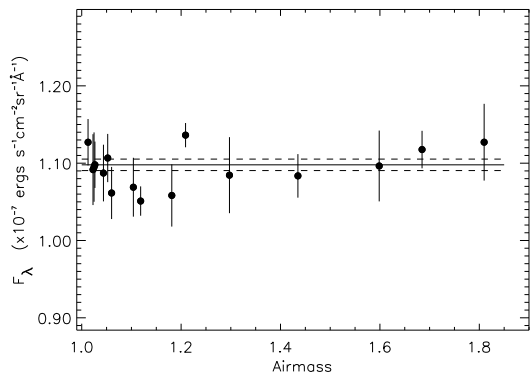}
\caption{\footnotesize 
The mean value of the ZL at 4600--4700\AA\ as 
measured in each of the 16 spectra taken on 1995 November 27 and 29,
adopting $C(3900,5100)=1.05$.  Each point represent the average of
the ZL flux measured in all 8 spectral features in a single
observation.  The horizontal line shows the mean ZL flux, which is
$109.7(\pm0.7)\times10^{-9}$ \escsa. Dashed lines show the one-sigma
statistical  error in the mean (0.6\%).
}
\label{fig:lco.results2}
\end{center}
\end{figure}

\begin{figure}[t] 
\begin{center}
\includegraphics[width=3in,angle=00]{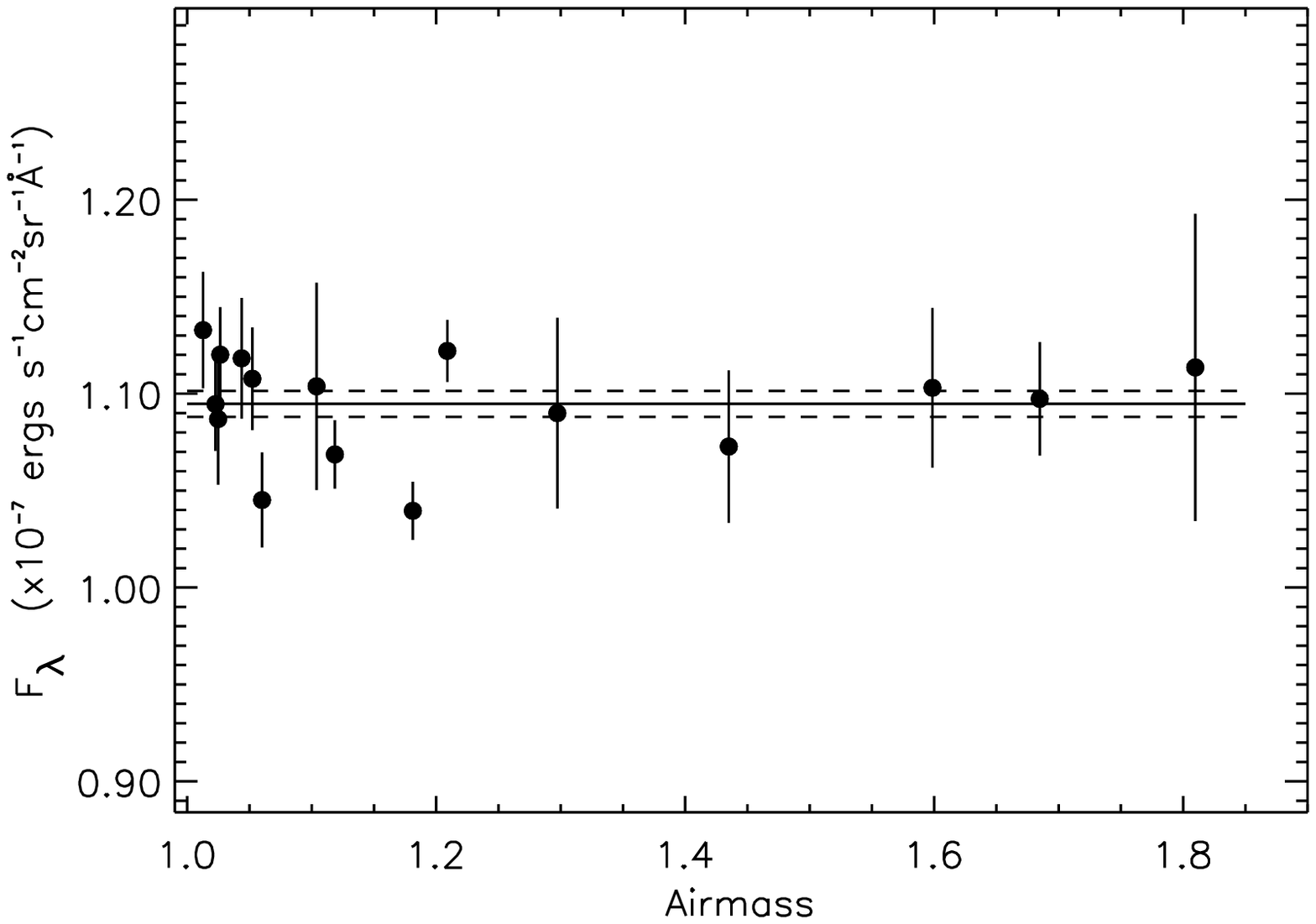}
\caption{\footnotesize 
The same as Figure \ref{fig:lco.results2}, but here only
the four features with the smallest error bars have been used 
to produce the mean value of the ZL at 4600--4700\AA\ in each spectrum
The horizontal line shows the mean ZL flux, which is
$109.4(\pm0.6)\times10^{-9}$ \escsa. Dashed lines show the one-sigma
statistical error in the mean (0.6\%).
}
\label{fig:lco.results3}
\end{center}
\end{figure}

The absolute flux of the ZL as measured from each of the 16 spectra
taken on 1995 November 27 and 29 are shown in Figures
\ref{fig:lco.results}--\ref{fig:lco.results3}.  In Figure
\ref{fig:lco.results}, we show the average $c(\lambda)$ as determined
for all 16 spectra over the run for each of the eight spectral
regions. The results are normalized to 1.0 at 4650\AA.
This plot illustrates two points: (1) from any single spectral
feature, we find a solution for the ZL flux with a standard deviation 
of only 1--6\% over 16 individual observations taken on two
nights and independently reduced; and (2), the solution
from independent spectral features are in excellent agreement, and
indicate that the ZL is roughly $5\pm1$\% redder than the solar
spectrum, or $C(3900,5100)=1.05\pm 1$ (excluding the point at
$\sim4700$\AA). This color is in excellent agreement with previous
estimates from the ground and space (see Leinert 1998).  It is also in
excellent agreement with our own measurement of the ZL color from our
simultaneous HST/FOS observations, which give  $C(4000,7000) =
1.044\pm1$ (see Paper I).

From the solution for the ZL as a function of wavelength we can
explore the impact of the scattered ISL flux on our measurements.  If
we increase (decrease) the total flux of the predicted scattered ISL,
$I_{\rm scat}(\lambda,t,\chi, {\rm ISL})$, the solution for the ZL
decreases (increases) {\it linearly} in response; a change of 50\% in
the ISL flux corresponds to a change of 6\% in the ZL solution at
$\sim3950$\AA, but $<2$\% at the other wavelengths. Thus, increasing
or decreasing $I_{\rm scat}(\lambda,t,\chi,{\rm ISL})$ by 50\% makes
the ZL solution at 3950\AA\ inconsistent at the two--sigma level with
solutions over the rest of the spectrum for a ZL color of $5\pm1\%$.
Also, increasing or decreasing the ISL flux consistently increases the
scatter in the ZL solution at all wavelengths; at $3950$\AA, the
scatter increases by 40\% in response to a change of 50\% in the ISL
flux at all airmasses.  Although this does not allow us to place a
stronger constraint on the error in $I_{\rm scat}(\lambda,t,\chi, {\rm
ISL})$ than those discussed in the Appendix, it does provide
independent verification that the predicted $I_{\rm
scat}(\lambda,t,\chi, {\rm ISL})$ values are in the right range.  It
also emphasizes that an error of 10\% in $I_{\rm
scat}(\lambda,t,\chi,{\rm ISL})$ changes the mean ZL solution by only
0.4\% at 4200--5100\AA.  As discussed in the Appendix, the 
uncertainty in $I_{\rm scat}(\lambda,t,\chi, {\rm ISL})$ contributes an
uncertainty to the ZL measurement in \S\ref{lco.resul} of
$<0.5$\% (see Table \ref{tab:lco.errorbudget}).

In Figures \ref{fig:lco.results2} and \ref{fig:lco.results3} we show
the mean ZL solution at 4600--4700\AA\ (for $C(3900,5100)=1.05$) as a
function of airmass from each of the 16 exposures.  Figure
\ref{fig:lco.results2} shows the solution obtained using all eight
spectral features solutions plotted in Figure \ref{fig:lco.results},
while Figure \ref{fig:lco.results3} shows the solution based on the
four spectral features with the smallest standard deviations in Figure
\ref{fig:lco.results}.  The horizontal line in each plot shows the
mean, the dashed lines show the one--sigma error in the mean. The
difference between the results in the two plots is less than 0.2\%.
No obvious trends appear in either plot as a function of airmass. In
fact, the mean value for points above and below 1.2 airmasses in
Figure \ref{fig:lco.results3} agrees to better than 0.3\% (half of the
error in the mean).  The error bars in Figure \ref{fig:lco.results3}
vary from 1--8\%, indicative of the small number of measurements (four
spectral features) being averaged together to produce the result at
each airmass. The standard deviations indicated by the error bars in
Figure \ref{fig:lco.results2} show less variation from point to point
(1.5--5\%), as eight measurements contribute to each point.

As discussed in the Appendix, we estimate that the uncertainty
in the calculated scattered ZL flux contributes an 
uncertainty to the ZL solution of 1.2\%.  The stability of our ZL
solution with airmass indicates that our calculated net extinction,
which incorperates the ZL scattering models from the Appendix, has the
correct behavior over the night.  However there is no way to
independently infer the accuracy of the mean net extinction from the
{\it rms} scatter in the ZL solution because the spectral shape of
$\tau_{\rm eff}(\lambda,t)$ does not change significantly over the
night (see Figure \ref{fig:eff_tau_ut2}); an error in the mean level
of $\tau_{\rm eff}$ will not affect the {\it rms} scatter in the
solution between exposures.  The errors in our result are
summarized in Table \ref{tab:lco.errorbudget}.  We discuss the
systematic uncertainties further in the next section.

\section{Discussion}\label{lco.discuss}

\begin{figure*}[t] 
\begin{center}
\includegraphics[width=5in,angle=-90]{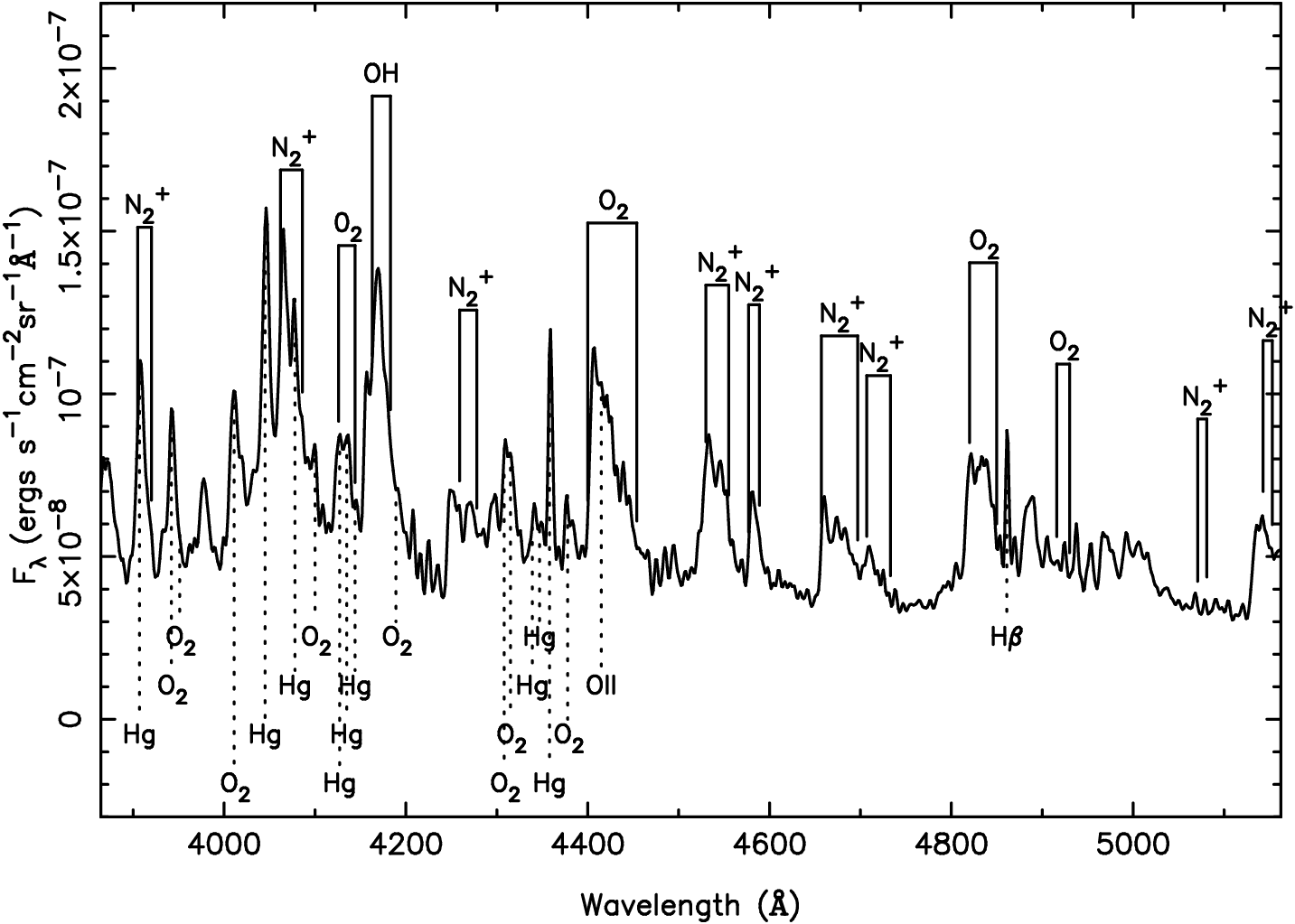}
\caption{\footnotesize Airglow emission lines in the observed spectrum
of the night sky after zodiacal light is subtracted. Broad emission
features (marked above) are molecular rotation--vibration bands of
O$_2$, N$_2$, H$_2$, OH, and NO$_2$.  Collisional de-excitation of
these molecules contributes a continuum as well.  Emission lines
(marked below) are also seen in this wavelength range from atomic
transitions (photoionized O and Hg).  }
\label{fig:lco.airglow}
\end{center}
\end{figure*}

The {\it rms} scatter in our ZL solution is less than 1\%.  This
demonstrates that statistical errors are quite small, be they a result
of instrumental effects or our analysis method.  Systematic
uncertainties due to instrumental effects are also quite small and are
straightforward to quantify (see Table \ref{tab:lco.errorbudget}).
This is demonstrated by the fact that we obtain consistent results to
within 0.3\% from data taken on two different nights, independently
reduced and calibrated.  The  uncertainties in the atmospheric
scattering model described in the Appendix have been considered very
carefully and we believe that the adopted uncertainties are
conservative.  However, the measurement presented here is obviously
complex and might be effected by systematic errors which are more
difficult to anticipate or quantify.

One such systematic effect might include moonlight or sunlight
scattered in the atmosphere.  Based on the scattering analysis in the
Appendix, it is clear that scattering into the line of sight from any
source, even the sun or moon, is negligible when that source is more
than 14 degrees below the horizon.  All of the observations in this
work took place when the sun was more than 18 degrees below the
horizon.  Although the observations took place several days after new
moon, the moon was below the horizon during all of our program
observations, and below 14 degrees for all but 1 exposure.  In
addition, the net effect of such solar-type scattering contributions,
if present, would be to {\it increase} our estimate of the zodiacal
light, and consequentally to artificially {\it decrease} the value of
the inferred EBL in Paper I.  We do not believe that such scattering
is likely to have influenced our results.

As discussed in \S\S \ref{lco.analy} and \ref{lco.resul}, the
contamination of solar features by airglow, while introducing a
systematic error, would not introduce a {\it stable} systematic error:
the flux of airglow features changes constantly through the night.
The stability of our ZL solution in the eight spectral regions we have
used demonstrates empirically that airglow is unlikely to have had a
significant effect on our results.  However, the possibility can't be
ruled out and may introduce a systematic error which we cannot quantify
and is not included in our estimate of the formal errors.

It is also possible that some Doppler shifting occurs in the ZL spectral
features relative to the solar spectrum due to motion of the dust in
the zodiacal plane.  For that reason, we allowed for a shift in
the central wavelength when calculating the correlation but found no
measurable offset.  We note, also, that the results of this method
would not be affected by the slight Doppler broadening which might
affect the spectral features of the ZL, because Doppler broadening
will not alter the total flux across a feature.  The correlation is
unaffected by the saw-tooth effect of subtracting features with
mismatched widths at the level of the 0.3\AA\ Doppler broadening which
is expected at the orientation of these observations (East \& Reay
1984).  Note also that while the resolution of the input spectra used
for calculating the scattered ISL flux is lower than the resolution of
our program observations (4\AA\ versus 2.6\AA), this will not affect
our analysis as long as regions with width $>>$4\AA\ are used in the
analysis. The smallest of our spectral regions is 15\AA.

Figure \ref{fig:lco.airglow} shows the airglow spectrum (the night sky
spectrum after zodiacal light is subtracted) we obtain by this method.
Emission lines from molecular rotation-vibration bands (O$_2$, N$_2$,
H$_2$, OH, and NO$_2$) are labeled, as are some atomic transmission
lines (O and Hg).  Identification of emission features in this
range of the spectrum is not complete (see Schmidtke \etal 1985,
Slanger \& Huestis 1981, Jones \etal 1985 and references therein).

Finally, we note that our measurement of both the mean
flux and ZL color are in very good agreement with typical values for
the similar viewing geometries quoted in the literature (see the
results of Levasseur--Regourd \& Dumont 1980, pictured in Figure
\ref{fig:ZLoverSky}, and Leinert 1998).

\section{Summary}

We have measured the mean surface brightness of the zodiacal light
along a single line of sight towards an extragalactic target using
ground--based spectrophotometry with a 300 arcsec$^2$ field of view.
The observations were made on on 27 and 29 November 1995, simultaneous
with HST observations of the same field. The goal of this coordinated
program is a measurement of the optical extragalactic background
light.  Because the zodiacal light at optical wavelengths results from
sunlight scattered by interplanetary dust concentrated in the ecliptic
plane, the flux towards an extragalactic field varies seasonally.
Variations in the interplanetary dust cloud with time and the solar
cycle may also affect the flux of zodiacal light.  The measurement of
zodiacal light present here is therefore uniquely relevant to the date
and target field.

Our results incorporate explicit calculation of net effect of
atmospheric scattering on terrestrial measurements of the zodiacal
light, and show that these effects are small ($<10\%$) for zodiacal
light measurements far from the Sun.  We find the mean flux to be
109.4\tto{-9} \escsa\ at 4650\AA\ (see Figure \ref{fig:lco.results3}),
and the color to be 5($\pm 1$)\% redder than the solar spectrum per
1000\AA.  The statistical uncertainty in the mean flux is 0.6\%
($1\sigma$), and the systematic uncertainty  is 1.1\%
($1\sigma$).  We discuss additional systematic effects which might
influence this measurement beyond those which are quantified here.
Our results are in good agreement with previous measurements of the ZL
at similar orientations with respect to the ecliptic plane and
scattering geometry (see Leinert \etal 1998 for a recent review).
This is the only optical measurement to date which isolates the ZL
from other uniform backgrounds, including diffuse Galactic light and
extragalactic background light. The color of the ZL as a function of
the line of sight through the interplanetary dust cloud
is further addressed in Paper I.

\acknowledgments

We thank an anonymous referee for comments which have significantly
improved the analysis.  We would also like to thank Carnegie Observatories
and specifically L.\ Searle, A.\ Oemler, and I.\ Thompson for generous
allocations of observing time at Las Campanas Observatory during the
course of this work. The observations would not have been successful
without the generous efforts of O.\ Duhalde, E.\ Cerda, I.\ Thompson
and especially night assistant H.\ Olivares.  We would also like to
thank R.\ Kurucz for kindly providing an electronic version of the
solar spectrum and for helpful discussions.  We have benefited greatly
from discussions with J.\ Dalcanton, S.\ Shectman, and A.\ Williams.
J.\ Dalcanton and I.\ Thompson provided the data for measurements of
the solid angle of the instrument.  RAB would like to thank R.\
Blandford, A.\ Readhead, and W.\ Sargent for financial support in the
early stages of this work.  This work was supported by NASA through
grants NAG LTSA 5-3254 and GO-05968.01-94A to WLF 
and 
through Hubble Fellowship grant HF-01088.01-97A awarded by STScI to RAB.

\appendix

\setcounter{figure}{0}

\section{Atmospheric Scattering}

\begin{figure}[t]
\begin{center}
\includegraphics[width=3.0in]{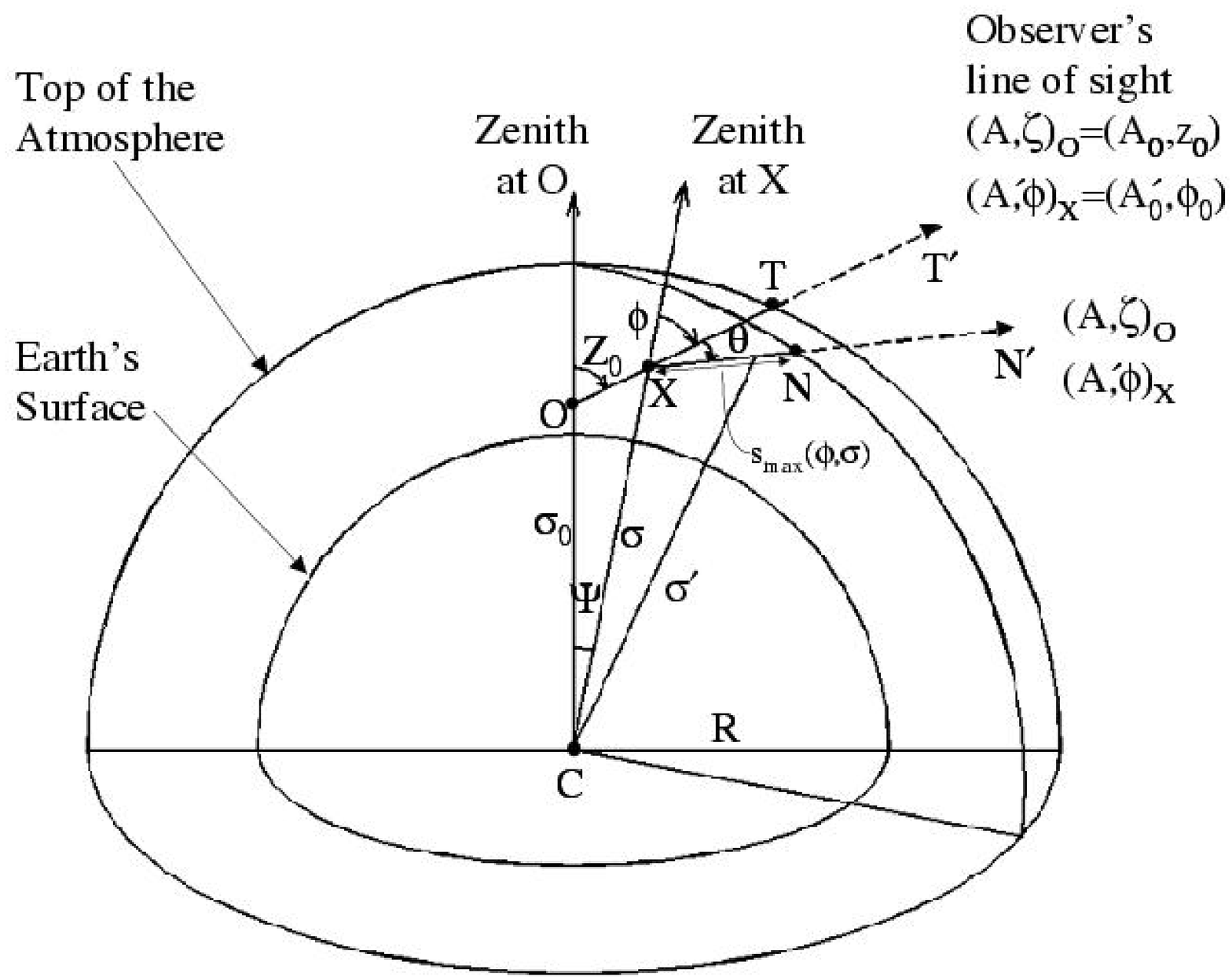}
\caption{\footnotesize Geometry for atmospheric scattering along the
line of sight $OXT$, with the observer at $O$ and the line of sight
exiting the atmosphere at $T$. In the case illustrated, scattering
occurs at $X$ from light entering the atmosphere at $N$ from $N^\prime$.}
\label{fig:ap.scatgeom}
\end{center}
\end{figure}

In this Appendix, we present all scattering calculations and results
used in Section \ref{lco.analy}. Scattering in the atmosphere is a
combination of Rayleigh scattering by molecules and Mie scattering by
particulates. The scattering due to these two components can be dealt
with individually.  We begin by describing a general model for the
scattering in a spherical atmosphere and then discuss the specific
parameters needed to calculate Rayleigh and Mie scattering affecting
observations from Las Campanas Observatory.  Finally, we present the
predicted contribution of scattered light to the program observations
analyzed in this paper. We address zodiacal light and the integrated
starlight as scattering sources separately.

\subsection{Generic Calculations}

\begin{figure}[t]
\begin{center}
\includegraphics[width=3.0in]{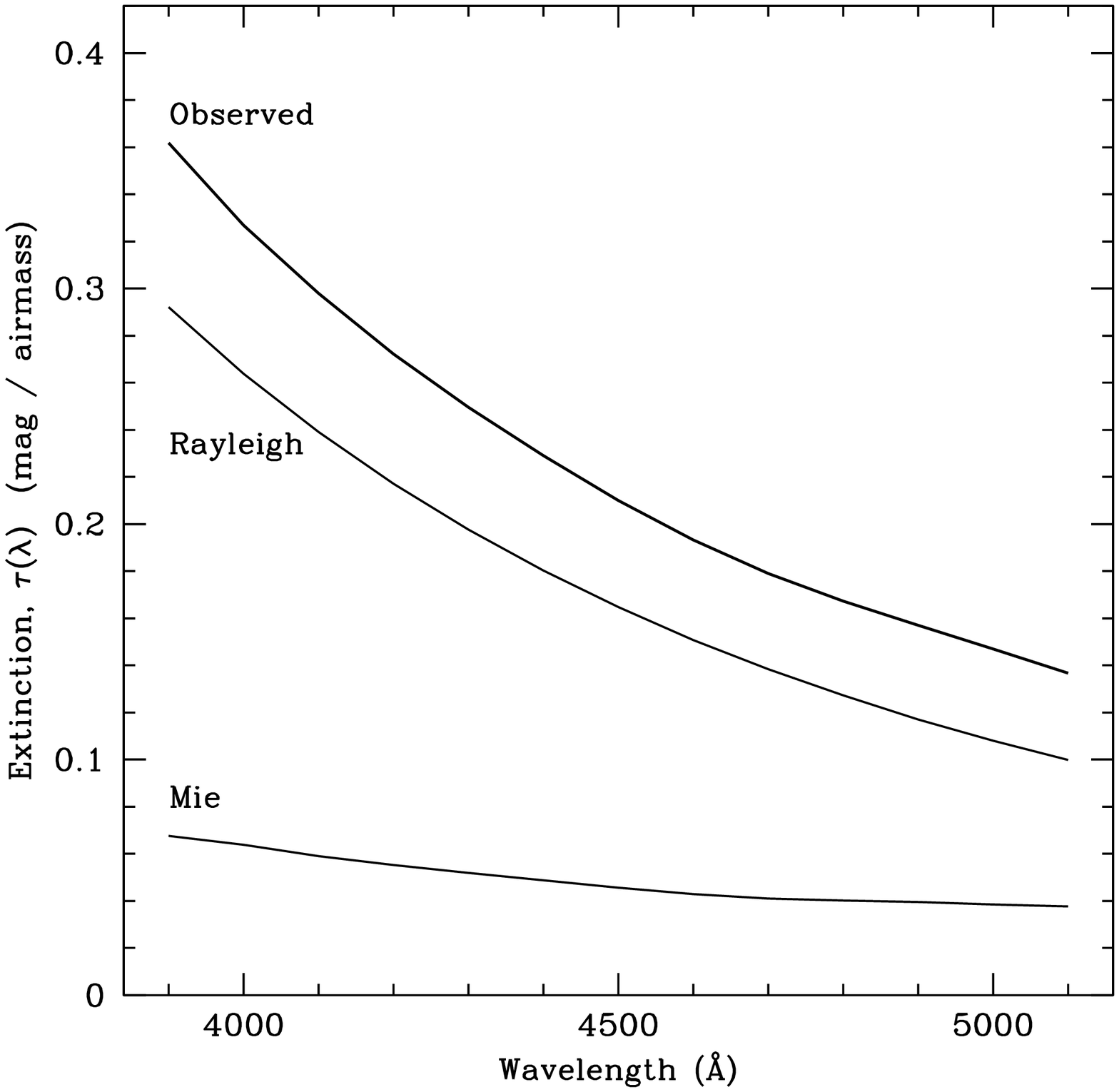}
\caption{\footnotesize We compare  the observed
extinction, $\tau(\lambda)$, derived from observations of standard
stars on 1995 November 27 and 29, with the predicted extinction due to
Rayleigh scattering, which is well known (see text).  The extinction
due to Mie scattering is the difference between the two.}
\label{fig:tau_all}
\end{center}
\end{figure}

The first calculations of radiative transfer in a Rayleigh scattering
atmosphere were published by Chandresekar in 1950.  Since then, a
number of authors have published radiative transfer calculations
addressing Rayleigh and Mie scattering in a curved atmosphere
(e.g. Sekera 1952, Sekera \& Ashburn 1953, and Ashburn 1954), and the
effects of multiple--scattering (Dave 1964,
de Bary \& Bullrich 1964, and de Bary 1964).  Careful measurements of
zodiacal light over the sky and intensity distributions of the daytime
sky have empirically demonstrated the accuracy of those calculations (e.g.
Elterman 1966; Green, Deepak, \& Lipofsky 1971; Weinberg 1964,
Dumont 1965).

To calculate the atmospheric scattering affecting the observations
described in this Paper, we begin by adopting the scattering geometry
and coordinate system definitions used by Wolstencroft \& van Breda
(1967, hereafter WvB67), illustrated in Figure~\ref{fig:ap.scatgeom}:
$(A,\zeta)_O$ and $(A^\prime, \phi)_X$ are azimuth/zenith--distance
coordinate systems centered on the observer at $O$ and a generic
point, $X$, along the line of sight, respectively.  The problem is
then to calculate the brightness observed at $O$, along the line of
sight $(A,\zeta)_O=(A_0, z_0)_O$.

Following WvB67, scattering occurs at the point $X$ for radiation
which entered the atmosphere at the point $N$ from the direction
$N^\prime$, given by $(A,\zeta)_O$ or $(A^\prime, \phi)_X$.  The light
arriving at $X$ from $N^\prime$ can be expressed as
\begin{equation}
I_X = {\sc L}(A^\prime, \phi) \ \sin\phi  \ 
e^{-C_{\rm ext}(\lambda)h_1(\phi,\sigma)} ds,
\label{eq:IX}
\end{equation}
in which the above-the-atmosphere source has flux ${\sc L}(A^\prime,
\phi)$, light is attenuated by $e^{-C_{\rm ext}(\lambda)
h_1(\phi,\sigma)}$ as it travels along $NX$, and $s$ is the distance
along that path.  Attenuation is a function of $C_{\rm ext}(\lambda)$,
the extinction cross section of the scattering particles in cm$^2$,
and of $h_1(\phi,\sigma)$, the effective column density of particles
along the line of sight.  The effective column density is defined by
the local zenith angle, $\phi$, and the distance, $\sigma$, which
defines the point $X$ relative to the center of the Earth (see Figure
\ref{fig:ap.scatgeom}):
\begin{equation}
h_1(\phi,\sigma)=\int_0^{s_{\rm max}(\phi,\sigma)} n(\sigma^\prime) ds^\prime,
\end{equation}
in which $s_{\rm max}(\phi,\sigma)$ is the distance from $X$ to the
top of the atmosphere at $N$, and $n(\sigma^\prime)$ is the
atmospheric number density of molecules in cm$^{-3}$ as a function of
distance from the center of the Earth, $\sigma^\prime$, and 
as a function of the distance $s^\prime$ along the line $XN$.

\begin{figure}[t]
\begin{center}
\includegraphics[width=3.0in]{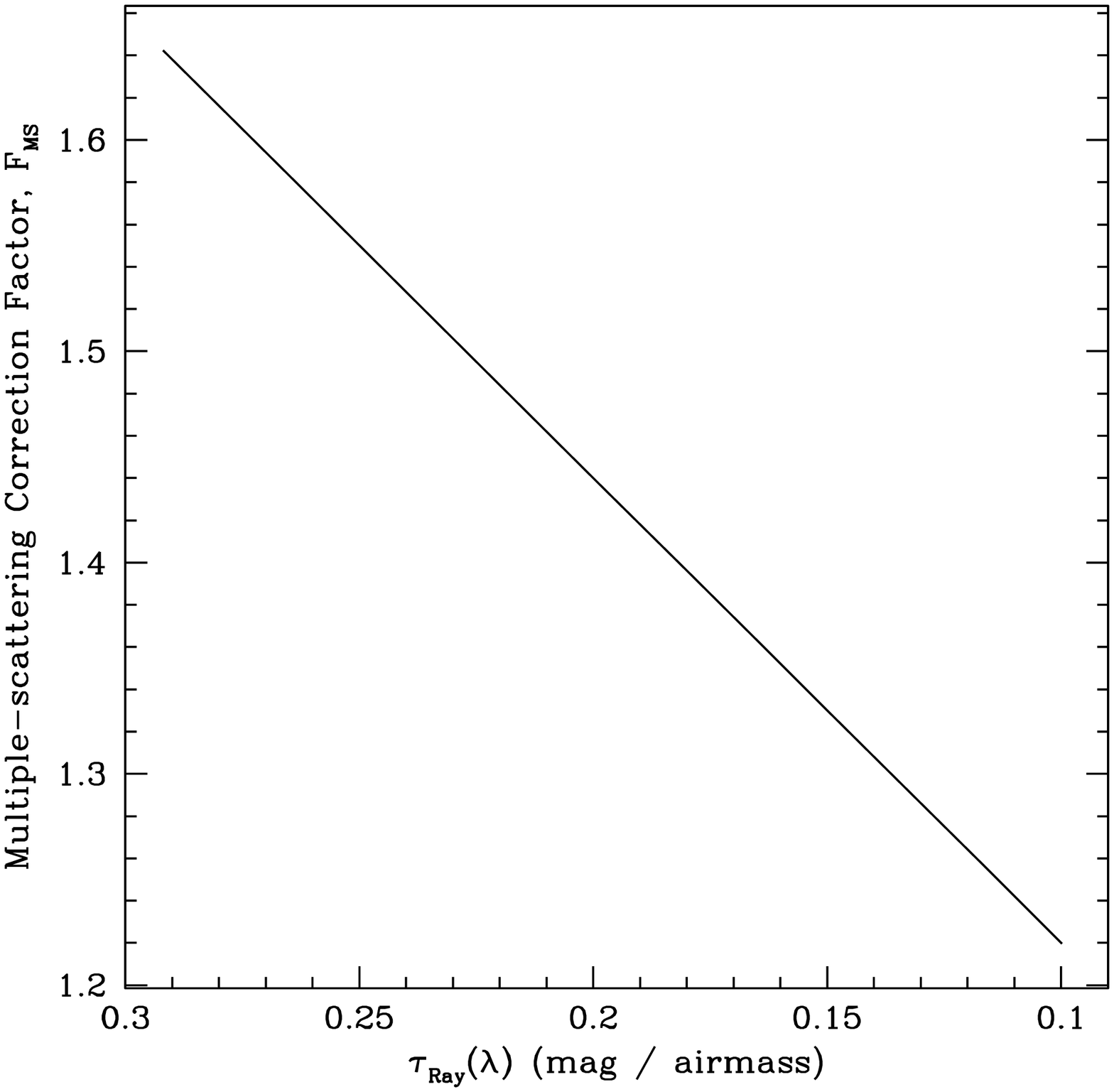}
\caption{\footnotesize Correction factor for multiple scattering,
$F_{\rm MS}$, as a function of the Rayleigh extinction.}
\label{fig:FmsRay}
\end{center}
\end{figure}

The light scattered towards the observer from $X$ is then
\begin{equation}
{{3} \over {16 \pi}} \ 
 C_{\rm scat}(\lambda) n(\sigma){\sc P}(\theta) \ 
I_X  d\phi\, dA^\prime\, ds,
\end{equation}
in which ${\sc P}(\theta)$ is the scattering phase function and $I_X$,
the flux arriving at point X is given in Equation
\ref{eq:IX}. Finally, the scattered light is further attenuated by the
factor $e^{-C_{\rm ext}(\lambda) h_2(z_0,\sigma)}$, in which 
\begin{equation}
h_2(z_0,\sigma)= h_1(z_0,\sigma_0) - h_1(z_0-\psi, \sigma).
\end{equation}
The total flux scattered into the line of sight $(z_0,A_0)$ from
sources distributed over entire visible hemisphere of the sky is then
\begin{eqnarray}
I_{\rm scat} (z_0,A_0)  \!\!\!\!  &=&\!\!\!\!
	{ 3 C_{\rm scat}(\lambda) \over 16 \pi}
	\int_{s=0}^{s_{\rm max}(z_0, \sigma_0)}
	\int_{\phi=0}^{\pi/2 + f(\sigma)}
	\int_{A^\prime=0}^{2\pi} \nonumber \\
	 & & n(\sigma) \ {\sc P}(\theta) \ 
         {\sc L}(A^\prime, \phi) \sin\phi\    \\   & & 
	e^{ -C_{\rm scat}(\lambda) \, [h_1(\phi, \sigma) + h_2(z_0,\sigma)]} \,
      d\phi\, dA^\prime\, ds.\nonumber 
\label{eq:scat_integral}
\end{eqnarray}
The visible sky at the point $X$ dips below the observer's horizon at
large values of $s$. Hence, the limit of the integral over $\phi$, is
greater than $\phi/2$ by the value $f(\sigma) = \cos^{-1}(R/\sigma)\le
14^\circ$, where $R$ is the radius of the Earth (6371 km).  The
equations needed to change variables between $(A^\prime, \phi)$ and
$(z_0,A_0)$ are given WvB67.

The phase function for Rayleigh scattering is $P(\theta) =
1+\cos^2(\theta)$.  The atmospheric density is given by the standard
barometric formula $n(\sigma) = n_0 e^{-H/H_0}$, where $H$ is the
altitude above sea level ($H=\sigma-R$), the scale height is $H_0 =
7.99$\,km, the density at sea level is $n_0=2.67\times10^{19}$
cm$^{-3}$ , and the effective scattering cross--section for air is
$C_{\rm scat}=7.78\times10^{-27} (\lambda / 4600\AA)^{-4}$\,cm$^2$
(see Schubert \& Walterscheid 1999 and van de Hulst 1952).  For
molecules in the atmosphere, extinction is entirely due to scattering,
so that $C_{\rm ext} = C_{\rm scat}$.  Atmospheric extinction due to
Rayleigh scattering is then $\tau^R(\lambda) = C_{\rm
ext}(\lambda)\int_R^\infty{n(\sigma)d\sigma}$.  For the duPont
telescope at Las Campanas, which is at an altitude of 2.28 km, the
expected Rayleigh extinction is $\tau^R(4600\AA)=0.12$.

\begin{figure}[t]
\begin{center}
\includegraphics[width=3.0in]{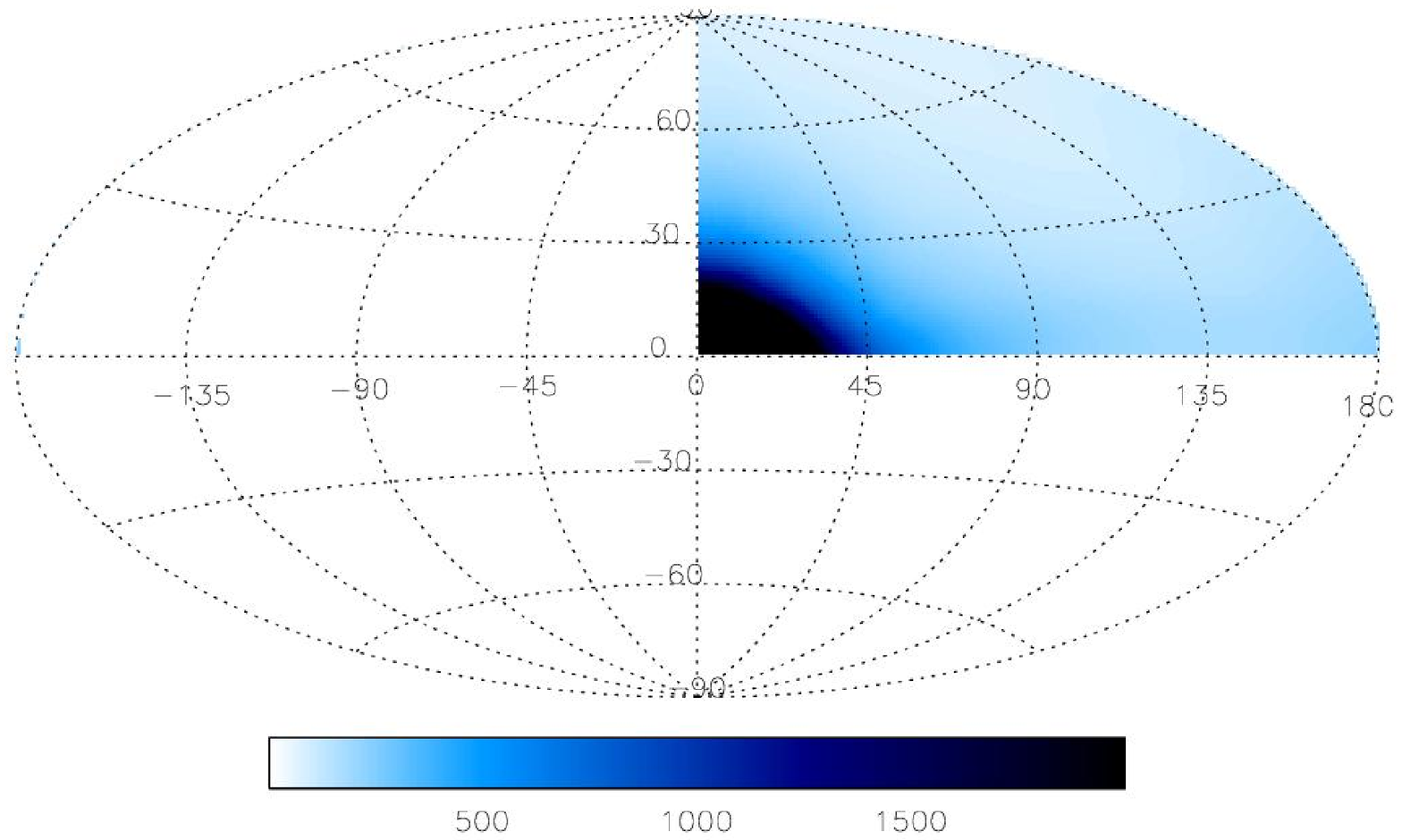}
\caption{\footnotesize Zodiacal light over the sky is plotted in
sun-centered ecliptic coordinates, so that the Sun is at
($\lambda-\lambda_\odot=0^\circ,\beta=0^\circ$), the anti-solar direction
is ($\lambda-\lambda_\odot=180^\circ,\beta=0^\circ$), and the ecliptic
poles are at $\beta=\pm 90^\circ$.  All four quadrants are assumed to be
the same, which is correct to within a few percent during the time of
year when our observations occur.}
\label{fig:ZLoverSky}
\end{center}
\end{figure}

The phase function for Mie scattering by particulates in the
atmosphere depends on the distribution of particle sizes, and must be
empirically determined.  We adopt the phase function measured by
Green, Deepak \& Lipofsky (1971) from their complete analysis of the
Mie (particulate) scattering and Rayleigh (molecular) scattering
components of the atmosphere based on the scattering of sunlight.
Their results are in good agreement with theoretical scattering models
and other estimates of the size-distribution of particles in the
atmosphere and have a negligible dependence on wavelength for our
purposes (see Elterman 1966, and Deepak \& Green 1970).  The
scattering and extinction coefficients for particulate scattering are
a function of the size distribution of particles and vary with time
and geography.  The extinction due to Mie scattering can, however, be
inferred from the observed extinction for a point source and the
calculated Rayleigh extinction coefficient: $\tau^{\rm M} = \tau_{\rm
obs} - \tau^{\rm R}\sim 0.05$ at 4500\AA\ (see Figure
\ref{fig:tau_all}).  This value is in good agreement with estimates
for Tenerife by Dumont (1965) and for Haleakala by Weinberg
(1964). This is not surprising as our observed $\tau_{\rm
obs}(\lambda)$ is consistent with the CTIO curves (Baldwin \& Stone
1984, Stone \& Baldwin 1983), and $\tau^R(\lambda)$ is simply a
function of the atmospheric density.

\begin{figure*}[t]
\begin{center}
\hbox{
\includegraphics[width=2.2in]{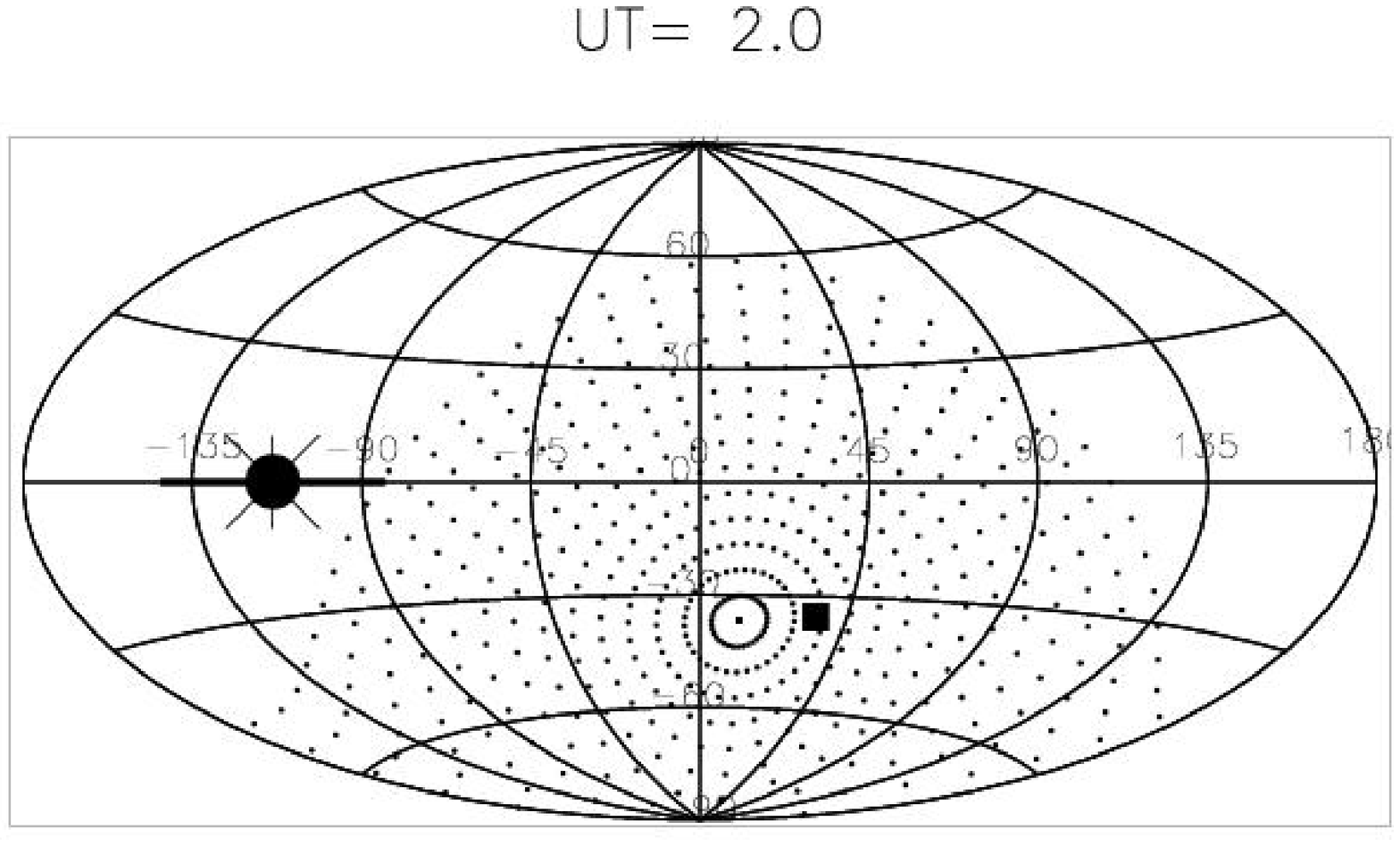}
\includegraphics[width=2.2in]{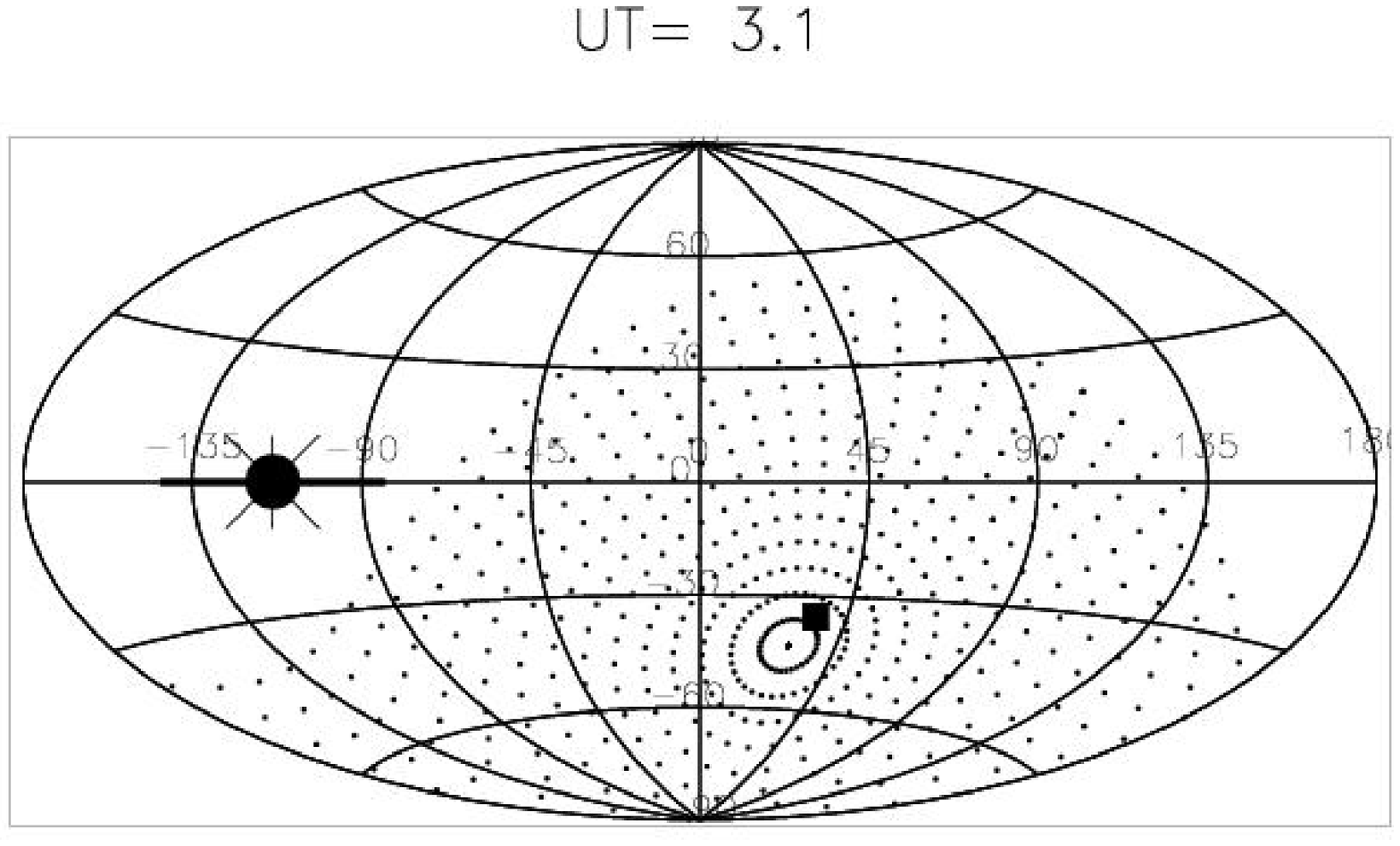}
\includegraphics[width=2.2in]{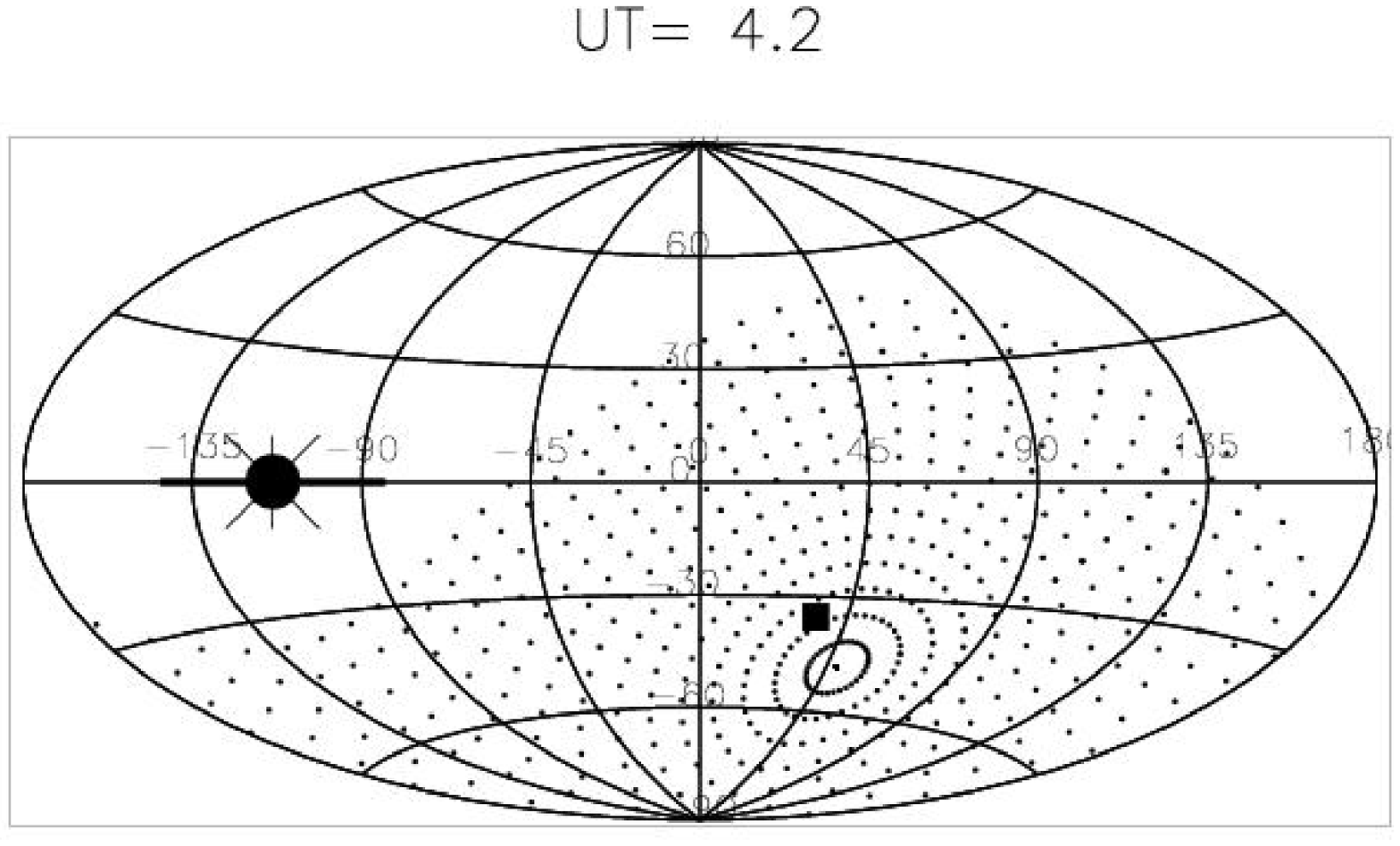}}
\hbox{
\includegraphics[width=2.2in]{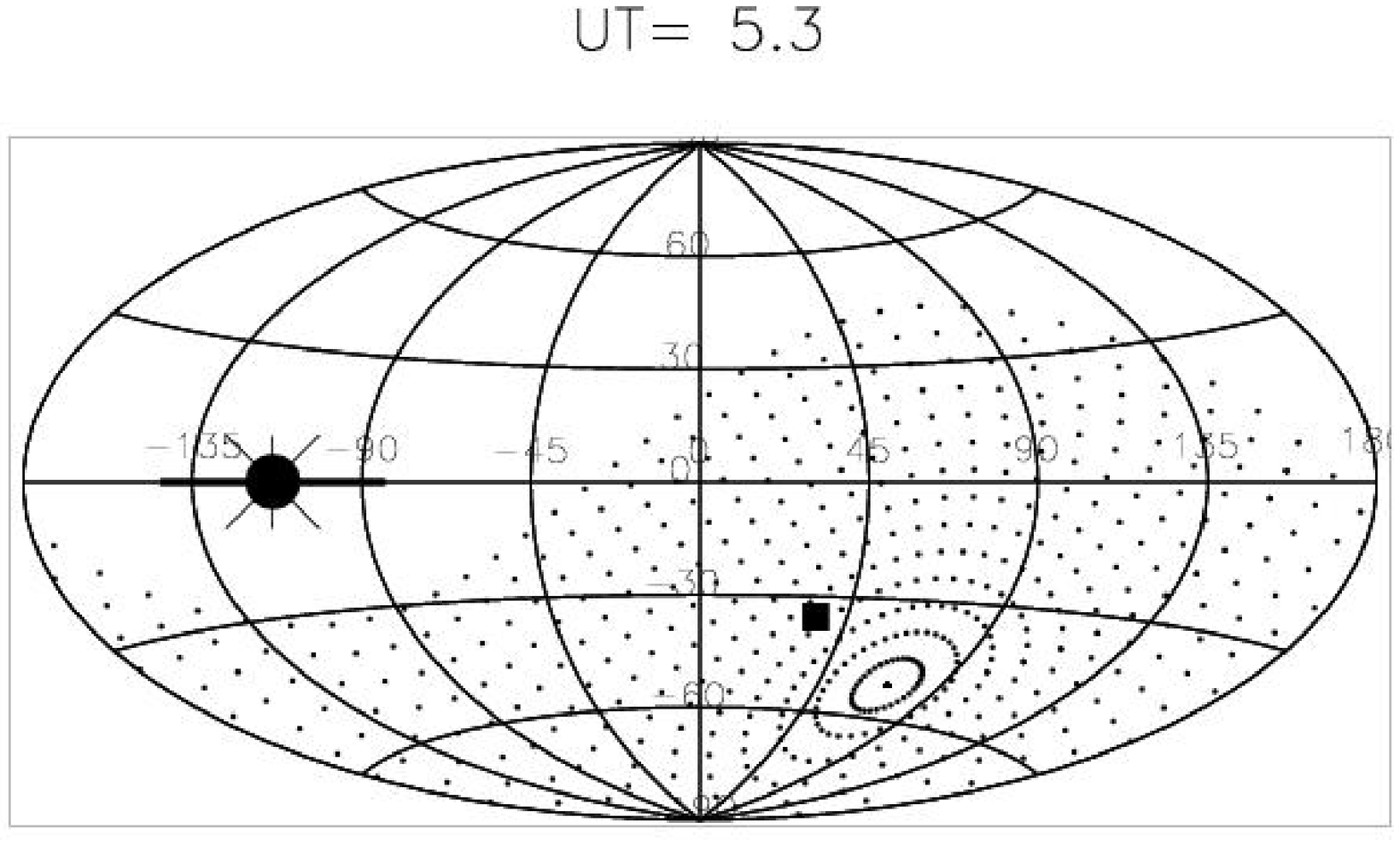}
\includegraphics[width=2.2in]{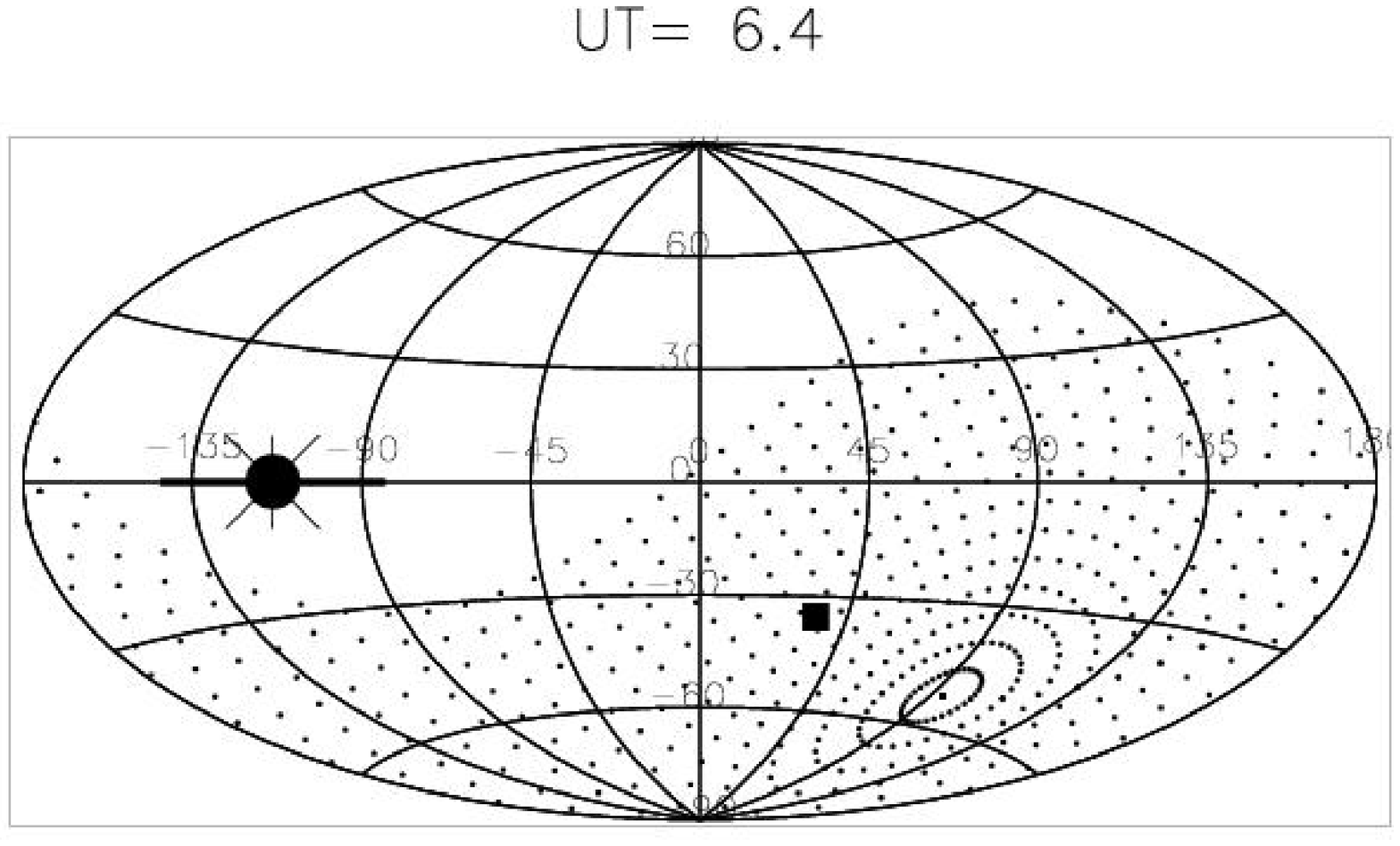}
\includegraphics[width=2.2in]{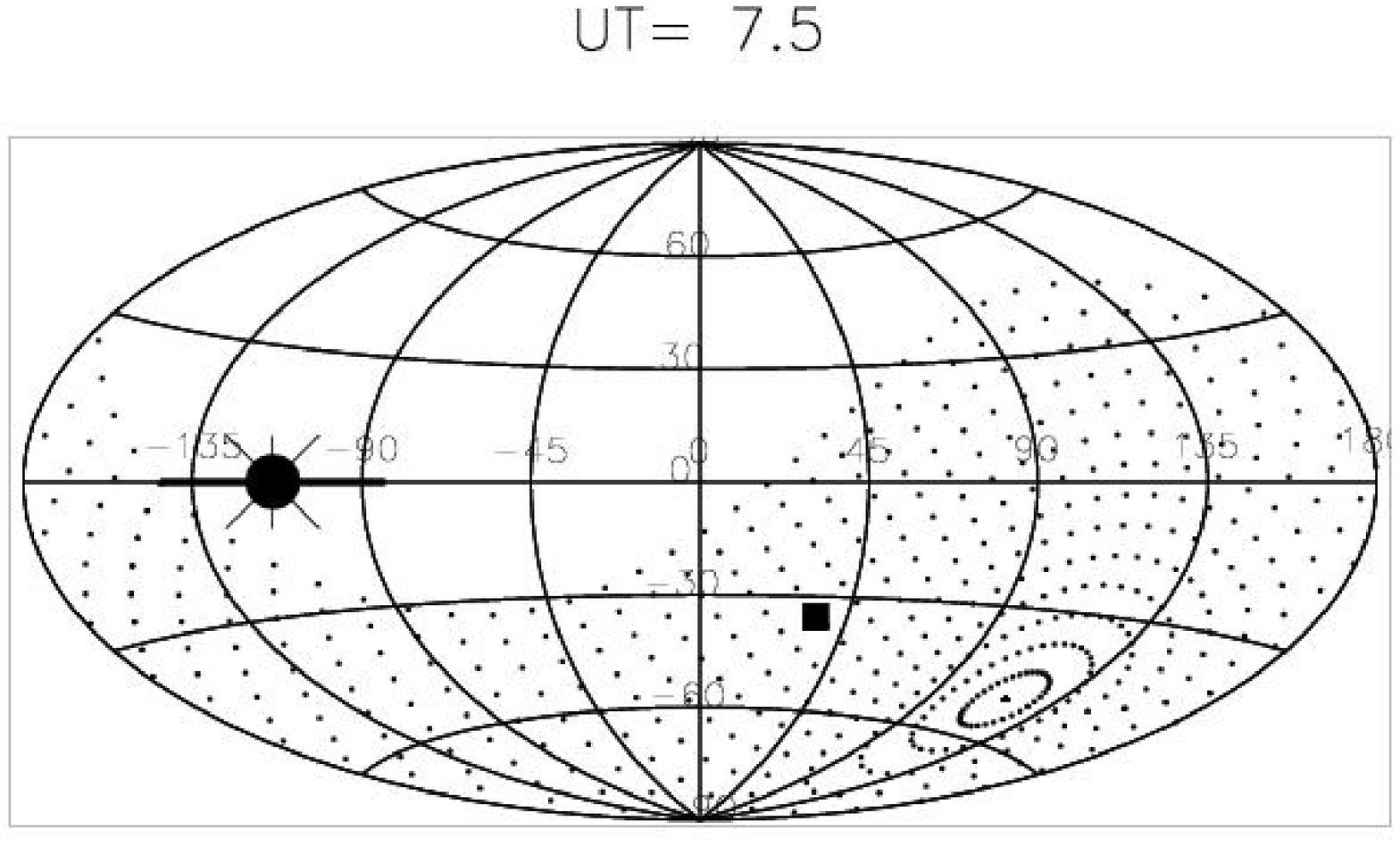}}
\caption{\footnotesize Integration pattern for calculating scattered
zodiacal light.  Each plot is an Aitoff projection of the sky in
Ecliptic coordinates. The position of the Sun and the ecliptic plane
are indicated by the Sun symbol with a bar through it.  Obviously, the
ecliptic plane is along the $0^\circ$ latitude line.  Our target field
is indicated by the square.  The ecliptic coordinates of the zenith
can be seen as the ``bullseye'' center of the integration pattern. The
motion of the target relative to the local zenith is clear as the
target moves from east to west of the zenith point in the integration
pattern.  The coordinates of the local horizon are traced by the edge
of the integration pattern.  The Sun is obviously located just
to the west of the horizon at the start of the night (UT=2.0 hr) and
just to the east of the horizon at the end of the observing night
(UT=7.5 hr).}
\label{fig:ecliptic_aitoff}
\end{center}
\end{figure*}

Unlike the case for molecules, the attenuation caused by particulates
is not entirely due to scattering.  Staude (1975) adopts values of
$C_{\rm scat} = 4.47\times 10^{-11}$ cm$^2$ and $C_{\rm ext} =
7.53\times 10^{-11}$ cm$^2$ for dry air at 4200\AA. With a sea level
density of $n_0=1.11\times10^4$\,cm$^{3}$, and a distribution scale
height of only $h_0 = 1.2$\,km, these parameters give $\tau^{\rm
M}\sim 0.01$ at 2 km.  We have scaled $C_{\rm scat}$
and $C_{\rm ext}$ to give values consistent with our observed value of
$\tau^{\rm M}$.  Scaling $H_0$ or assuming a different value of 
$H$ would have the same effect on $I^M_{\rm scat}(\lambda,{\rm ZL})$.

The scattering model discussed above describes a single scattering event.
However, multiple scattering events become significant for scattering
angles $\theta \gta 30^\circ$ (de Bary 1964, de Bary \& Bullrich
1964).  Consequently, we apply a multiple scattering correction for
Rayleigh scattering which is adopted from Dave (1964) and plotted in
Figure \ref{fig:FmsRay}. The correction factor plotted in Figure
\ref{fig:FmsRay} is simply the factor by which the intensity of
scattered light increases over the single-scattering case. Multiple
scattering does not occur due to particulates (Mie case) because
of the small scattering angles which dominate that process and 
very small values of $\tau^M(\lambda)$.

To confirm the accuracy of our calculations, we checked our scattering
model against published results of Staude (1975), WvB67, and Ashburn (1954)
for a uniform, sky--filling source of unit flux. We find that our
results are consistent to within 4\% for zenith angles $z\le80^\circ$
before the multiple scattering correction is applied.  (WvB67 predates
evidence for the effects of multiple scattering, and Staude adopts the
same corrections used here.)  The uncertainty in the multiple
scattering correction is roughly 4--7\%, increasing with larger values
of $\tau$.

Using the expressions above, we calculate the scattered light flux,
$I_{\rm scat}(\lambda)$, resulting from Mie scattering by particulates
and Rayleigh scattering by molecules throughout the nights of our
observations. The results depend explicitly on the absolute position
of the Sun and the Galactic center relative to the observatory and
relative to the target field. In the following sections, we consider
the cases of zodiacal light (ZL) and integrated starlight (ISL) as the
extra-terrestrial source of flux separately.

\subsection{Zodiacal Light}\label{append.zl}

\begin{figure}[t] 
\begin{center}
\includegraphics[width=2.75in,angle=-90]{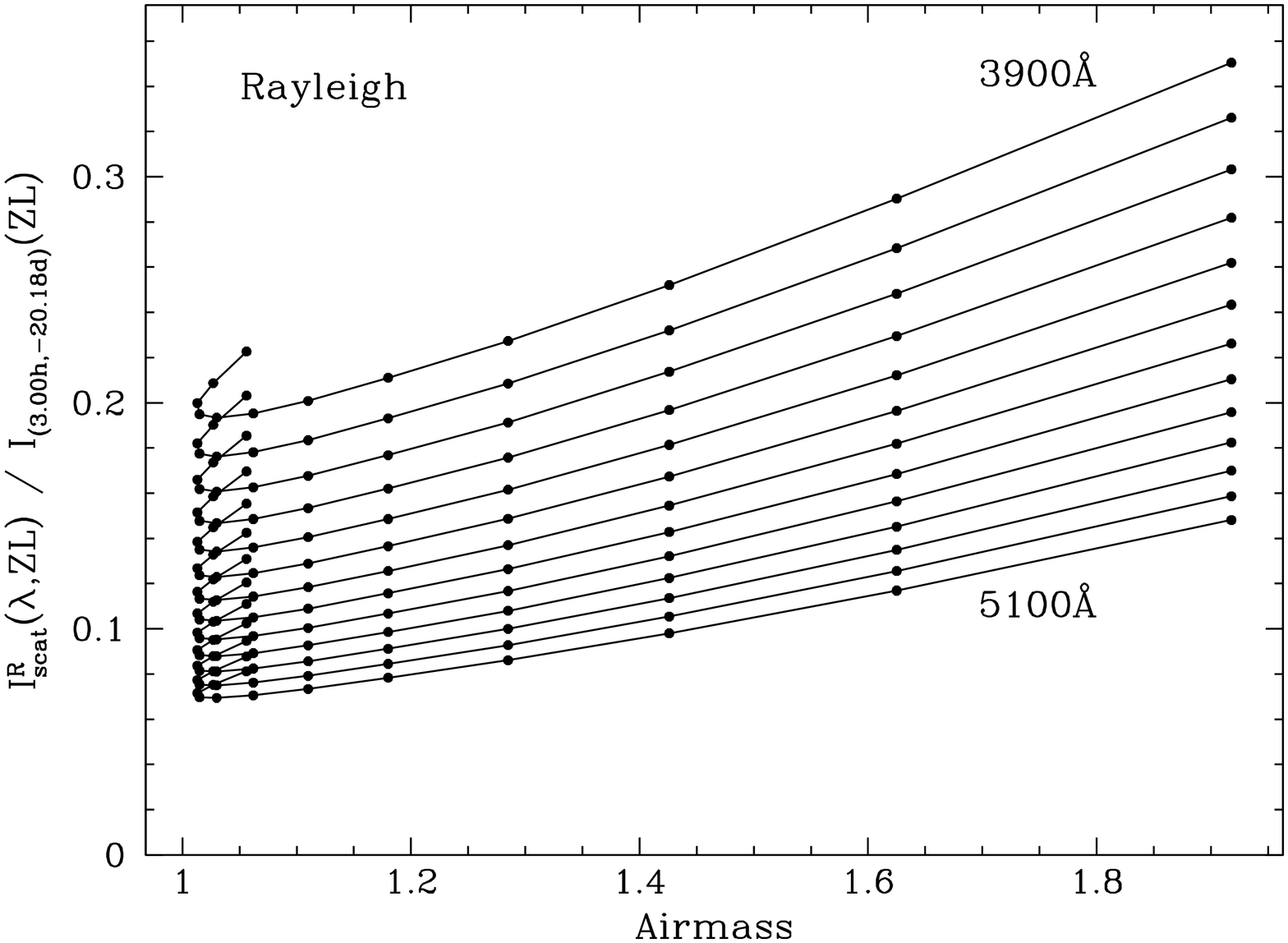}
\caption{\footnotesize The Rayleigh scattered zodiacal light
contribution to our observed night sky flux 
on the nights of 1995 November 27-29 as a function of airmass at 100\AA\ intervals from
3900\AA\ to 5100\AA. Flux decreases monatonically with wavelength at any
airmass, following the behavior of $\tau^{\rm R}(\lambda)$.
$I^R_{\rm scat}(\lambda, {\rm ZL})$
is given as a fraction of the above-the-atmosphere zodiacal light
flux in our target field; this removes the spectrum of the zodiacal light
itself and emphasizes the wavelength dependence of the scattering.
The each line shows the change in $I^R_{\rm scat}(\lambda, {\rm ZL})$
as the target field goes from slightly east of zenith, through
zenith, to far west of zenith during the night. Dots mark half-hour
intervals between UT=2.0 hr and UT=7.5 hr.}
\label{fig:ray_gain_airmass}
\end{center}
\end{figure}

\begin{figure}[t]
\begin{center}
\includegraphics[width=2.75in,angle=-90]{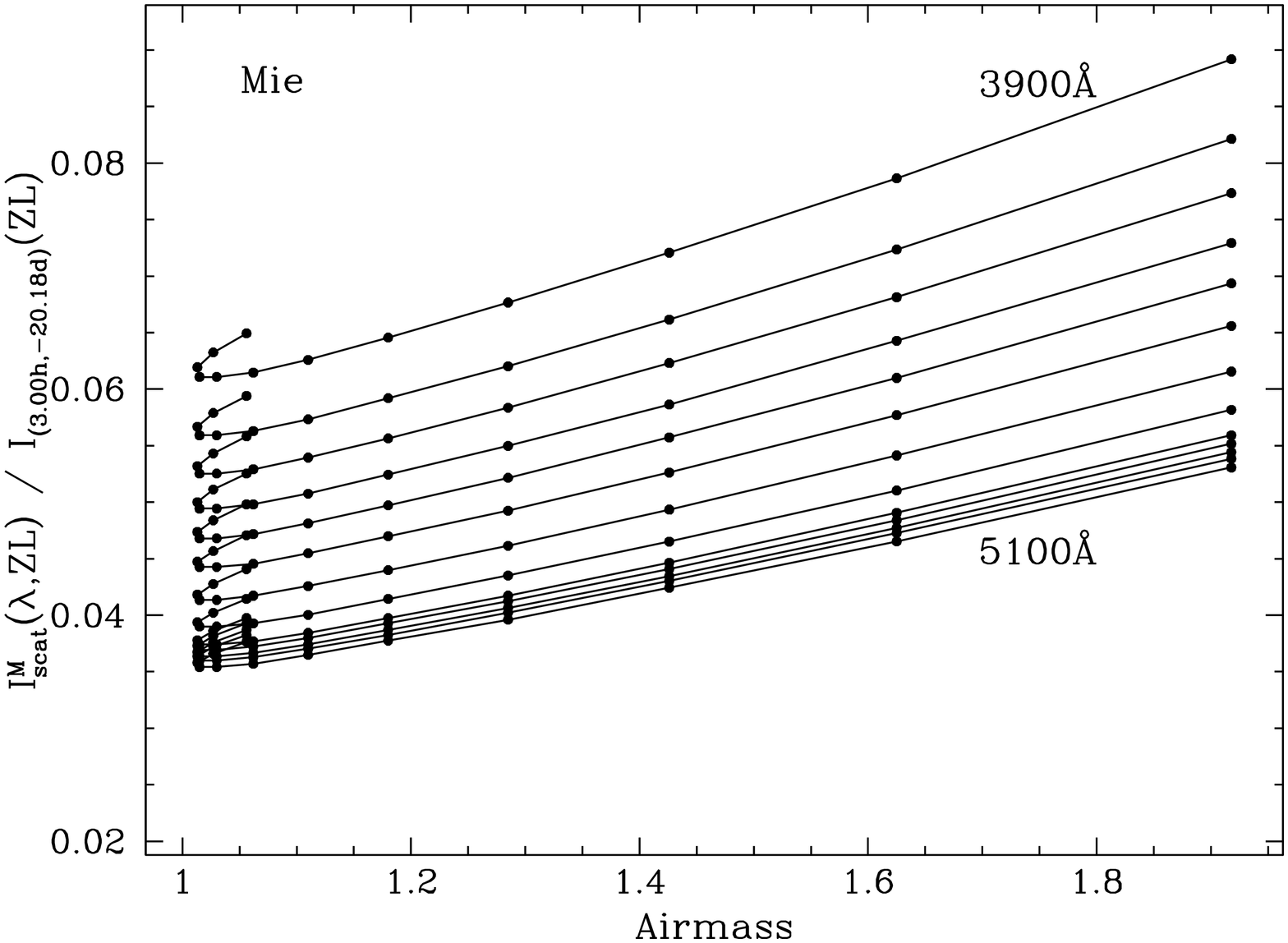}
\caption{\footnotesize Same as Figure \ref{fig:ray_gain_airmass}, but
here we plot the contribution of Mie scattered zodiacal light along
the line of sight. Again, the scattered light flux is plotted as a
fraction of the above-the-atmosphere zodiacal light flux in our target
field.}
\label{fig:mie_gain_airmass}
\end{center}
\end{figure}

To calculate the scattered zodiacal light along the line of sight of
our observations, we adopt values for the zodiacal flux given in
Leinert \etal (1998), which are taken from Levasseur-Regourd \& Dumont
(1980) with values at elongations $\epsilon<30^\circ$ added from
space--based observations.  These ZL values are above-the-atmosphere
fluxes and are in excellent agreement with later space--based results
(see Leinert \etal 1981).  To obtain a smooth flux distribution of ZL
on the sky (see Figure~\ref{fig:ZLoverSky}), we use the spherical
interpolation method developed by Renka (1997).

In Figure \ref{fig:ecliptic_aitoff}, we show the integration pattern
in $\phi$ (zenith angle) and $A^\prime$ (azimuth) used to calculate
$I_{\rm scat}(\lambda, {\rm ZL})$ (Equation \ref{eq:scat_integral}) at the indicated
times during the nights of our observations.  Actual calculations were
done with twice the number of integration points shown in the figures.
The visible part of the sky (shown by the integration pattern) is
at least 30 degrees from the Sun during our observations.

\begin{figure}[t]
\begin{center}
\includegraphics[width=2.75in,angle=-90]{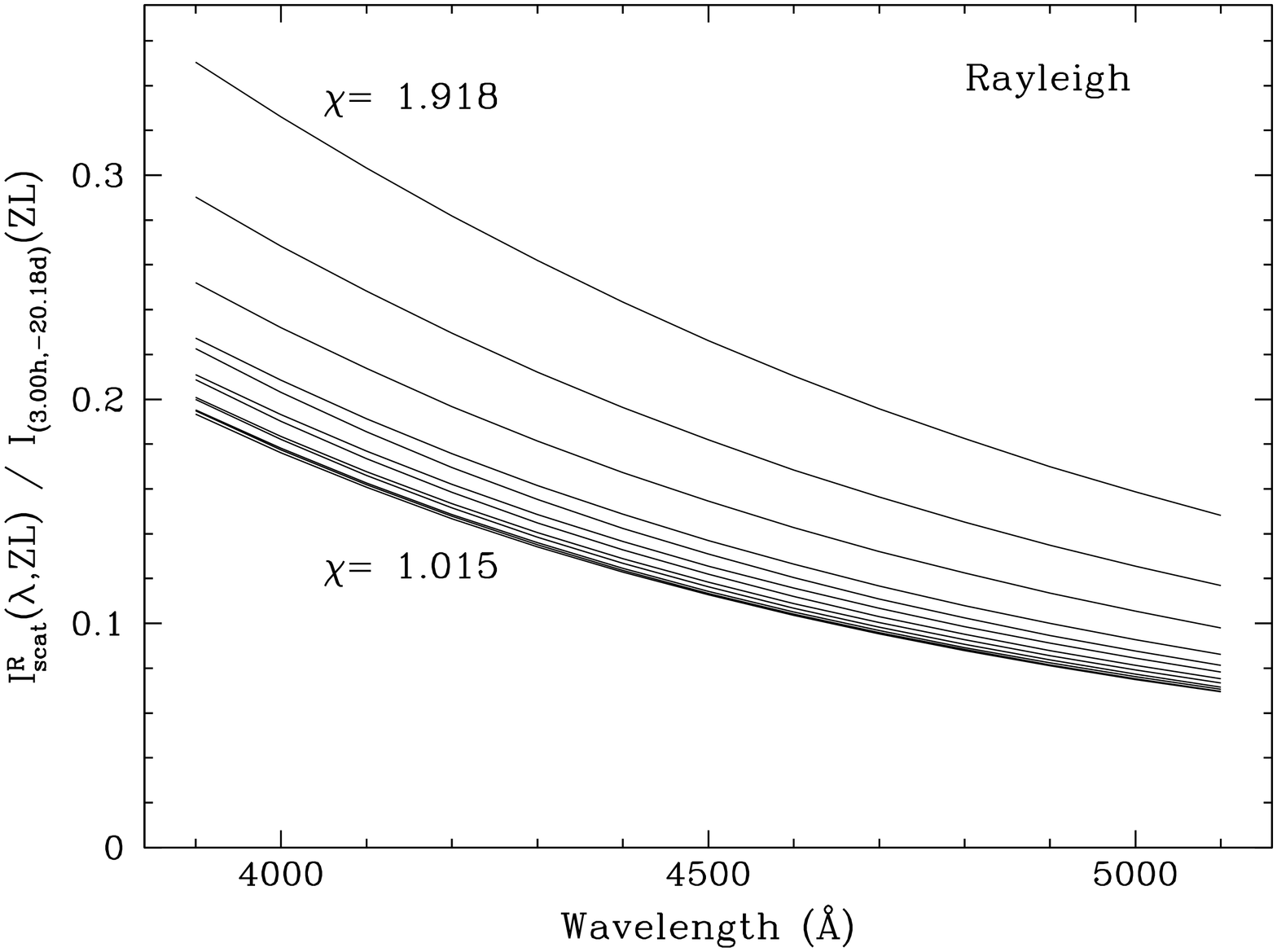}
\caption{\footnotesize Same as Figure \ref{fig:ray_gain_airmass}, but
here each line indicates $I^R_{\rm scat}(\lambda, {\rm ZL})$ as a function of
wavelength at discrete times.  Scattering is maximized at high airmass
(far west of zenith), and minimized slightly west of zenith, when the
airmass is still relatively low and the Sun has had time to set far
enough that highest flux regions of zodiacal light are no longer in
the visible hemisphere of the sky.}
\label{fig:ray_gain_wavel}
\end{center}
\end{figure}

\begin{figure}[t]
\begin{center}
\includegraphics[width=2.75in,angle=-90]{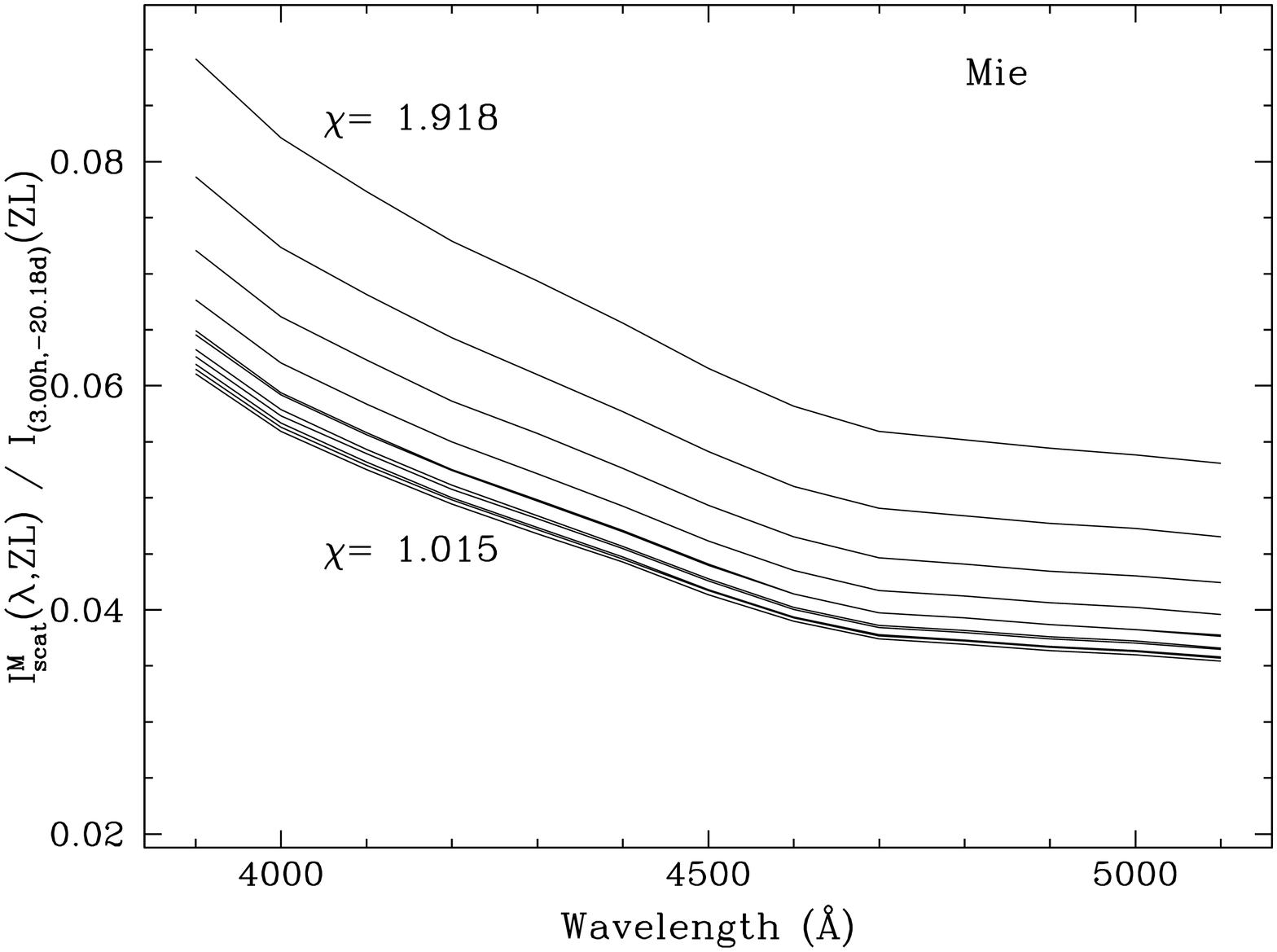}
\caption{\footnotesize Same as Figure \ref{fig:ray_gain_wavel}, but
showing the Mie scattered zodiacal light.}
\label{fig:mie_gain_wavel}
\end{center}
\end{figure}

As a technical detail, we have made the simplifying assumption that
the spectral shape of the ZL over the visible hemisphere is
uniform. That is, only the mean flux of the ZL changes.  Although
there are variations in the color (defined in Equation \ref{eq:color})
of the ZL over the range 3900-5100\AA\ from $\epsilon =30^\circ$ to
$\epsilon =180^\circ$, the total change is empirically less than $\sim
8\%$ and our target is in the center of the expected color range
(e.g., Frey \etal 1974, Leinert \etal 1981).  We have run trail
scattering models in which we change the flux with $\epsilon$ over the
sky by $\pm 4\%$, and we find that the effect on the predicted
scattered flux is changed by 0.4\% at airmasses higher than 1.6, and
0.2\% at the lowest airmass.  In other words, by ignoring the color
variation in ZL over the sky, the scattered light model will be wrong
by 0.2\% at 3900\AA\ relative to the value at 5100\AA, or $\pm 0.1$\%
over the range 3900-5100\AA\ for our observing situation 
(positions of the Sun relative to the target and the horizon). 

\begin{figure}[t] 
\begin{center}
\includegraphics[width=2.75in,angle=-90]{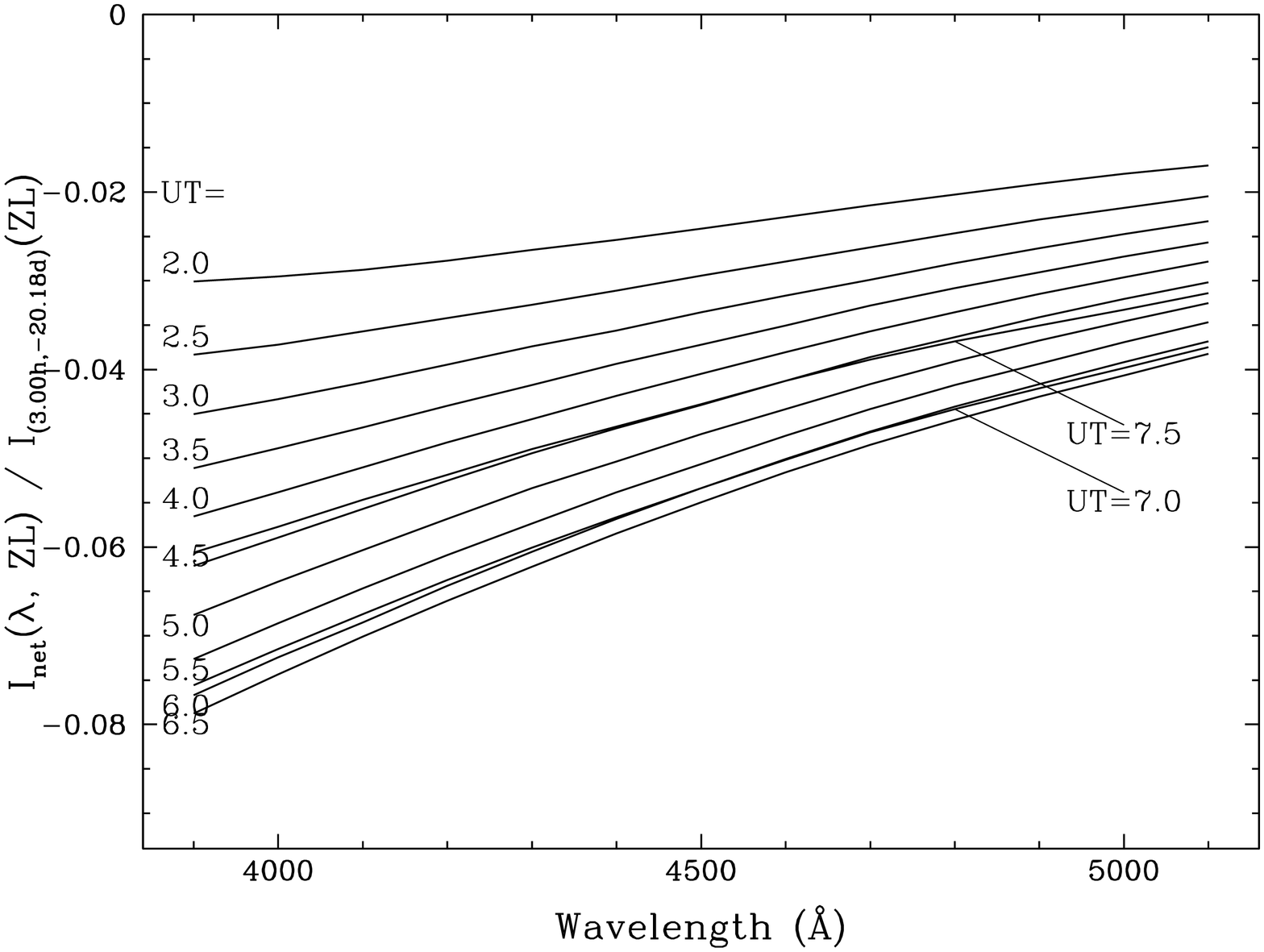}
\caption{\footnotesize The net effect of the atmosphere on the
observed spectrum of zodiacal light.  $I_{\rm net}({\rm ZL})$ is the flux
received from the target field plus the scattered light coming from
the entire visible hemisphere of the sky: $I_{\rm net}(\lambda,ZL)= I_{\rm
scat}(\lambda,{\rm ZL}) - I_{(3h,-20d)}({\rm ZL})\,(1-e^{-\chi\tau(\lambda,{\rm
obs})})$.  $I_{\rm net}(\lambda,{\rm ZL})$ is plotted as a function of wavelength at
discrete times, as labeled.  Flux units are the same as in Figures
\ref{fig:ray_gain_airmass} -- \ref{fig:mie_gain_wavel}.}
\label{fig:net_wavel}
\end{center}
\end{figure}

\begin{figure}[t] 
\begin{center}
\includegraphics[width=2.75in,angle=-90]{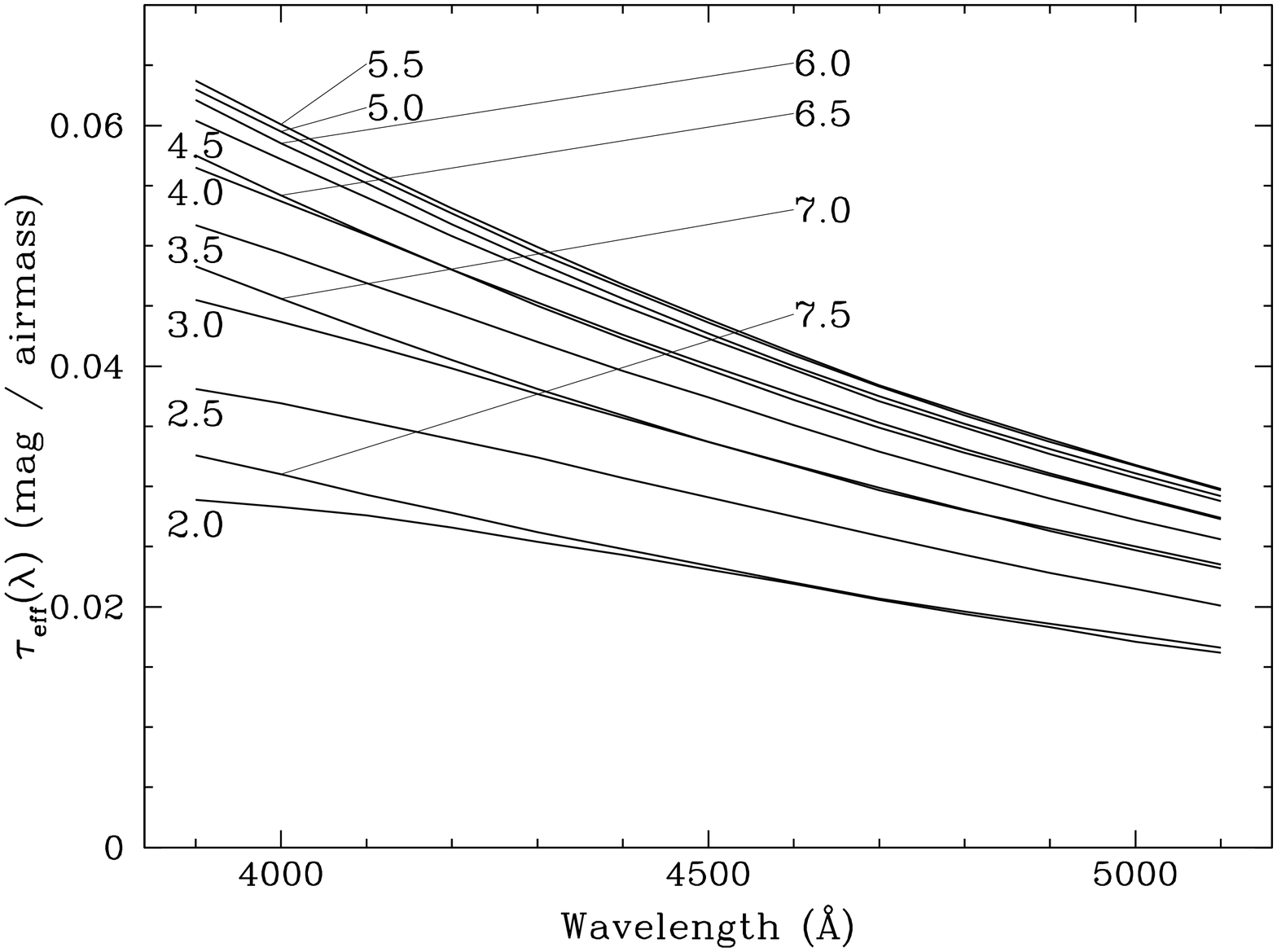}
\caption{\footnotesize Each line shows the effective extinction for
the zodiacal light as a function of wavelength for our observations at
the indicated UT. The effective extinction corresponds to the net loss of
light relative to the above--the--atmosphere flux of the zodiacal
light in our field of view (see Figure \ref{fig:net_wavel}). For
comparison with the total observed extinction derived from standard
stars see Figure \ref{fig:lco.extin}. }
\label{fig:eff_tau_ut2}
\end{center}
\end{figure}

The predicted Rayleigh and Mie scattering flux of ZL, $I^{\rm R}_{\rm
scat}$ and $I^{\rm M}_{\rm scat}$, respectively, along the line of
sight to our target field throughout our observations is shown in
Figures \ref{fig:ray_gain_airmass}--\ref{fig:mie_gain_wavel}.  In
those Figures, we show the scattered light as a function of the
above-the-atmosphere ZL flux in target field at
($\alpha=3.00$h,$\delta=-20.18$d), $I_{\rm (3h,-20d)}({\rm ZL})$.
This removes the spectrum of the ZL
and highlights the wavelength dependence of $I_{\rm scat}$.
The predicted scattered flux is not symmetric about the zenith because
the distribution of ZL over the sky is not symmetric: the scattered
light will be smaller at the same airmass if the Sun is further below
the horizon, i.e. the middle of the night.  The scattered flux is
therefore minimized near UT$\sim$4, when the field is still at low
airmass and the brightest regions of the ZL are below the horizon. In
Figure \ref{fig:net_wavel}, we show the total combined effect of the
atmosphere on the ZL flux received from the target field:
\begin{eqnarray}
I_{\rm net}(\lambda,t,\chi,{\rm ZL})\!\!\!&=&\!\!\!I^{\rm R}_{\rm scat}(\lambda,t,\chi,{\rm ZL})
+ I^{\rm M}_{\rm scat}(\lambda,t,\chi,{\rm ZL})  \nonumber \\
&&\!\!\!\!\!\!\!\!  - I_{\rm (3h,-20d)}(\lambda,{\rm ZL}) ( 1-e^{-\tau_{\rm obs}(\lambda)\chi}).
\end{eqnarray}
Finally, from $I_{\rm net}(\lambda,{\rm ZL})$ we can calculate an {\it effective}
extinction for the ZL from our target field at the specific times at
which our observations occurred.  The effective extinction is defined
by the equation
\begin{equation}
I_{\rm net}(\lambda,t,\chi, {\rm ZL}) = (1-e^{\tau_{\rm eff}(\lambda, t)\chi})
I_{\rm (3h,-20d)}(\lambda, {\rm ZL})
\label{eq:Inet}
\end{equation}
The effective extinction is plotted in Figure \ref{fig:eff_tau_ut2},
and is specific to our target field, times of observation, observed
extinction, geographic latitude and longitude, and altitude.

The result which is applied to our ZL measurement from the modeling
discussed here is $\tau_{\rm eff}(\lambda,t)$, which corresponds to
$I_{\rm net}(\lambda,t,\chi,{\rm ZL})$ rather than $I_{\rm
scat}(\lambda,t,\chi, {\rm ZL})$.  The virtue of this approach is that
the absolute flux accuracy of the adopted ZL over the sky
(Fig. \ref{fig:ZLoverSky}) does not affect our results; only the
accuracy of the {\it relative} flux distribution over the sky matters.
In the regions of the sky which dominate the scattering for our
observations (solar elongations of $\epsilon>30^\circ$), the relative
flux errors for the ZL over the visible hemisphere of the sky are
$\le$5\% over large areas ($>30^\circ$), and better on small scales.
Such errors will propagate into final measurement of the ZL at the
level of $<1$\% at high airmass, and $<0.4$\% at low
airmass. Nevertheless, we note that the above-the-atmosphere value of
the ZL from Levasseur--Regourd \& Dumont (1980) agrees with our
measurement in our target field to within 2\%.

To evaluate the accuracy of our calculated values of $I_{\rm net}({\rm
ZL})$ (see Equation \ref{eq:Inet}), we estimate that the 
uncertainty in our scattering calculations is 8\% at the low zenith
angles ($<30^\circ$) where the bulk of our observations occur.  This
estimate is based on the comparisons between scattering models and
atmospheric measurements presented in Green \etal (1971), Dave (1964),
and Staude (1975), and is consistent with the uncertainties discussed
in WvB67, Ashburn (1954), and Sekera \& Ashburn (1953).  The
time--weighted average of $I_{\rm scat}({\rm ZL})$ over the course of
our observations is $\sim0.15\times I_{\rm (3h,-20d)}(\lambda, {\rm
ZL})$, so $I_{\rm scat}({\rm ZL})$ has an uncertainty of
1.2\% of the ZL flux in our target field.
The uncertainty in the observed extinction is much less than 1\% and
adds negligibly to this error.  See \S\ref{lco.resul} for further
discussion of the accuracies of the zodiacal light measurement.

We can also assess the accuracy of $\tau_{\rm eff}(\lambda)$ 
independently from our own data, as discussed in \S5.  Notice that
$I_{\rm net}({\rm ZL})$ changes with time in a way which is only weakly
dependent on wavelength.  A consistent solution for the ZL with both
wavelength and airmass will be strong confirmation of the accuracy of
the values for $\tau_{\rm eff}(\lambda,t)$ calculated here.

\subsection{Integrated Starlight}\label{append.isl}

Unlike the scattered ZL, the scattering which results from integrated
starlight (ISL) must be incorporated into our analysis of the
observed night sky spectrum as an absolute flux value.  We must
therefore first derive a spectrum for the ISL as a function of position
over the sky.  To do so, we have followed the method suggested by
Mattila (1980a, 1980b), which we briefly summarize  here.

The spatial and flux distribution of stars of all spectral types can be
described by exponential distributions perpendicular to the Galactic
plane (in the $z$ direction) and narrow Gaussian distributions in
intrinsic magnitude.  The mean emission per pc$^3$ from stars of type
$i$  as a function of distance from the Galactic plane, $z$, can
be written as
\begin{equation}
j_i(z) = j_i(0) e^{-|z|/h_i} = D_i(0) 10^{-0.4 M_i} e^{-|z|/h_i} ,
\end{equation}
in which $D_i(0)$ is the number density of stars per cubic parsec in
the plane, $h_i$ is the scale height of the vertical distribution,
 and $M_i$ is the mean absolute magnitude of the spectral
type $i$.  The observed flux is also attenuated by interstellar
extinction, which can be expressed by a two--component extinction law
characterized by a total extinction $a_0(\lambda)= a_1(\lambda)+
a_2(\lambda)$, with $a_1(\lambda):a_2(\lambda)$ in the ratio
$1.84:0.62$ (Neckel 1965).  The $z$--dependence of extinction can be written as
\begin{equation}
a(z,\lambda) = a_1(\lambda) / [1 + (z/20)^2] + a_2(\lambda) / [1 + (z/100)^2]
\end{equation}
for $z$ given in parsecs (Neckel 1965, Neckel 1980). 
We find a good
fit to the observed ISL by adopting standard values for $a_0(\lambda)$
(Zombeck 1990), scaled to $a_0(V)=1.5$ mag kpc$^{-1}$.  

\begin{figure}[t] 
\begin{center}
\hbox{
\includegraphics[width=3.0in]{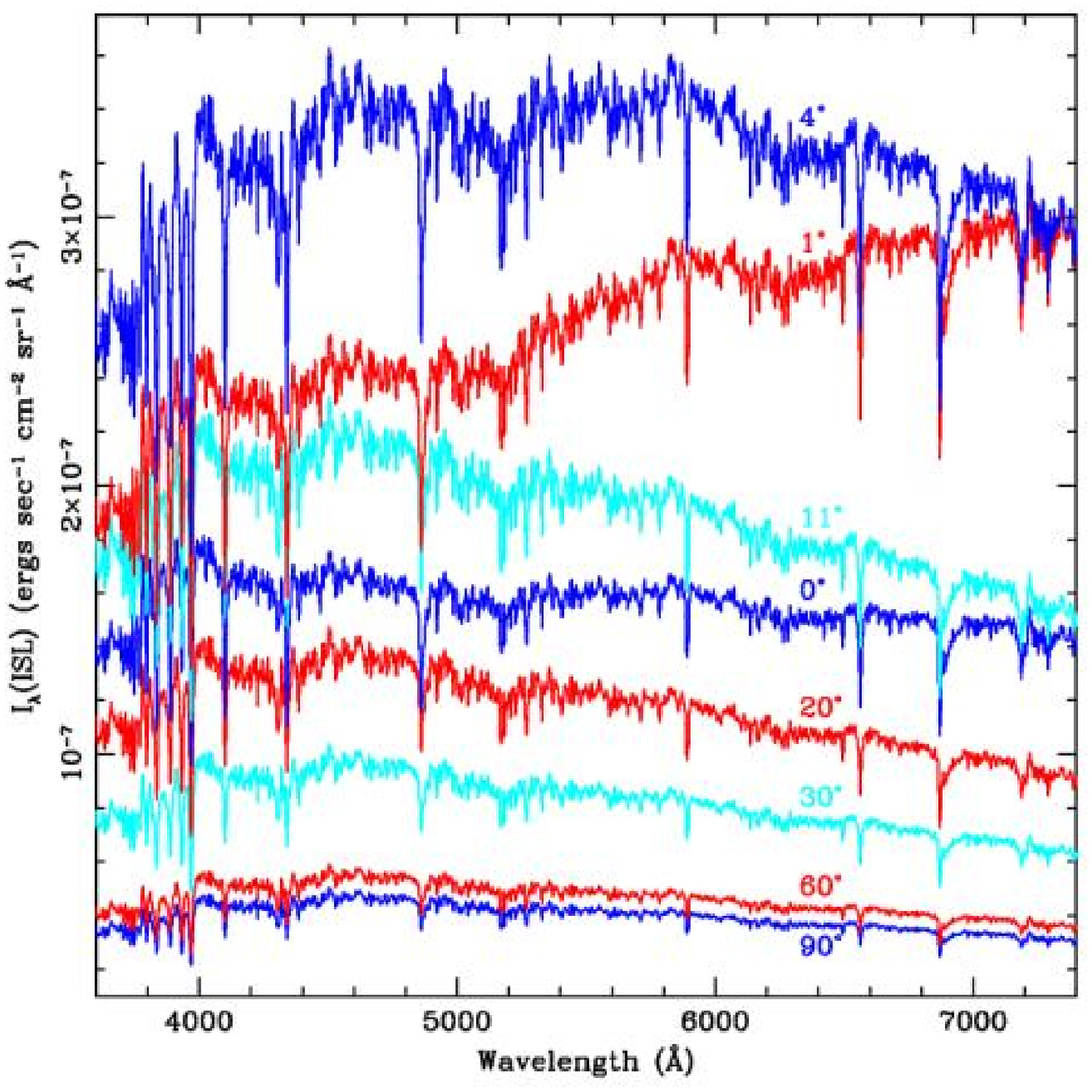}}
\caption{\footnotesize The mean integrated starlight (ISL) spectra as
a function of Galactic latitude (as labeled) produced by the model
described \S\ref{append.isl}. }
\label{fig:islspec}
\end{center}
\end{figure}

In cylindrical coordinates, the flux per unit solid angle (\escsa)
from stars fainter than $m_0$ along the line of sight at
Galactic latitude $b$ can be expressed by the volume integral
\begin{equation}
I_\lambda(b) = \sum{ {j_i(0) f_i(\lambda) \over 4 \pi \sin{b} }
		\int_{z_0}^{\infty} { e^{-z/h_i} 10^{-0.4
		A_\lambda(z)/\sin{b}} dz }},
\end{equation}
in which $r$ is the distance along the line of sight from the observer
in parsecs and $f_i$ is the spectral energy density of a star of type
$i$ in \esca. In the derivation of the above integral, the $1/4 \pi
r^2$ loss of flux from each star along the line of sight has canceled
with the $r^2 d\Omega $ in the volume integral, and we have changed
variables using the relation $r=z/\sin{b}$.  The lower limit of
integration is simply the distance modulus for stars of each type
corresponding to the bright magnitude cut--off, $m_0$, so that $z_0 = 10^{0.2
(m_0-M_i +5)}$ in parsecs. Finally, the extinction from $z$ to the
observer is 
\begin{eqnarray}
\lefteqn{A_\lambda(z)= \int_0^z{ a(z)dz} } \\
&& = 20 a_1(\lambda) \arctan (z/20)  + 100 a_2(\lambda) \arctan(z/100).\nonumber
\end{eqnarray}
Using 33 individual stellar types described by the parameters $M_i$,
$D_i$, and $h_i$ from Wainscoat \etal (1992), we obtain integrated
spectra which agree with the observed star counts at $V$ and $B$
(Roach \& Megill 1961, see also Allen 1973) to $m_0= 6\ V$ mag at
$|b|>5^\circ$ to better than 10\%, which is more than adequate for our
purposes.  The spectral energy densities for each stellar type,
$f_i(\lambda)$ were obtained from the STScI archive (Jacoby, Hunter, \&
Christian 1984) and have a resolution of roughly 4\AA. We include
stars by type with $m_0<6\ V$ mag from the SAO star catalog
by hand.  We felt this was necessary as the statistical variation in
stellar density on small scales around the solar neighborhood can have
a significant impact on the accuracy of the model, while variations
are apparently averaged out in stellar populations at large distances.
In Figure \ref{fig:islspec}, we plot the total integrated starlight
(ISL) with no bright magnitude cut off at $0^\circ <|b|< 90^\circ$.
The total flux at $|b|=90^\circ$ is roughly 20\tto{-9} \escsa, while
the flux near the plane is as high as 300\tto{-9} \escsa.
Interstellar extinction limits the ISL flux in the plane in our
model, probably more than is appropriate.  However the flux rises
rapidly even $1^\circ$ degree above the plane to more realistic
values. The limited sky area at $b=0^\circ$ precludes this from
impacting the accuracy of the models.

\begin{figure}[t]
\begin{center}
\includegraphics[width=3.0in]{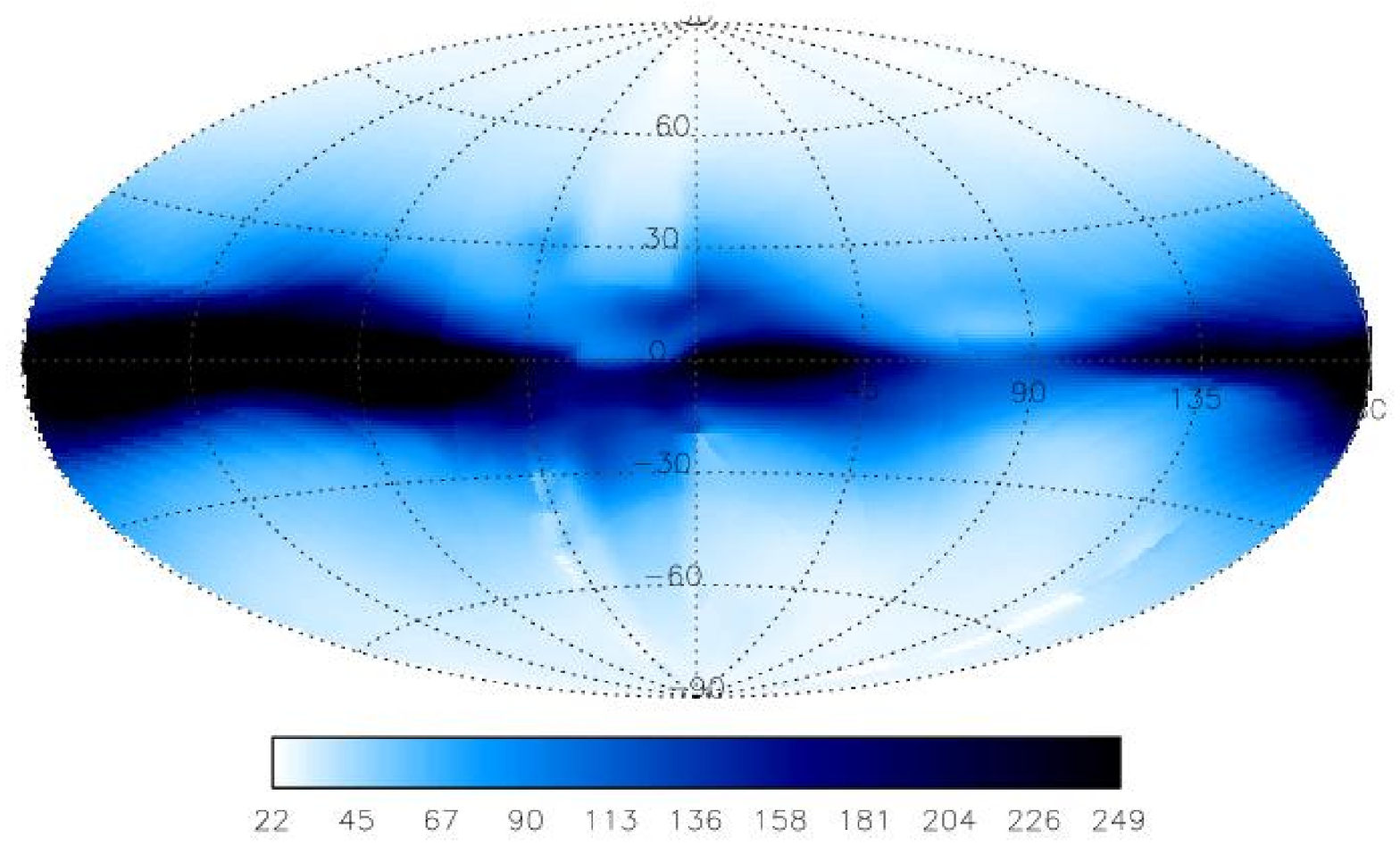}
\caption{\footnotesize Integrated starlight over the sky from star
counts at $V$. Flux units are 1\tto{-9} \escsa.}
\label{fig:ISLoverSky}
\end{center}
\end{figure}

\begin{figure*}[t] 
\begin{center}
\hbox{
\includegraphics[width=2.2in]{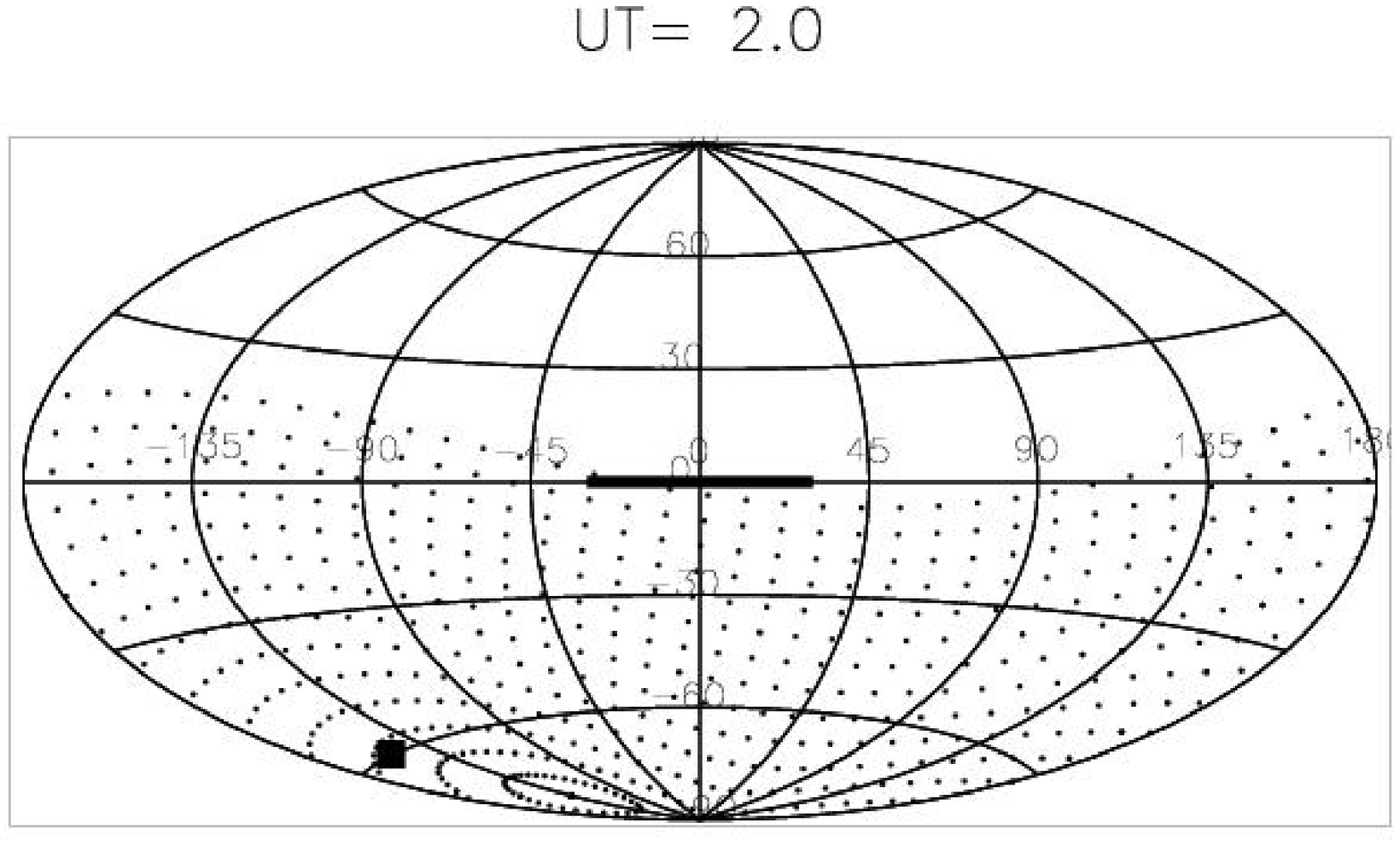}
\includegraphics[width=2.2in]{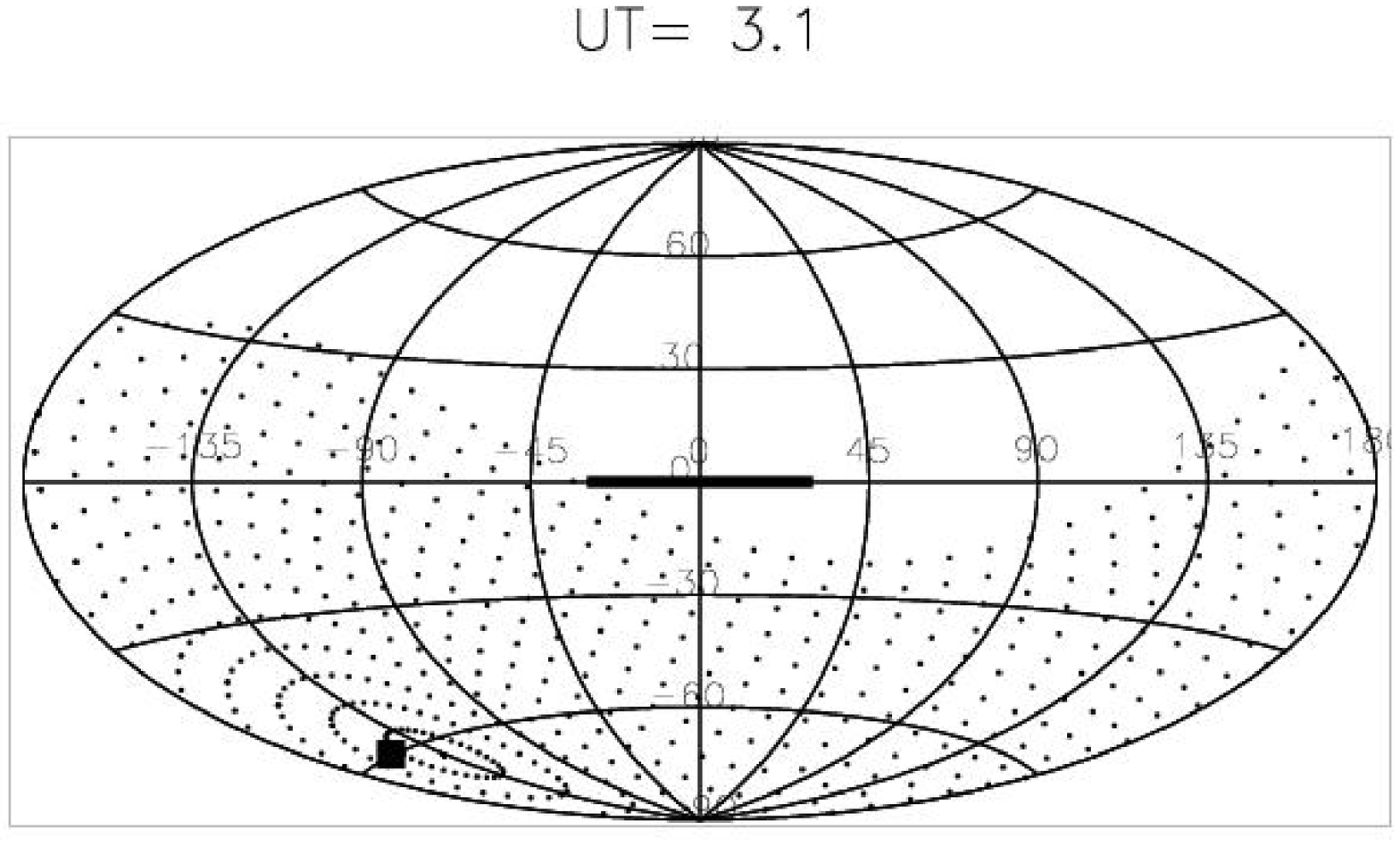}
\includegraphics[width=2.2in]{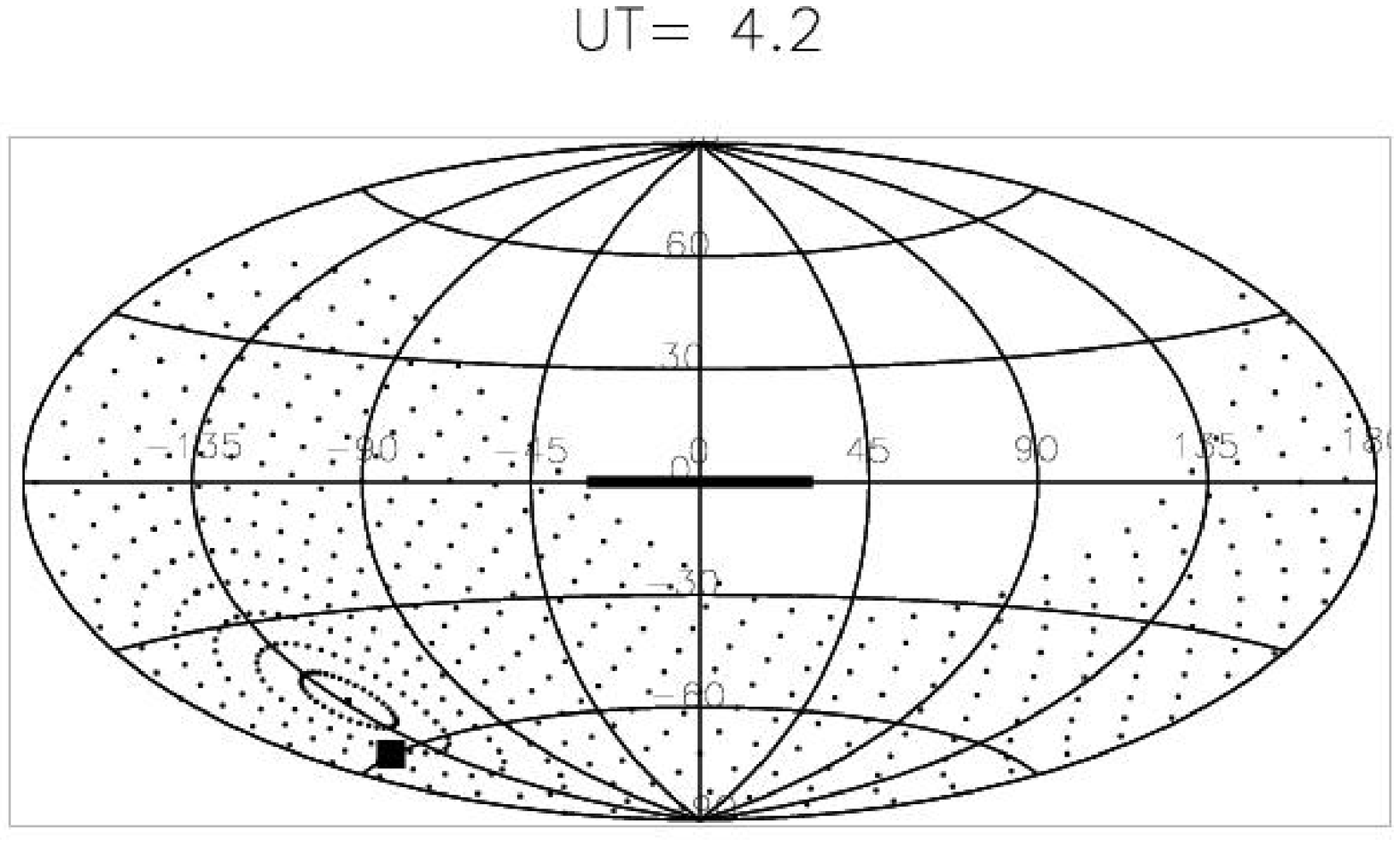}}
\hbox{
\includegraphics[width=2.2in]{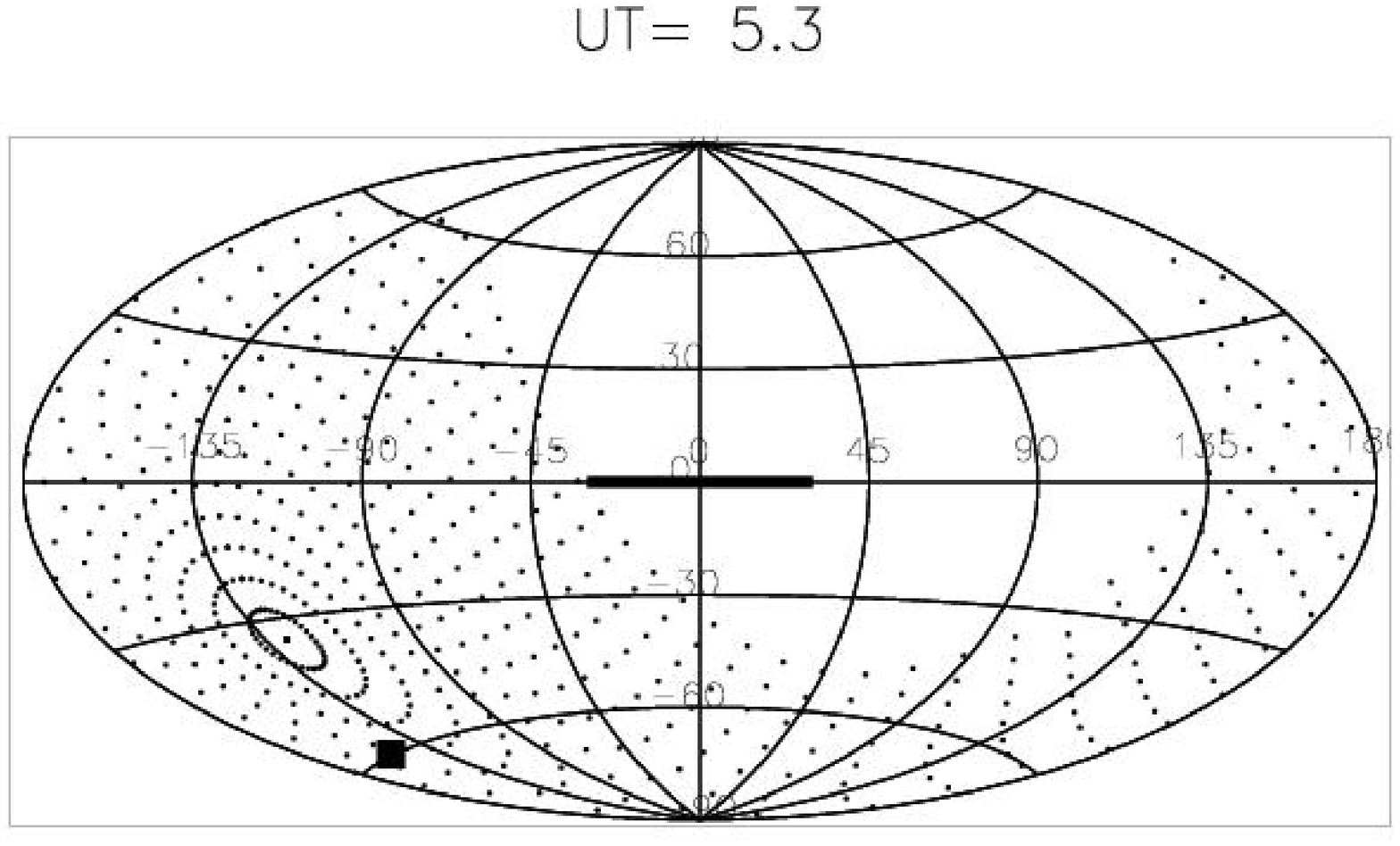}
\includegraphics[width=2.2in]{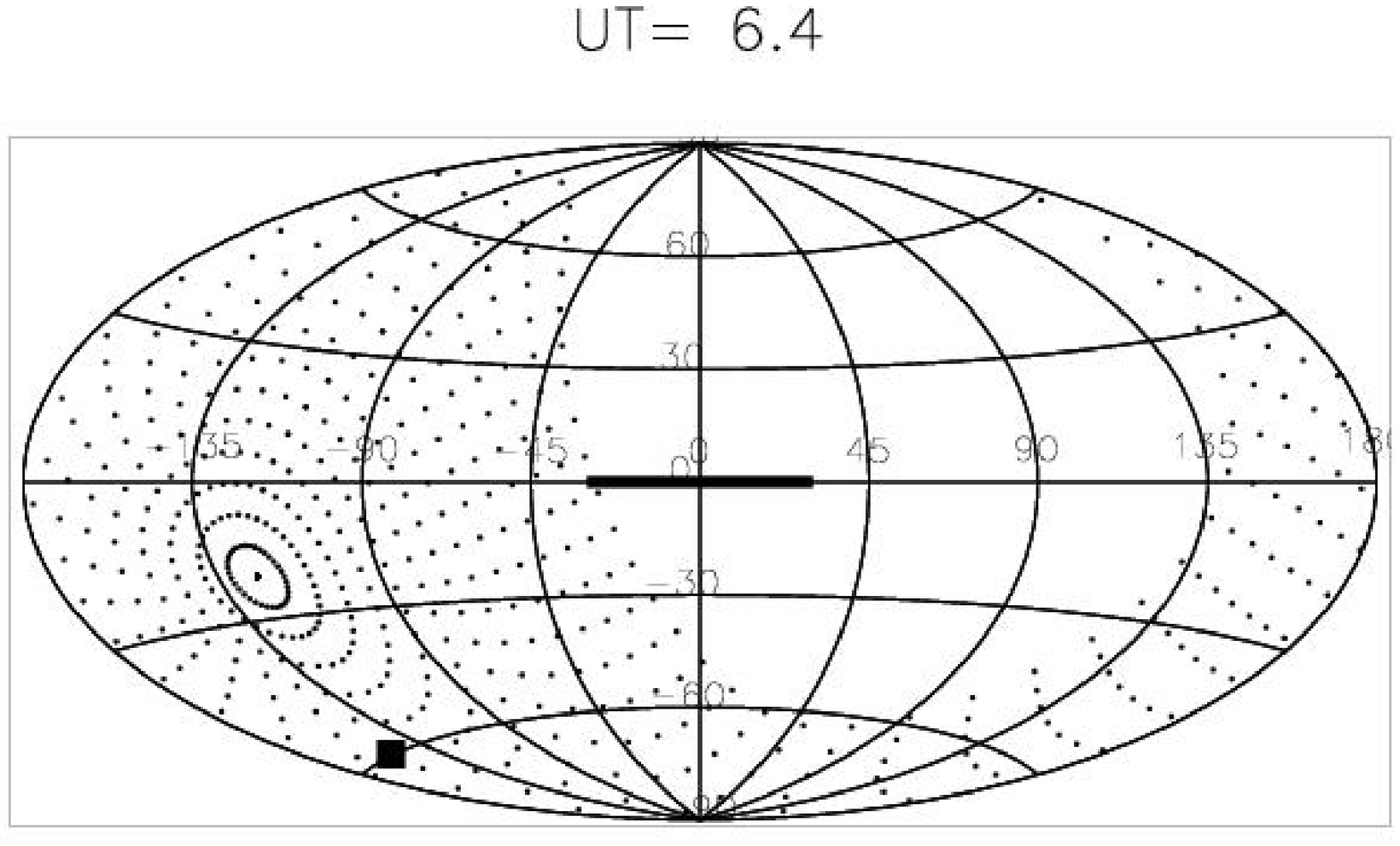}
\includegraphics[width=2.2in]{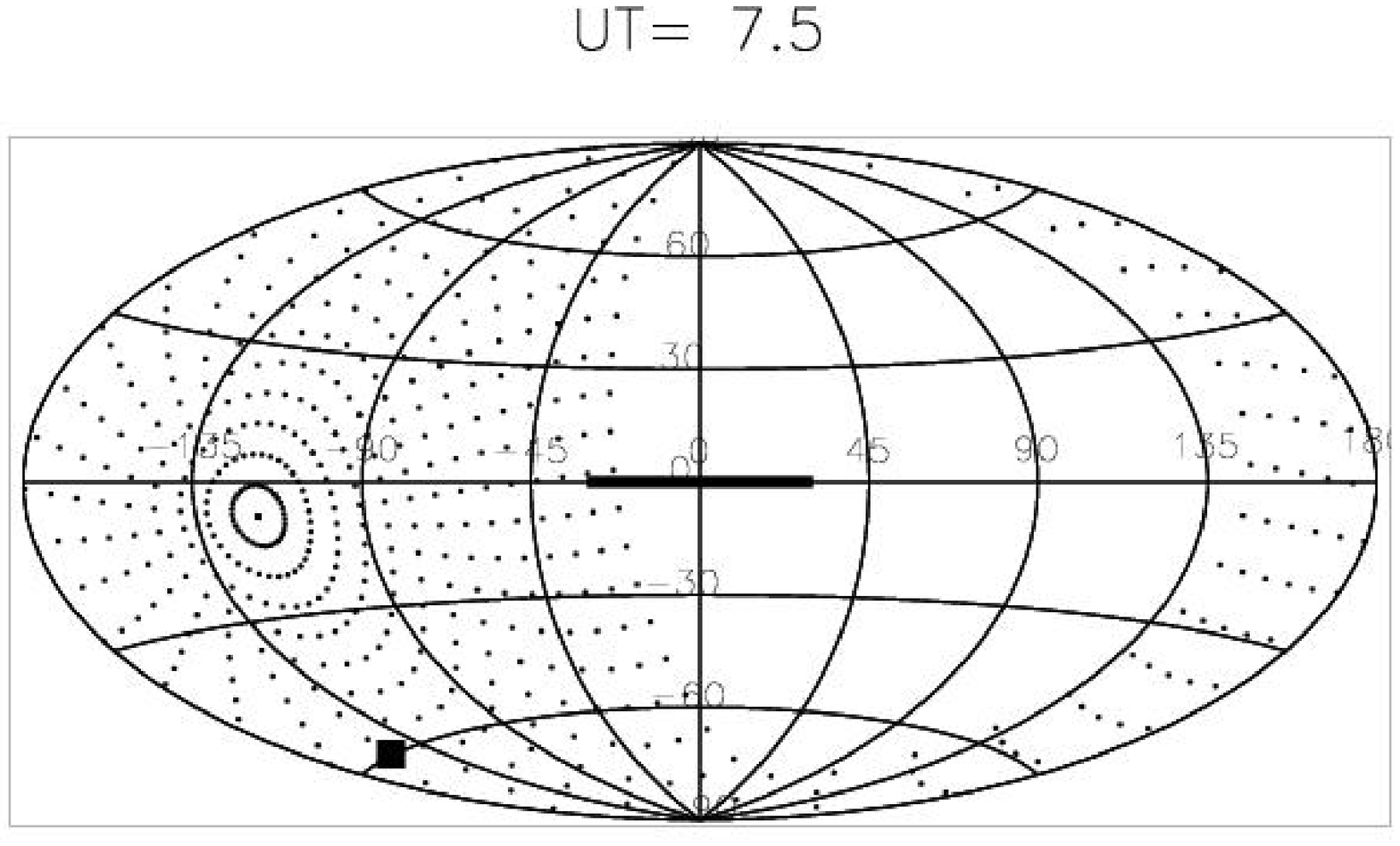}}
\caption{\footnotesize Integration pattern for the calculation of
scattered ISL flux.  Each plot is an Aitoff projection of the sky in
Galactic coordinates. The center of the Galaxy is at the center of
each plot ($l=0^\circ, b=0^\circ$). The Galactic plane within 30
degrees of the Galactic center is marked by the thick line.  Our
target field is indicated by the square.  The Galactic coordinates of
the zenith can be seen as the ``bullseye'' center of the integration
pattern.  The coordinates of the local horizon are shown by the edge
of the integration pattern.  The Galactic plane is running along the
horizon at the start of the night (UT=2.0 hr), and is perpendicular to
the horizon at the end of the night (UT=7.5 hr). The Galactic center is
never above the horizon.  See Figure \ref{fig:ecliptic_aitoff} for
further discussion.}
\label{fig:galaxy_aitoff}
\end{center}
\end{figure*}

\begin{figure}[t] 
\begin{center}
\hbox{
\includegraphics[width=2.75in,angle=-90]{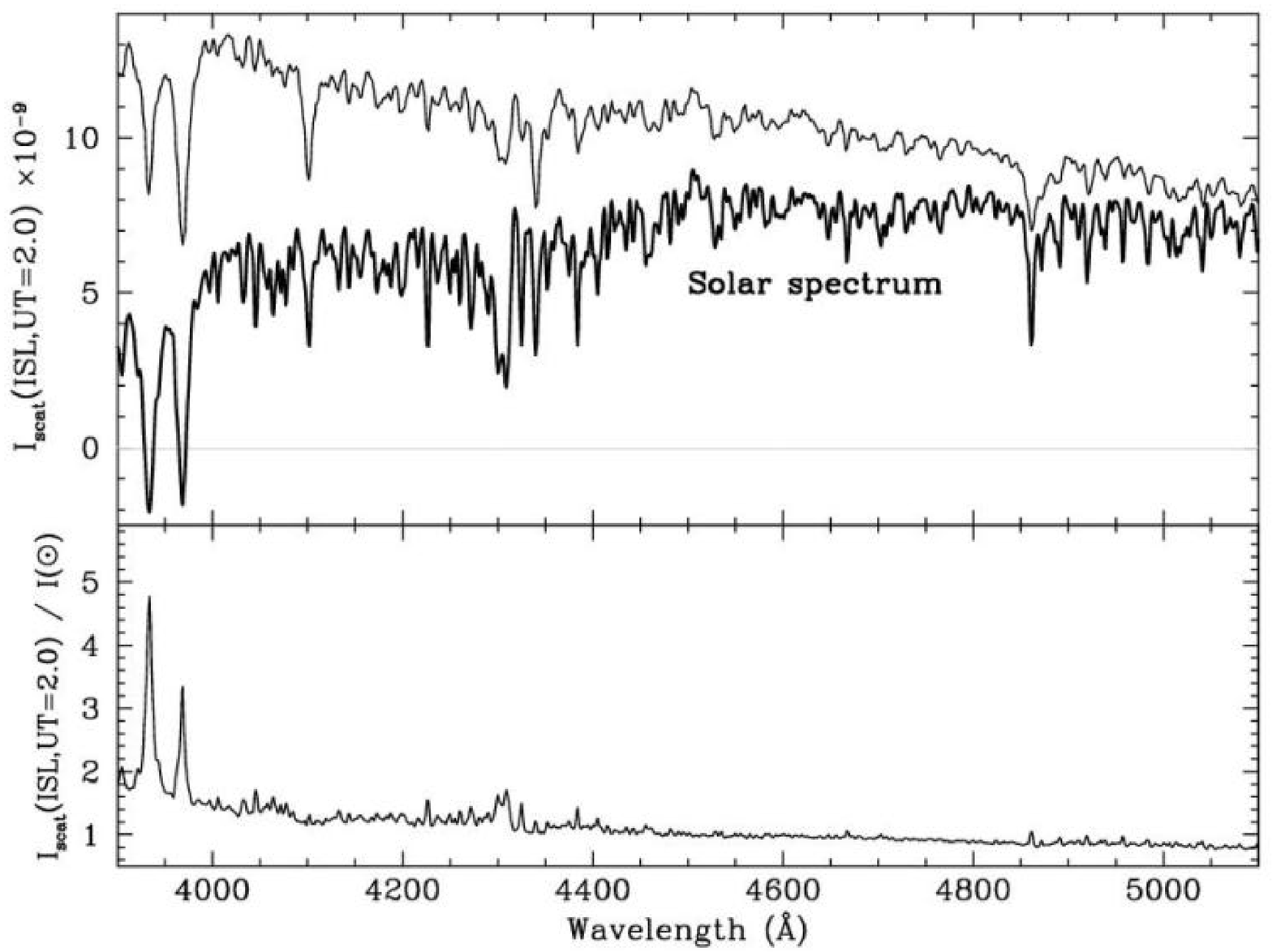}}
\caption{\footnotesize The upper plot shows the predicted spectrum of
the scattered ISL along the line of sight to the
EBL field at UT=2.0 hr for our observations. Units are \escsa. The
solar spectrum is also shown, scaled to the same flux and offset,
to allow  visual comparison of the spectral features.  The lower
plot shows the ratio of that spectrum to the solar spectrum,
normalized at 4600\AA, the center of our observed wavelength range.
Note that the total flux from the scattered ISL in this case is $<5$\%
of the total flux of the combined zodiacal light and airglow,
and the spectral features are weaker in the scattered ISL.}
\label{fig:isl_ut2_sun}
\end{center}
\end{figure}

\begin{figure}[t] 
\begin{center}
\hbox{
\includegraphics[width=2.75in,angle=-90]{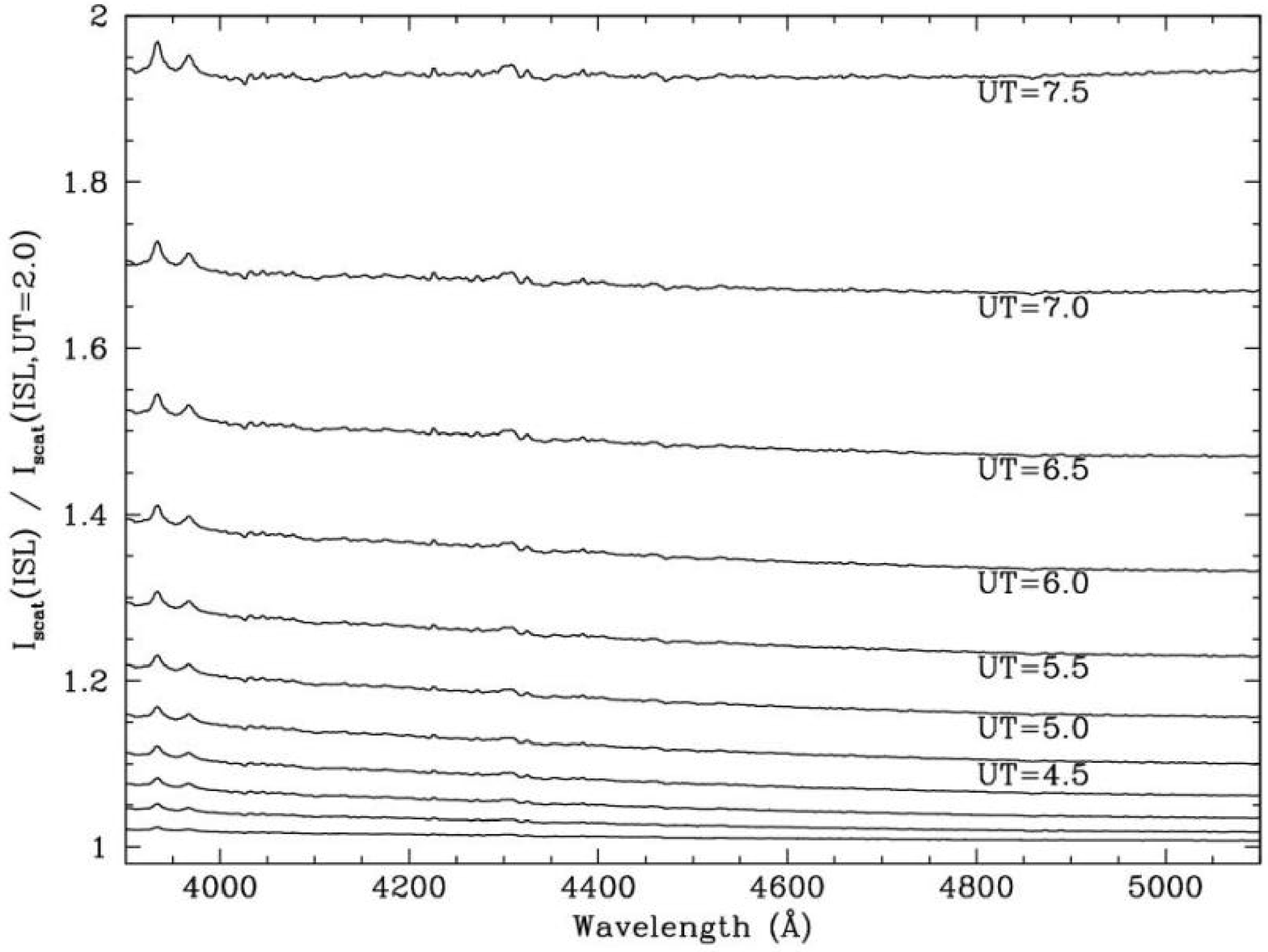}}
\caption{\footnotesize The ratio of the predicted scattered ISL
spectrum at the indicated UT on the night of our observations,
relative to the predicted spectrum at UT=2.0 hr.  The plot demonstrates
that relatively small changes ($<4$\%) occur in the strength of
individual spectral features over the night, while changes in
broad--band color and mean flux are significant.}
\label{fig:islut2_islutX}
\end{center}
\end{figure}

Figure \ref{fig:isl_ut2_sun} shows the total scattered ISL flux due to
Rayleigh and Mie scattering which contributes to observations of our
target field at the beginning of the observing night (UT=2.0 hr). The
total flux ($\sim 12\times10^{-9}$ \escsa) is roughly 12\% of the ZL
flux above the atmosphere in our target field. The ISL flux in the
observed sky spectrum will impact our measurement of the ZL flux only
if the ISL spectrum has the same features as the solar (zodiacal
light) spectrum.  In the lower plot of Figure \ref{fig:isl_ut2_sun},
we therefore plot ratio of the $I_{\rm scat}(\lambda,t,\chi,{\rm
ISL})$ to the solar spectrum, normalized at 4600\AA.  It is clear from
this plot that the scattered ISL and solar spectra differ by 5--10\%
at $>4500$\AA, but by a factor of 3--5 in the strength of spectral features
at less than 4500\AA.  In Figure \ref{fig:islut2_islutX}, we plot the
ratio of $I_{\rm scat}(\lambda,t, \chi, {\rm ISL})$ throughout the
night to $I_{\rm scat}(\lambda,t,\chi,{\rm ISL})$ at UT=2.0 hr. From this
plot it is clear that strength of spectral features changes only very
weakly throughout the night, by $<1$\% over the majority of the spectrum
and by $<4\%$ at 3900-4000\AA\ (CaI H \& K). The consistency of our ZL
measurement (\S\ref{lco.resul}) over the full wavelength range
3900-5100\AA\ is, therefore, a strong test of the accuracy of the
predicted contribution of scattered ISL.  As discussed in
\S\ref{lco.analy}, the predicted scattered ISL flux is entirely
consistent with our observations.  See \S\ref{lco.analy} for further
discussion.

Obviously, the model we describe above makes no allowance for
variation in the ISL with Galactic longitude.  For comparison, we show
in Figure \ref{fig:ISLoverSky} an Aitoff projection of the ISL from
star counts over the sky, which shows that the ISL has only minor
dependence on longitude at $b>20^\circ$.  At lower latitudes where the
variation is greater with longitude, the spectroscopic model we employ
does give a good approximation to the average ISL.  Because the
contribution to the scattering comes from a wide spread in longitude
(compare Figures \ref{fig:galaxy_aitoff} and \ref{fig:ISLoverSky}),
the mean value is adequate for our purposes.  To test this, we ran
simulations in which we maintained the mean ISL flux with latitude,
but varied the ISL flux with longitude by $\pm25$\%. We find that the
total scattered ISL is affected by less than 9\% throughout the night
due to longitudinal variations around the mean. 

The mean flux in our models for the ISL is consistent with the star
counts of Roach \& Megill (1961) to within $\pm 10$\% at both $V$ and
$B$.  As in the previous section, we estimate that the 
uncertainty in our scattering calculations is 8\% at the low zenith
angles ($<30^\circ$) where the bulk of our observations occur.
Combining these, we estimate the uncertainty in $I_{\rm
scat}(\lambda,t, \chi, {\rm ISL})$ to be 13\%.  As the relative mean
strength of spectral features in starlight is 0.6-4\% of the total ZL
flux in our target field at the $4000-5200$\AA, this corresponds to an
uncertainty of $<0.5$\%.

Any significant errors in our model, either in mean flux as might be
caused by longitudinal variations in the ISL, or in the spectral
energy distribution, would show up as inconsistencies in the solution
for the ZL flux as a function of wavelength. Furthermore, errors would
be worst at higher airmass, where Figure \ref{fig:galaxy_aitoff} shows
that the low-galactic-latitude sky has a greater impact on the
scattered ISL, the mean flux is greater, and the stellar-type mix is
more sensitive.  No such variations with wavelength are found, as 
we have discussed in \S\ref{lco.analy}.   See \S\ref{lco.resul} for further
discussion of the accuracies of the zodiacal light measurement.


\begin{references}
\hspace{1cm}

\reference{}{Allen, C.W.\ 1973, Astrophysical Quantities, (London: Athlone Press), 244}

\reference{}{Ashburn, E.V.\ 1954, J.\ Atm.\ Terrest.\ Phys., 5, 83}

\reference{}{Baldwin, J.H.\ \& Stone, R.P.S.\ 1984,  MN,  206, 241}

\reference{}{Beggs, D.W., Blackwell, D.E., Dewhirst, D.W., \& Wolstencroft, R.D.\ 1964, MN, 127, 329}

\reference{}{Bernstein, R.A., Freedman, W.L., Madore, B.F.\ 2002a, ApJ, in press (Paper I)}

\reference{}{Bernstein, R.A., Freedman, W.L., Madore, B.F.\ 2002c, ApJ, in press (Paper III)}

\reference{}{Berriman, G.B., Boggess, N.W., Hauser, M.G., Kelsall, T., Lisse, C.M., Moseley, S.H., Reach, W.T., \& Silverberg, R.F.\ 1994, ApJ, L63}

\reference{}{Brownlee, D.E.\ 1978, in  Cosmic Dust, ed.\ A.M.\ McDonnell (New York: Wiley and Sons), 295}

\reference{}{Chandrasekhar, S.\ 1950, Radiative Transfer, (Oxford: Oxford Univ. Press).}

\reference{}{Dave, J.V.\ 1964, J. Opt. Soc. America, 54, 307} 

\reference{}{de Bary, E.\  \& Bullrich K.\ 1964, J.\ Opt.\ Soc.\ America, 54, 1413} 

\reference{}{de Bary, E.\  1964 Appl.~Opt., 3,1293} 

\reference{}{Deepak, A., Green, A.E.S.\ 1970, Appl.~Opt., 9, 2362} 

\reference{}{Dermott, S.F., Jayaraman, S., Xu, Y.L., Grogan, K.\  \& Gustafson, B.A.S.\ 1996, in AIP Conference Proceedings No. 348, Unveiling the Cosmic Infrared Background, ed.\ E.\ Dwek (Woodburry, NY: AIP Press), 25}

\reference{}{Dube, R.R., Wickes, W.C.\  \& Wilkinson, D.T.\ 1979, ApJ, 232, 333}

\reference{}{Dumont, R.\ 1965, Ann. Astrophys., 28, 265}

\reference{}{East, I.R. \& Reay, N.K.\  1984, AA, 139, 512}

\reference{}{Elterman, L.\ 1966, \ao, 5, 1769} 

\reference{}{Fechtig, H., Hartung, J.B., Nagel, K., Neukum, G., \& Storzer, D.\ 1974, in Proc.\ Fifth Lunar Sci.\ Conf., Geochem.\ Cosmochem.\ Sup., 5, 3, 2463}

\reference{}{Filippenko, A.V.\ 1982, PASP, 94,  244}

\reference{}{Frey, A., Hofmann, W., Lemke, D.\ \& Thum, C.\ 1974,  A\&A, 36, 447}

\reference{}{Gilliland, R.L.\ 1992, in ASP Conference Series Vol.\ 23, Astronomical CCD Observing and Reduction Techniques, ed.\ S.B.\ Howell (San Fransisco: BookCrafters, Inc.), 68}

\reference{}{Girard, T.M., Grundy, W.M., Lopez, C.E., \& Van Altena, W.F.\ 1989, AJ, 98, 227}

\reference{}{Green, A.E.S., Deepak, A., Lipofsky, B.J.\ 1971, Appl.~Opt., 10, 1263}

\reference{}{Hamuy, M., Walker, A.R., Suntzeff, N.B., Gigoux, P., Heathcote, S.R.\  \& Phillips, M.M.\ 1992, PASP, 104, 553}

\reference{}{Hayes, D.S.\ \& Laytham, D.W.\ 1975, ApJ, 197, 587}

\reference{}{Hayes, D.S.\ 1985, in IAU Symposium No.\ 111, Calibration of Fundamental Quantities,  eds.\ D.S.\ Hayes, L.E.\ Pasinetti, \&  A.G.\ Davis Phillip (Reidel, Dordrecht), 225}

\reference{}{Jacoby, G.H., Hunter, D.A., Christian, C.A.\ 1984, ApJS, 56, 257}

\reference{}{Johnson, H.L.\ \& Harris III, D.L.\ 1954, ApJ, 120, 196}

\reference{}{Jones, A.V., Meier, R.R., Shefov, N.N. 1985, J. Atmosph. Terrest. Phys., 47, 623}

\reference{}{Kurucz, R.L., Furenlid, I., Brault, J., \& Testerman, L.\ 1984, Solar Flux Atlas from 296 to 1300 nm, National Solar Observatory Atlas No.\ 1 (Sunspot: National Solar Observatory)}

\reference{}{Leinert, Ch., Richter, I., Pitz, E., Planck, B.\ 1981, A\&A, 103, 177}

\reference{}{Leinert, Ch., \etal 1998, AAS, 127, 1}

\reference{}{Levasseur--Regourd, A.C.\ \& Dumont, R.\ 1980, AA, 84, 277}

\reference{}{Matsuura, S., Matsumoto, T., \& Matsuhara, H.\ 1995, Icarus 115, 199}

\reference{}{Mattila, K.\ 1980, AA\&S, 39, 53 (1980a)}

\reference{}{Mattila, K.\ 1980, A\&A, 82, 373 (1980b)}

\reference{}{Martin, C., Hurwitz, M.\  \&  Bowyer, S.\ 1991, ApJ, 379, 549}

\reference{}{Neckel, Th.\ 1965, ZA, 63, 221}

\reference{}{Neckel, Th.,  Klare, G. ,Sarcander, M.\ 1980, A\&AS, 42, 251}

\reference{}{Neckel, H.\ \& Labs, D.\ 1984, Sol.\ Phys., 90, 205}

\reference{}{Press, W., Teukolsky, S., Vetterling, W., \& Flannery, B.\ 1992, Numerical Recipes, (2d ed.; Cambridge: Cambridge Univ. Press)}

\reference{}{Reach, W.T., Abergel, A., Boulanger, F., D\'{e}sert, Perault, M., Bernard, J.P., Bloommaert, J., Desarsky, C.,Cesarsky, D., Metcalfe, L., Puget, J.L., Sibille, F., Vigroux, L.\ 1996, AA, 315, L381}

\reference{}{Renka, R.J.\ 1997, ACM Transactions on Mathematical Software, 23, 435}

\reference{}{Reynolds, R.J.\ 1990, in IAU Symposium No.\ 139, Galactic and Extragalactic Background Radiation, ed.\ S.\ Bowyer \& Ch.\ Leinert (Dordrecht: Kluwer), 159}

\reference{}{Richter, I., Leinert, Ch., \& Planck, B.\ 1982, AA, 110, 115} 

\reference{}{Roach, F.E.\ \& Meinel, A.\ 1955,  ApJ, 84, 120}

\reference{}{Roach, F.E.\ \& Meinel, A.\ 1961,  ApJ, 133, 228} 

\reference{}{R\"oser, S.\ \& Staude, J.\ 1978, AA, 67, 381}

\reference{}{Schubert, G.\ \& Walterscheid, R.L.\ 1999, in Allen's Astrophysical Quantities,  Ed.\ A.N.\ Cox, (Springer-Verlag: New York), 257}

\reference{}{Schiffer, R.\ 1985, AA, 148, 347}

\reference{}{Schmidtke, G.\ 1985, J. Atmosph. Terrest. Phys., 47, 147}

\reference{}{Sekera, Z.\ 1952, Tables Relating to Rayleigh Scattering of Light in the Atmosphere,  UCLA, Dept. of Meteorology, Sci. Report No. 3.}

\reference{}{Sekera, Z.\ \& Ashburn, E.V.\ 1953, Tables Relating to Rayleigh Scattering in the Atmosphere,  U.S. Navel Ordinance Test Station, Inyokern, Nav. Ord. Report 2061} 

\reference{}{Slanger, T.G.\ \& Huestis, D.L. 1981, J.~Geophys.~Res., 86, 3551}

\reference{}{Staude, H.J.\ 1975, A\&A, 39, 325}

\reference{}{Stone, R.P.S.\ \& Baldwin, J.H. 1983,  MN,  204, 347}

\reference{}{Taylor, B.J.\ 1984, ApJS, 54, 259}

\reference{}{van de Hulst, H.C.\ 1952, in The Atmospheres of Earth and Planets, Ed, G.P. Kuiper, (Univ of Chicago Press: Chicago), 52}

\reference{}{van Rhijn, P.J.\ 1924, Bull Astr.\ Inst.\ Netherlands, 2, 75}

\reference{}{van Rhijn, P.J.\ 1925, Pub.\ Gron.\ Ast.\ Obs., No.\ 43}

\reference{}{Wainscoat, R.J., Cohen, M., Volk, K., Walker, H.J.\ \& Schwartz, D.E.\ 1992, 83, 111}

\reference{}{Weinberg, J.L.\ 1964, Ann. Astrophys., 27, 718}

\reference{}{Weiss--Wrana, K.\ 1983, AA, 126, 240}

\reference{}{Wolstencroft, R.D.\ \& van Breda, I.G.\ 1967, ApJ, 147, 255} 

\reference{}{Zombeck, M.V.\ 1990, Handbook of Space Astronomy and Astrophysic, 2nd ed., (Cambridge: Cabridge Univ. Press), 104}

\end{references}
\end{document}